\input harvmac


%
%
\def\unredoffs{} \def\redoffs{\voffset=-.31truein\hoffset=-.48truein}
\def\speclscape{}
%
%
%
%
%

\newbox\leftpage \newdimen\fullhsize \newdimen\hstitle \newdimen\hsbody
\tolerance=1000\hfuzz=2pt
\catcode`\@=11 
\ifx\hyperdef\UNd@FiNeD\def\hyperdef#1#2#3#4{#4}\def\hyperref#1#2#3#4{#4}\fi
\def\bigans{b }
\def\answ{b }
%
\ifx\answ\bigans\message{(This will come out unreduced.}
\magnification=1200\unredoffs\baselineskip=16pt plus 2pt minus 1pt
\hsbody=\hsize \hstitle=\hsize 
\else\message{(This will be reduced.} \let\l@r=L
\magnification=1000\baselineskip=16pt plus 2pt minus 1pt \vsize=7truein
\redoffs \hstitle=8truein\hsbody=4.75truein\fullhsize=10truein\hsize=\hsbody
\output={\ifnum\pageno=0 
  \shipout\vbox{\speclscape{\hsize\fullhsize\makeheadline}
    \hbox to \fullhsize{\hfill\pagebody\hfill}}\advancepageno
  \else
  \almostshipout{\leftline{\vbox{\pagebody\makefootline}}}\advancepageno
  \fi}
\def\almostshipout#1{\if L\l@r \count1=1 \message{[\the\count0.\the\count1]}
      \global\setbox\leftpage=#1 \global\let\l@r=R
 \else \count1=2
  \shipout\vbox{\speclscape{\hsize\fullhsize\makeheadline}
      \hbox to\fullhsize{\box\leftpage\hfil#1}}  \global\let\l@r=L\fi}
\fi
%
\newcount\yearltd\yearltd=\year\advance\yearltd by -2000

\def\Title#1#2{\nopagenumbers\abstractfont\hsize=\hstitle\rightline{#1}%
\vskip 1in\centerline{\titlefont #2}\abstractfont\vskip .5in\pageno=0}
\def\Date#1{\vfill\leftline{#1}\tenpoint\supereject\global\hsize=\hsbody%
\footline={\hss\tenrm\hyperdef\hypernoname{page}\folio\folio\hss}}%
%

\def\draftmode{\message{ DRAFTMODE }\def\draftdate{{\rm preliminary draft:
\number\month/\number\day/\number\yearltd\ \ \hourmin}}%
\headline={\hfil\draftdate}\writelabels\baselineskip=20pt plus 2pt minus 2pt
 {\count255=\time\divide\count255 by 60 \xdef\hourmin{\number\count255}
  \multiply\count255 by-60\advance\count255 by\time
  \xdef\hourmin{\hourmin:\ifnum\count255<10 0\fi\the\count255}}}
\def\nolabels{\def\wrlabeL##1{}\def\eqlabeL##1{}\def\reflabeL##1{}}
\def\writelabels{\def\wrlabeL##1{\leavevmode\vadjust{\rlap{\smash%
{\line{{\escapechar=` \hfill\rlap{\sevenrm\hskip.03in\string##1}}}}}}}%
\def\eqlabeL##1{{\escapechar-1\rlap{\sevenrm\hskip.05in\string##1}}}%
\def\reflabeL##1{\noexpand\llap{\noexpand\sevenrm\string\string\string##1}}}
\nolabels
%
\global\newcount\secno \global\secno=0
\global\newcount\meqno \global\meqno=1
\def\s@csym{}
\def\newsec#1{\global\advance\secno by1%
{\toks0{#1}\message{(\the\secno. \the\toks0)}}%
\global\subsecno=0\eqnres@t\let\s@csym\secsym\xdef\secn@m{\the\secno}\noindent
{\bf\hyperdef\hypernoname{section}{\the\secno}{\the\secno.} #1}%
\writetoca{{\string\hyperref{}{section}{\the\secno}{\the\secno.}} {#1}}%
\par\nobreak\medskip\nobreak}
\def\eqnres@t{\xdef\secsym{\the\secno.}\global\meqno=1\bigbreak\bigskip}
\def\sequentialequations{\def\eqnres@t{\bigbreak}}\xdef\secsym{}
\global\newcount\subsecno \global\subsecno=0
\def\subsec#1{\global\advance\subsecno by1%
{\toks0{#1}\message{(\s@csym\the\subsecno. \the\toks0)}}%
\ifnum\lastpenalty>9000\else\bigbreak\fi
\noindent{\it\hyperdef\hypernoname{subsection}{\secn@m.\the\subsecno}%
{\secn@m.\the\subsecno.} #1}\writetoca{\string\quad
{\string\hyperref{}{subsection}{\secn@m.\the\subsecno}{\secn@m.\the\subsecno.}}
{#1}}\par\nobreak\medskip\nobreak}
\def\appendix#1#2{\global\meqno=1\global\subsecno=0\xdef\secsym{\hbox{#1.}}%
\bigbreak\bigskip\noindent{\bf Appendix \hyperdef\hypernoname{appendix}{#1}%
{#1.} #2}{\toks0{(#1. #2)}\message{\the\toks0}}%
\xdef\s@csym{#1.}\xdef\secn@m{#1}%
\writetoca{\string\hyperref{}{appendix}{#1}{Appendix {#1.}} {#2}}%
\par\nobreak\medskip\nobreak}
%
%
\def\checkm@de#1#2{\ifmmode{\def\f@rst##1{##1}\hyperdef\hypernoname{equation}%
{#1}{#2}}\else\hyperref{}{equation}{#1}{#2}\fi}
\def\eqnn#1{\DefWarn#1\xdef #1{(\noexpand\relax\noexpand\checkm@de%
{\s@csym\the\meqno}{\secsym\the\meqno})}%
\wrlabeL#1\writedef{#1\leftbracket#1}\global\advance\meqno by1}
\def\f@rst#1{\c@t#1a\em@ark}\def\c@t#1#2\em@ark{#1}
\def\eqna#1{\DefWarn#1\wrlabeL{#1$\{\}$}%
\xdef #1##1{(\noexpand\relax\noexpand\checkm@de%
{\s@csym\the\meqno\noexpand\f@rst{##1}}{\hbox{$\secsym\the\meqno##1$}})}
\writedef{#1\numbersign1\leftbracket#1{\numbersign1}}\global\advance\meqno by1}
\def\eqn#1#2{\DefWarn#1%
\xdef #1{(\noexpand\hyperref{}{equation}{\s@csym\the\meqno}%
{\secsym\the\meqno})}$$#2\eqno(\hyperdef\hypernoname{equation}%
{\s@csym\the\meqno}{\secsym\the\meqno})\eqlabeL#1$$%
\writedef{#1\leftbracket#1}\global\advance\meqno by1}
\def\xeqn{\expandafter\xe@n}\def\xe@n(#1){#1}
\def\xeqna#1{\expandafter\xe@n#1}
\def\eqns#1{(\e@ns #1{\hbox{}})}
\def\e@ns#1{\ifx\UNd@FiNeD#1\message{eqnlabel \string#1 is undefined.}%
\xdef#1{(?.?)}\fi{\let\hyperref=\relax\xdef\next{#1}}%
\ifx\next\em@rk\def\next{}\else%
\ifx\next#1\xeqn#1\else\def\n@xt{#1}\ifx\n@xt\next#1\else\xeqna#1\fi
\fi\let\next=\e@ns\fi\next}

\def\DefWarn#1{\ifx\UNd@FiNeD#1\else
\immediate\write16{*** WARNING: the label \string#1 is already defined ***}\fi}
%
\newskip\footskip\footskip14pt plus 1pt minus 1pt 
\def\footnotefont{\ninepoint}\def\f@t#1{\footnotefont #1\@foot}
\def\f@@t{\baselineskip\footskip\bgroup\footnotefont\aftergroup\@foot\let\next}
\setbox\strutbox=\hbox{\vrule height9.5pt depth4.5pt width0pt}
\global\newcount\ftno \global\ftno=0
\def\foot{\global\advance\ftno by1\def\foot@rg{\hyperref{}{footnote}%
{\the\ftno}{\the\ftno}\xdef\foot@rg{\noexpand\hyperdef\noexpand\hypernoname%
{footnote}{\the\ftno}{\the\ftno}}}\footnote{$^{\foot@rg}$}}
%
\newwrite\ftfile
\def\footend{\def\foot{\global\advance\ftno by1\chardef\wfile=\ftfile
\hyperref{}{footnote}{\the\ftno}{$^{\the\ftno}$}%
\ifnum\ftno=1\immediate\openout\ftfile=\jobname.fts\fi%
\immediate\write\ftfile{\noexpand\smallskip%
\noexpand\item{\noexpand\hyperdef\noexpand\hypernoname{footnote}
{\the\ftno}{f\the\ftno}:\ }\pctsign}\findarg}%
\def\footatend{\vfill\eject\immediate\closeout\ftfile{\parindent=20pt
\centerline{\bf Footnotes}\nobreak\bigskip\input \jobname.fts }}}
\def\footatend{}
%
%
\global\newcount\refno \global\refno=1
\newwrite\rfile
\def\ref{[\hyperref{}{reference}{\the\refno}{\the\refno}]\nref}
\def\nref#1{\DefWarn#1%
\xdef#1{[\noexpand\hyperref{}{reference}{\the\refno}{\the\refno}]}%
\writedef{#1\leftbracket#1}%
\ifnum\refno=1\immediate\openout\rfile=\jobname.refs\fi
\chardef\wfile=\rfile\immediate\write\rfile{\noexpand\item{[\noexpand\hyperdef%
\noexpand\hypernoname{reference}{\the\refno}{\the\refno}]\ }%
\reflabeL{#1\hskip.31in}\pctsign}\global\advance\refno by1\findarg}
\def\findarg#1#{\begingroup\obeylines\newlinechar=`\^^M\pass@rg}
{\obeylines\gdef\pass@rg#1{\writ@line\relax #1^^M\hbox{}^^M}%
\gdef\writ@line#1^^M{\expandafter\toks0\expandafter{\striprel@x #1}%
\edef\next{\the\toks0}\ifx\next\em@rk\let\next=\endgroup\else\ifx\next\empty%
\else\immediate\write\wfile{\the\toks0}\fi\let\next=\writ@line\fi\next\relax}}
\def\striprel@x#1{} \def\em@rk{\hbox{}}
\def\lref{\begingroup\obeylines\lr@f}
\def\lr@f#1#2{\DefWarn#1\gdef#1{\let#1=\UNd@FiNeD\ref#1{#2}}\endgroup\unskip}

\def\addref#1{\immediate\write\rfile{\noexpand\item{}#1}} 
\def\listrefs{\footatend\vfill\supereject\immediate\closeout\rfile\writestoppt
\baselineskip=\footskip\centerline{{\bf References}}\bigskip{\parindent=20pt%
\frenchspacing\escapechar=` \input \jobname.refs\vfill\eject}\nonfrenchspacing}
\def\startrefs#1{\immediate\openout\rfile=\jobname.refs\refno=#1}
\def\xref{\expandafter\xr@f}\def\xr@f[#1]{#1}
\def\refs#1{\count255=1[\r@fs #1{\hbox{}}]}
\def\r@fs#1{\ifx\UNd@FiNeD#1\message{reflabel \string#1 is undefined.}%
\nref#1{need to supply reference \string#1.}\fi%
\vphantom{\hphantom{#1}}{\let\hyperref=\relax\xdef\next{#1}}%
\ifx\next\em@rk\def\next{}%
\else\ifx\next#1\ifodd\count255\relax\xref#1\count255=0\fi%
\else#1\count255=1\fi\let\next=\r@fs\fi\next}
%

%
\newwrite\ffile\global\newcount\figno \global\figno=1
\def\fig{fig.~\hyperref{}{figure}{\the\figno}{\the\figno}\nfig}
\def\nfig#1{\DefWarn#1%
\xdef#1{fig.~\noexpand\hyperref{}{figure}{\the\figno}{\the\figno}}%
\writedef{#1\leftbracket fig.\noexpand~\xfig#1}%
\ifnum\figno=1\immediate\openout\ffile=\jobname.figs\fi\chardef\wfile=\ffile%
{\let\hyperref=\relax
\immediate\write\ffile{\noexpand\medskip\noexpand\item{Fig.\ %
\noexpand\hyperdef\noexpand\hypernoname{figure}{\the\figno}{\the\figno}. }
\reflabeL{#1\hskip.55in}\pctsign}}\global\advance\figno by1\findarg}
\def\listfigs{\vfill\eject\immediate\closeout\ffile{\parindent40pt
\baselineskip14pt\centerline{{\bf Figure Captions}}\nobreak\medskip
\escapechar=` \input \jobname.figs\vfill\eject}}
\def\xfig{\expandafter\xf@g}\def\xf@g fig.\penalty\@M\ {}
\def\figs#1{figs.~\f@gs #1{\hbox{}}}
\def\f@gs#1{{\let\hyperref=\relax\xdef\next{#1}}\ifx\next\em@rk\def\next{}\else
\ifx\next#1\xfig #1\else#1\fi\let\next=\f@gs\fi\next}
\def\figin{\epsfcheck\figin}\def\figins{\epsfcheck\figins}
\def\epsfcheck{\ifx\epsfbox\UNd@FiNeD
\message{(NO epsf.tex, FIGURES WILL BE IGNORED)}
\gdef\figin##1{\vskip2in}\gdef\figins##1{\hskip.5in}
\else\message{(FIGURES WILL BE INCLUDED)}%
\gdef\figin##1{##1}\gdef\figins##1{##1}\fi}
\def\DefWarn#1{}
\def\figinsert{\goodbreak\midinsert}
\def\ifig#1#2#3{\DefWarn#1\xdef#1{fig.~\noexpand\hyperref{}{figure}%
{\the\figno}{\the\figno}}\writedef{#1\leftbracket fig.\noexpand~\xfig#1}%
\figinsert\figin{\centerline{#3}}\medskip\centerline{\vbox{\baselineskip12pt
\advance\hsize by -1truein\noindent\wrlabeL{#1=#1}\footnotefont%
{\bf Fig.~\hyperdef\hypernoname{figure}{\the\figno}{\the\figno}:} #2}}
\bigskip\endinsert\global\advance\figno by1}
\newwrite\lfile
{\escapechar-1\xdef\pctsign{\string\%}\xdef\leftbracket{\string\{}
\xdef\rightbracket{\string\}}\xdef\numbersign{\string\#}}
\def\writedefs{\immediate\openout\lfile=\jobname.defs \def\writedef##1{%
{\let\hyperref=\relax\let\hyperdef=\relax\let\hypernoname=\relax
 \immediate\write\lfile{\string\def\string##1\rightbracket}}}}%
\def\writestop{\def\writestoppt{\immediate\write\lfile{\string\pageno
 \the\pageno\string\startrefs\leftbracket\the\refno\rightbracket
 \string\def\string\secsym\leftbracket\secsym\rightbracket
 \string\secno\the\secno\string\meqno\the\meqno}\immediate\closeout\lfile}}
\def\writestoppt{}\def\writedef#1{}
\def\seclab#1{\DefWarn#1%
\xdef #1{\noexpand\hyperref{}{section}{\the\secno}{\the\secno}}%
\writedef{#1\leftbracket#1}\wrlabeL{#1=#1}}
\def\subseclab#1{\DefWarn#1%
\xdef #1{\noexpand\hyperref{}{subsection}{\secn@m.\the\subsecno}%
{\secn@m.\the\subsecno}}\writedef{#1\leftbracket#1}\wrlabeL{#1=#1}}
\def\applab#1{\DefWarn#1%
\xdef #1{\noexpand\hyperref{}{appendix}{\secn@m}{\secn@m}}%
\writedef{#1\leftbracket#1}\wrlabeL{#1=#1}}
\newwrite\tfile \def\writetoca#1{}
\def\leaderfill{\leaders\hbox to 1em{\hss.\hss}\hfill}
\def\writetoc{\immediate\openout\tfile=\jobname.toc
   \def\writetoca##1{{\edef\next{\write\tfile{\noindent ##1
   \string\leaderfill {\string\hyperref{}{page}{\noexpand\number\pageno}%
                       {\noexpand\number\pageno}} \par}}\next}}}
\newread\ch@ckfile
\def\listtoc{\immediate\closeout\tfile\immediate\openin\ch@ckfile=\jobname.toc
\ifeof\ch@ckfile\message{no file \jobname.toc, no table of contents this pass}%
\else\closein\ch@ckfile\centerline{\bf Contents}\nobreak\medskip%
{\baselineskip=12pt\footnotefont\parskip=0pt\catcode`\@=11\input\jobname.toc
\catcode`\@=12\bigbreak\bigskip}\fi}
\catcode`\@=12 
%
\edef\tfontsize{\ifx\answ\bigans scaled\magstep3\else scaled\magstep4\fi}
\font\titlerm=cmr10 \tfontsize \font\titlerms=cmr7 \tfontsize
\font\titlermss=cmr5 \tfontsize \font\titlei=cmmi10 \tfontsize
\font\titleis=cmmi7 \tfontsize \font\titleiss=cmmi5 \tfontsize
\font\titlesy=cmsy10 \tfontsize \font\titlesys=cmsy7 \tfontsize
\font\titlesyss=cmsy5 \tfontsize \font\titleit=cmti10 \tfontsize
\skewchar\titlei='177 \skewchar\titleis='177 \skewchar\titleiss='177
\skewchar\titlesy='60 \skewchar\titlesys='60 \skewchar\titlesyss='60
\def\titlefont{\def\rm{\fam0\titlerm}
\textfont0=\titlerm \scriptfont0=\titlerms \scriptscriptfont0=\titlermss
\textfont1=\titlei \scriptfont1=\titleis \scriptscriptfont1=\titleiss
\textfont2=\titlesy \scriptfont2=\titlesys \scriptscriptfont2=\titlesyss
\textfont\itfam=\titleit \def\it{\fam\itfam\titleit}\rm}
 \ifx\answ\bigans\else scaled\magstep1\fi
\ifx\answ\bigans\def\abstractfont{\tenpoint}\else
\font\absit=cmti10 scaled \magstep1
\font\abssl=cmsl10 scaled \magstep1
\font\absrm=cmr10 scaled\magstep1 \font\absrms=cmr7 scaled\magstep1
\font\absrmss=cmr5 scaled\magstep1 \font\absi=cmmi10 scaled\magstep1
\font\absis=cmmi7 scaled\magstep1 \font\absiss=cmmi5 scaled\magstep1
\font\abssy=cmsy10 scaled\magstep1 \font\abssys=cmsy7 scaled\magstep1
\font\abssyss=cmsy5 scaled\magstep1 \font\absbf=cmbx10 scaled\magstep1
\skewchar\absi='177 \skewchar\absis='177 \skewchar\absiss='177
\skewchar\abssy='60 \skewchar\abssys='60 \skewchar\abssyss='60
\def\abstractfont{\def\rm{\fam0\absrm}
\textfont0=\absrm \scriptfont0=\absrms \scriptscriptfont0=\absrmss
\textfont1=\absi \scriptfont1=\absis \scriptscriptfont1=\absiss
\textfont2=\abssy \scriptfont2=\abssys \scriptscriptfont2=\abssyss
\textfont\itfam=\absit \def\it{\fam\itfam\absit}\def\footnotefont{\tenpoint}%
\textfont\slfam=\abssl \def\sl{\fam\slfam\abssl}%
\textfont\bffam=\absbf \def\bf{\fam\bffam\absbf}\rm}\fi
\def\tenpoint{\def\rm{\fam0\tenrm}
\textfont0=\tenrm \scriptfont0=\sevenrm \scriptscriptfont0=\fiverm
\textfont1=\teni  \scriptfont1=\seveni  \scriptscriptfont1=\fivei
\textfont2=\tensy \scriptfont2=\sevensy \scriptscriptfont2=\fivesy
\textfont\itfam=\tenit \def\it{\fam\itfam\tenit}\def\footnotefont{\ninepoint}%
\textfont\bffam=\tenbf \def\bf{\fam\bffam\tenbf}\def\sl{\fam\slfam\tensl}\rm}
\font\ninerm=cmr9 \font\sixrm=cmr6 \font\ninei=cmmi9 \font\sixi=cmmi6
\font\ninesy=cmsy9 \font\sixsy=cmsy6 \font\ninebf=cmbx9
\font\nineit=cmti9 \font\ninesl=cmsl9 \skewchar\ninei='177
\skewchar\sixi='177 \skewchar\ninesy='60 \skewchar\sixsy='60
\def\ninepoint{\def\rm{\fam0\ninerm}
\textfont0=\ninerm \scriptfont0=\sixrm \scriptscriptfont0=\fiverm
\textfont1=\ninei \scriptfont1=\sixi \scriptscriptfont1=\fivei
\textfont2=\ninesy \scriptfont2=\sixsy \scriptscriptfont2=\fivesy
\textfont\itfam=\ninei \def\it{\fam\itfam\nineit}\def\sl{\fam\slfam\ninesl}%
\textfont\bffam=\ninebf \def\bf{\fam\bffam\ninebf}\rm}
%
%

\hyphenation{anom-aly anom-alies coun-ter-term coun-ter-terms}
\def\inv{^{\raise.15ex\hbox{${\scriptscriptstyle -}$}\kern-.05em 1}}

\def\Dsl{\,\raise.15ex\hbox{/}\mkern-13.5mu D} 
\def\dsl{\raise.15ex\hbox{/}\kern-.57em\partial}

\def\lspace{\ifx\answ\bigans{}\else\qquad\fi}
\def\lbspace{\ifx\answ\bigans{}\else\hskip-.2in\fi} 
\def\boxeqn#1{\vcenter{\vbox{\hrule\hbox{\vrule\kern3pt\vbox{\kern3pt
	\hbox{${\displaystyle #1}$}\kern3pt}\kern3pt\vrule}\hrule}}}
\def\mbox#1#2{\vcenter{\hrule \hbox{\vrule height#2in
		\kern#1in \vrule} \hrule}}  
%

\def\darr#1{\raise1.5ex\hbox{$\leftrightarrow$}\mkern-16.5mu #1}

\def\roughly#1{\raise.3ex\hbox{$#1$\kern-.75em\lower1ex\hbox{$\sim$}}}

\def\smallfig#1#2#3{\DefWarn#1\xdef#1{fig.~\the\figno}
\writedef{#1\leftbracket fig.\noexpand~\the\figno}%
\figinsert\figin{\centerline{#3}}\medskip\centerline{\vbox{
\baselineskip12pt\advance\hsize by -1truein
\noindent\footnotefont{\bf Fig.~\the\figno:} #2}}
\endinsert\global\advance\figno by1}

\def\bb{
\font\tenmsb=msbm10
\font\sevenmsb=msbm7
\font\fivemsb=msbm5
\textfont1=\tenmsb
\scriptfont1=\sevenmsb
\scriptscriptfont1=\fivemsb
}

\input amssym

%
%
\ifx\pdfoutput\undefined
\input epsf
\def\fig#1{\epsfbox{#1.eps}}
\def\figscale#1#2{\epsfxsize=#2\epsfbox{#1.eps}}
%
%
\else
\def\fig#1{\pdfximage {#1.pdf}\pdfrefximage\pdflastximage}
\def\figscale#1#2{\pdfximage width#2 {#1.pdf}\pdfrefximage\pdflastximage}
\fi

\def\IZ{\relax\ifmmode\mathchoice
{\hbox{\cmss Z\kern-.4em Z}}{\hbox{\cmss Z\kern-.4em Z}} {\lower.9pt\hbox{\cmsss Z\kern-.4em Z}}
{\lower1.2pt\hbox{\cmsss Z\kern-.4em Z}}\else{\cmss Z\kern-.4em Z}\fi}

\newif\ifdraft\draftfalse
\newif\ifinter\interfalse
\ifdraft\draftmode\else\interfalse\fi
\def\journal#1&#2(#3){\unskip, \sl #1\ \bf #2 \rm(19#3) }
\def\andjournal#1&#2(#3){\sl #1~\bf #2 \rm (19#3) }

\def\frac#1#2{{#1\over#2}}

\def\inbar{\,\vrule height1.5ex width.4pt depth0pt}
\def\IC{\relax\hbox{$\inbar\kern-.3em{\rm C}$}}
\def\IR{\relax{\rm I\kern-.18em R}}
\def\IP{\relax{\rm I\kern-.18em P}}
\def\Z{{\bf Z}}

%
%


%
\catcode`\@=11
\def\slash#1{\mathord{\mathpalette\c@ncel{#1}}}
\overfullrule=0pt

\def\underrel#1\over#2{\mathrel{\mathop{\kern\z@#1}\limit\ell_{#2}}}

\catcode`\@=12


%


\def\[{[}
\def\]{]}

\def\comment#1{ }

\def\g{\gamma}

%
\def\draftnote#1{\ifdraft{\baselineskip2ex
                 \vbox{\kern1em\hrule\hbox{\vrule\kern1em\vbox{\kern1ex
                 \noindent \underbar{NOTE}: #1
             \vskip1ex}\kern1em\vrule}\hrule}}\fi}
\def\internote#1{\ifinter{\baselineskip2ex
                 \vbox{\kern1em\hrule\hbox{\vrule\kern1em\vbox{\kern1ex
                 \noindent \underbar{Internal Note}: #1
             \vskip1ex}\kern1em\vrule}\hrule}}\fi}

%
%



%
%
%
%

%

\def\inv{^{-1}}
\def\pt{\partial}


\def\b{\beta}

\def\1{{\ds 1}}
\def\R{\hbox{$\bb R$}}
\def\C{\hbox{$\bb C$}}

\def\Z{\hbox{$\bb Z$}}

\def\P{\hbox{$\bb P$}}

\def\S{\hbox{$\bb S$}}

\newfam\frakfam
\font\teneufm=eufm10
\font\seveneufm=eufm7
\font\fiveeufm=eufm5
\textfont\frakfam=\teneufm
\scriptfont\frakfam=\seveneufm
\scriptscriptfont\frakfam=\fiveeufm
\def\frak{\fam\frakfam \teneufm}

\lref\NiarchosAH{
  V.~Niarchos,
  ``Seiberg dualities and the 3d/4d connection,''
JHEP {\bf 1207}, 075 (2012).
[arXiv:1205.2086 [hep-th]].
}

\lref\AharonyGP{
  O.~Aharony,
  ``IR duality in d = 3 N=2 supersymmetric USp(2N(c)) and U(N(c)) gauge theories,''
Phys.\ Lett.\ B {\bf 404}, 71 (1997).
[hep-th/9703215].
}

\lref\AffleckAS{
  I.~Affleck, J.~A.~Harvey and E.~Witten,
  ``Instantons and (Super)Symmetry Breaking in (2+1)-Dimensions,''
Nucl.\ Phys.\ B {\bf 206}, 413 (1982)..
}

\lref\IntriligatorID{
  K.~A.~Intriligator and N.~Seiberg,
  ``Duality, monopoles, dyons, confinement and oblique confinement in supersymmetric SO(N(c)) gauge theories,''
Nucl.\ Phys.\ B {\bf 444}, 125 (1995).
[hep-th/9503179].
}

\lref\PasquettiFJ{
  S.~Pasquetti,
  ``Factorisation of N = 2 Theories on the Squashed 3-Sphere,''
JHEP {\bf 1204}, 120 (2012).
[arXiv:1111.6905 [hep-th]].
}

\lref\HarveyIT{
  J.~A.~Harvey,
 ``TASI 2003 lectures on anomalies,''
[hep-th/0509097].
}

\lref\BeemMB{
  C.~Beem, T.~Dimofte and S.~Pasquetti,
  ``Holomorphic Blocks in Three Dimensions,''
[arXiv:1211.1986 [hep-th]].
}

\lref\RastelliTBZ{
  L.~Rastelli and S.~S.~Razamat,
  ``The supersymmetric index in four dimensions,''
[arXiv:1608.02965 [hep-th]].
}

\lref\DiPietroBCA{
  L.~Di Pietro and Z.~Komargodski,
  ``Cardy formulae for SUSY theories in $d =$ 4 and $d =$ 6,''
JHEP {\bf 1412}, 031 (2014).
[arXiv:1407.6061 [hep-th]].
}

\lref\SeibergPQ{
  N.~Seiberg,
  ``Electric - magnetic duality in supersymmetric nonAbelian gauge theories,''
Nucl.\ Phys.\ B {\bf 435}, 129 (1995).
[hep-th/9411149].
}

\lref\AharonyBX{
  O.~Aharony, A.~Hanany, K.~A.~Intriligator, N.~Seiberg and M.~J.~Strassler,
  ``Aspects of N=2 supersymmetric gauge theories in three-dimensions,''
Nucl.\ Phys.\ B {\bf 499}, 67 (1997).
[hep-th/9703110].
}

\lref\IntriligatorNE{
  K.~A.~Intriligator and P.~Pouliot,
  ``Exact superpotentials, quantum vacua and duality in supersymmetric SP(N(c)) gauge theories,''
Phys.\ Lett.\ B {\bf 353}, 471 (1995).
[hep-th/9505006].
}

\lref\KarchUX{
  A.~Karch,
  ``Seiberg duality in three-dimensions,''
Phys.\ Lett.\ B {\bf 405}, 79 (1997).
[hep-th/9703172].
}

\lref\SafdiRE{
  B.~R.~Safdi, I.~R.~Klebanov and J.~Lee,
  ``A Crack in the Conformal Window,''
[arXiv:1212.4502 [hep-th]].
}

\lref\SchweigertTG{
  C.~Schweigert,
  ``On moduli spaces of flat connections with nonsimply connected structure group,''
Nucl.\ Phys.\ B {\bf 492}, 743 (1997).
[hep-th/9611092].
}

\lref\GiveonZN{
  A.~Giveon and D.~Kutasov,
  ``Seiberg Duality in Chern-Simons Theory,''
Nucl.\ Phys.\ B {\bf 812}, 1 (2009).
[arXiv:0808.0360 [hep-th]].
}

\lref\Spiridonov{
  Spiridonov, V.~P.,
  ``Aspects of elliptic hypergeometric functions,''
[arXiv:1307.2876 [math.CA]].
}

\lref\GaiottoBE{
  D.~Gaiotto, G.~W.~Moore and A.~Neitzke,
  ``Framed BPS States,''
[arXiv:1006.0146 [hep-th]].
}

\lref\AldayRS{
  L.~F.~Alday, M.~Bullimore and M.~Fluder,
  ``On S-duality of the Superconformal Index on Lens Spaces and 2d TQFT,''
JHEP {\bf 1305}, 122 (2013).
[arXiv:1301.7486 [hep-th]].
}

\lref\ArdehaliBLA{
  A.~Arabi Ardehali,
  ``High-temperature asymptotics of supersymmetric partition functions,''
JHEP {\bf 1607}, 025 (2016).
[arXiv:1512.03376 [hep-th]].
}

\lref\RazamatJXA{
  S.~S.~Razamat and M.~Yamazaki,
  ``S-duality and the N=2 Lens Space Index,''
[arXiv:1306.1543 [hep-th]].
}

\lref\NiarchosAH{
  V.~Niarchos,
  ``Seiberg dualities and the 3d/4d connection,''
JHEP {\bf 1207}, 075 (2012).
[arXiv:1205.2086 [hep-th]].
}

\lref\almost{
  A.~Borel, R.~Friedman, J.~W.~Morgan,
  ``Almost commuting elements in compact Lie groups,''
arXiv:math/9907007.
}

\lref\BobevKZA{
  N.~Bobev, M.~Bullimore and H.~C.~Kim,
  ``Supersymmetric Casimir Energy and the Anomaly Polynomial,''
JHEP {\bf 1509}, 142 (2015).
[arXiv:1507.08553 [hep-th]].
}

\lref\KapustinJM{
  A.~Kapustin and B.~Willett,
  ``Generalized Superconformal Index for Three Dimensional Field Theories,''
[arXiv:1106.2484 [hep-th]].
}

\lref\AharonyGP{
  O.~Aharony,
  ``IR duality in d = 3 N=2 supersymmetric USp(2N(c)) and U(N(c)) gauge theories,''
Phys.\ Lett.\ B {\bf 404}, 71 (1997).
[hep-th/9703215].
}

\lref\FestucciaWS{
  G.~Festuccia and N.~Seiberg,
  ``Rigid Supersymmetric Theories in Curved Superspace,''
JHEP {\bf 1106}, 114 (2011).
[arXiv:1105.0689 [hep-th]].
}

\lref\RomelsbergerEG{
  C.~Romelsberger,
  ``Counting chiral primaries in N = 1, d=4 superconformal field theories,''
Nucl.\ Phys.\ B {\bf 747}, 329 (2006).
[hep-th/0510060].
}

\lref\KapustinKZ{
  A.~Kapustin, B.~Willett and I.~Yaakov,
  ``Exact Results for Wilson Loops in Superconformal Chern-Simons Theories with Matter,''
JHEP {\bf 1003}, 089 (2010).
[arXiv:0909.4559 [hep-th]].
}

\lref\DolanQI{
  F.~A.~Dolan and H.~Osborn,
  ``Applications of the Superconformal Index for Protected Operators and q-Hypergeometric Identities to N=1 Dual Theories,''
Nucl.\ Phys.\ B {\bf 818}, 137 (2009).
[arXiv:0801.4947 [hep-th]].
}

\lref\GaddeIA{
  A.~Gadde and W.~Yan,
  ``Reducing the 4d Index to the $S^3$ Partition Function,''
JHEP {\bf 1212}, 003 (2012).
[arXiv:1104.2592 [hep-th]].
}

\lref\DolanRP{
  F.~A.~H.~Dolan, V.~P.~Spiridonov and G.~S.~Vartanov,
  ``From 4d superconformal indices to 3d partition functions,''
Phys.\ Lett.\ B {\bf 704}, 234 (2011).
[arXiv:1104.1787 [hep-th]].
}

\lref\ImamuraUW{
  Y.~Imamura,
 ``Relation between the 4d superconformal index and the $S^3$ partition function,''
JHEP {\bf 1109}, 133 (2011).
[arXiv:1104.4482 [hep-th]].
}

\lref\BeemYN{
  C.~Beem and A.~Gadde,
  ``The $N=1$ superconformal index for class $S$ fixed points,''
JHEP {\bf 1404}, 036 (2014).
[arXiv:1212.1467 [hep-th]].
}

\lref\HamaEA{
  N.~Hama, K.~Hosomichi and S.~Lee,
  ``SUSY Gauge Theories on Squashed Three-Spheres,''
JHEP {\bf 1105}, 014 (2011).
[arXiv:1102.4716 [hep-th]].
}

\lref\GaddeEN{
  A.~Gadde, L.~Rastelli, S.~S.~Razamat and W.~Yan,
  ``On the Superconformal Index of N=1 IR Fixed Points: A Holographic Check,''
JHEP {\bf 1103}, 041 (2011).
[arXiv:1011.5278 [hep-th]].
}

\lref\EagerHX{
  R.~Eager, J.~Schmude and Y.~Tachikawa,
  ``Superconformal Indices, Sasaki-Einstein Manifolds, and Cyclic Homologies,''
[arXiv:1207.0573 [hep-th]].
}

\lref\AffleckAS{
  I.~Affleck, J.~A.~Harvey and E.~Witten,
  ``Instantons and (Super)Symmetry Breaking in (2+1)-Dimensions,''
Nucl.\ Phys.\ B {\bf 206}, 413 (1982)..
}

\lref\SeibergPQ{
  N.~Seiberg,
  ``Electric - magnetic duality in supersymmetric nonAbelian gauge theories,''
Nucl.\ Phys.\ B {\bf 435}, 129 (1995).
[hep-th/9411149].
}

\lref\BahDG{
  I.~Bah, C.~Beem, N.~Bobev and B.~Wecht,
  ``Four-Dimensional SCFTs from M5-Branes,''
JHEP {\bf 1206}, 005 (2012).
[arXiv:1203.0303 [hep-th]].
}

\lref\debult{
  F.~van~de~Bult,
  ``Hyperbolic Hypergeometric Functions,''
University of Amsterdam Ph.D. thesis
}

\lref\OhmoriAMP{
  K.~Ohmori, H.~Shimizu, Y.~Tachikawa and K.~Yonekura,
  ``Anomaly polynomial of general $6d$ SCFTs,''
PTEP {\bf 2014}, 103B07 (2014).
[arXiv:1408.5572 [hep-th]].}

\lref\Shamirthesis{
  I.~Shamir,
  ``Aspects of three dimensional Seiberg duality,''
  M. Sc. thesis submitted to the Weizmann Institute of Science, April 2010.
  }

\lref\slthreeZ{
  J.~Felder, A.~Varchenko,
  ``The elliptic gamma function and $SL(3,Z) \times Z^3$,'' $\;\;$
[arXiv:math/0001184].
}

\lref\BeniniNC{
  F.~Benini, T.~Nishioka and M.~Yamazaki,
  ``4d Index to 3d Index and 2d TQFT,''
Phys.\ Rev.\ D {\bf 86}, 065015 (2012).
[arXiv:1109.0283 [hep-th]].
}

\lref\GaiottoWE{
  D.~Gaiotto,
  ``N=2 dualities,''
  JHEP {\bf 1208}, 034 (2012).
  [arXiv:0904.2715 [hep-th]].
}

\lref\SpiridonovZA{
  V.~P.~Spiridonov and G.~S.~Vartanov,
  ``Elliptic Hypergeometry of Supersymmetric Dualities,''
Commun.\ Math.\ Phys.\  {\bf 304}, 797 (2011).
[arXiv:0910.5944 [hep-th]].
}

\lref\BeniniMF{
  F.~Benini, C.~Closset and S.~Cremonesi,
  ``Comments on 3d Seiberg-like dualities,''
JHEP {\bf 1110}, 075 (2011).
[arXiv:1108.5373 [hep-th]].
}

\lref\BeniniGI{
  F.~Benini, S.~Benvenuti and Y.~Tachikawa,
  ``Webs of five-branes and N=2 superconformal field theories,''
JHEP {\bf 0909}, 052 (2009).
[arXiv:0906.0359 [hep-th]].
}

\lref\ClossetVP{
  C.~Closset, T.~T.~Dumitrescu, G.~Festuccia, Z.~Komargodski and N.~Seiberg,
  ``Comments on Chern-Simons Contact Terms in Three Dimensions,''
JHEP {\bf 1209}, 091 (2012).
[arXiv:1206.5218 [hep-th]].
}

\lref\SpiridonovHF{
  V.~P.~Spiridonov and G.~S.~Vartanov,
  ``Elliptic hypergeometry of supersymmetric dualities II. Orthogonal groups, knots, and vortices,''
[arXiv:1107.5788 [hep-th]].
}

\lref\RazamatQFA{
  S.~S.~Razamat,
  ``On the $\cal{N} =$ 2 superconformal index and eigenfunctions of the elliptic RS model,''
Lett.\ Math.\ Phys.\  {\bf 104}, 673 (2014).
[arXiv:1309.0278 [hep-th]].
}

\lref\SpiridonovWW{
  V.~P.~Spiridonov and G.~S.~Vartanov,
  ``Elliptic hypergeometric integrals and 't Hooft anomaly matching conditions,''
JHEP {\bf 1206}, 016 (2012).
[arXiv:1203.5677 [hep-th]].
}

\lref\HeckmanXDL{
  J.~J.~Heckman, P.~Jefferson, T.~Rudelius and C.~Vafa,
  ``Punctures for Theories of Class ${\cal{S}}_\Gamma$,''
[arXiv:1609.01281 [hep-th]].
}

\lref\DimoftePY{
  T.~Dimofte, D.~Gaiotto and S.~Gukov,
  ``3-Manifolds and 3d Indices,''
[arXiv:1112.5179 [hep-th]].
}

\lref\DumitrescuLTQ{
  T.~T.~Dumitrescu,
  ``An introduction to supersymmetric field theories in curved space,''
[arXiv:1608.02957 [hep-th]].
}

\lref\KimWB{
  S.~Kim,
  ``The Complete superconformal index for N=6 Chern-Simons theory,''
Nucl.\ Phys.\ B {\bf 821}, 241 (2009), [Erratum-ibid.\ B {\bf 864}, 884 (2012)].
[arXiv:0903.4172 [hep-th]].
}

\lref\WillettGP{
  B.~Willett and I.~Yaakov,
  ``N=2 Dualities and Z Extremization in Three Dimensions,''
[arXiv:1104.0487 [hep-th]].
}

\lref\BeniniNOA{
  F.~Benini and A.~Zaffaroni,
  ``A topologically twisted index for three-dimensional supersymmetric theories,''
JHEP {\bf 1507}, 127 (2015).
[arXiv:1504.03698 [hep-th]].
}

\lref\ImamuraSU{
  Y.~Imamura and S.~Yokoyama,
  ``Index for three dimensional superconformal field theories with general R-charge assignments,''
JHEP {\bf 1104}, 007 (2011).
[arXiv:1101.0557 [hep-th]].
}

\lref\BershadskyVM{
  M.~Bershadsky, A.~Johansen, V.~Sadov and C.~Vafa,
  ``Topological reduction of 4-d SYM to 2-d sigma models,''
Nucl.\ Phys.\ B {\bf 448}, 166 (1995).
[hep-th/9501096].
}

\lref\FreedYA{
  D.~S.~Freed, G.~W.~Moore and G.~Segal,
  ``The Uncertainty of Fluxes,''
Commun.\ Math.\ Phys.\  {\bf 271}, 247 (2007).
[hep-th/0605198].
}

\lref\HwangQT{
  C.~Hwang, H.~Kim, K.~-J.~Park and J.~Park,
  ``Index computation for 3d Chern-Simons matter theory: test of Seiberg-like duality,''
JHEP {\bf 1109}, 037 (2011).
[arXiv:1107.4942 [hep-th]].
}

\lref\GreenDA{
  D.~Green, Z.~Komargodski, N.~Seiberg, Y.~Tachikawa and B.~Wecht,
  ``Exactly Marginal Deformations and Global Symmetries,''
JHEP {\bf 1006}, 106 (2010).
[arXiv:1005.3546 [hep-th]].
}

\lref\IntriligatorJJ{
  K.~A.~Intriligator and B.~Wecht,
  ``The Exact superconformal R symmetry maximizes a,''
Nucl.\ Phys.\ B {\bf 667}, 183 (2003).
[hep-th/0304128].
}

\lref\IntriligatorID{
  K.~A.~Intriligator and N.~Seiberg,
  ``Duality, monopoles, dyons, confinement and oblique confinement in supersymmetric SO(N(c)) gauge theories,''
Nucl.\ Phys.\ B {\bf 444}, 125 (1995).
[hep-th/9503179].
}

\lref\SeibergNZ{
  N.~Seiberg and E.~Witten,
  ``Gauge dynamics and compactification to three-dimensions,''
In *Saclay 1996, The mathematical beauty of physics* 333-366.
[hep-th/9607163].
}

\lref\KinneyEJ{
  J.~Kinney, J.~M.~Maldacena, S.~Minwalla and S.~Raju,
  ``An Index for 4 dimensional super conformal theories,''
  Commun.\ Math.\ Phys.\  {\bf 275}, 209 (2007).
  [hep-th/0510251].
}

\lref\NakayamaUR{
  Y.~Nakayama,
  ``Index for supergravity on AdS(5) x T**1,1 and conifold gauge theory,''
Nucl.\ Phys.\ B {\bf 755}, 295 (2006).
[hep-th/0602284].
}

\lref\GaddeKB{
  A.~Gadde, E.~Pomoni, L.~Rastelli and S.~S.~Razamat,
  ``S-duality and 2d Topological QFT,''
JHEP {\bf 1003}, 032 (2010).
[arXiv:0910.2225 [hep-th]].
}

\lref\GaddeTE{
  A.~Gadde, L.~Rastelli, S.~S.~Razamat and W.~Yan,
  ``The Superconformal Index of the $E_6$ SCFT,''
JHEP {\bf 1008}, 107 (2010).
[arXiv:1003.4244 [hep-th]].
}

\lref\AharonyCI{
  O.~Aharony and I.~Shamir,
  ``On $O(N_c)$ d=3 N=2 supersymmetric QCD Theories,''
JHEP {\bf 1112}, 043 (2011).
[arXiv:1109.5081 [hep-th]].
}

\lref\GiveonSR{
  A.~Giveon and D.~Kutasov,
  ``Brane dynamics and gauge theory,''
Rev.\ Mod.\ Phys.\  {\bf 71}, 983 (1999).
[hep-th/9802067].
}

\lref\MaruyoshiTQK{
  K.~Maruyoshi and J.~Song,
  ``The Full Superconformal Index of the Argyres-Douglas Theory,''
[arXiv:1606.05632 [hep-th]].
}

\lref\SpiridonovQV{
  V.~P.~Spiridonov and G.~S.~Vartanov,
  ``Superconformal indices of ${\cal N}=4$ SYM field theories,''
Lett.\ Math.\ Phys.\  {\bf 100}, 97 (2012).
[arXiv:1005.4196 [hep-th]].
}
\lref\GaddeUV{
  A.~Gadde, L.~Rastelli, S.~S.~Razamat and W.~Yan,
  ``Gauge Theories and Macdonald Polynomials,''
Commun.\ Math.\ Phys.\  {\bf 319}, 147 (2013).
[arXiv:1110.3740 [hep-th]].
}

\lref\AlvarezGaumeIG{
  L.~Alvarez-Gaume and E.~Witten,
 ``Gravitational Anomalies,''
Nucl.\ Phys.\ B {\bf 234}, 269 (1984).
}

\lref\KapustinGH{
  A.~Kapustin,
  ``Seiberg-like duality in three dimensions for orthogonal gauge groups,''
[arXiv:1104.0466 [hep-th]].
}
\lref\ClossetSXA{
  C.~Closset and I.~Shamir,
  ``The ${\cal N}=1$ Chiral Multiplet on $T^2\times S^2$ and Supersymmetric Localization,''
JHEP {\bf 1403}, 040 (2014).
[arXiv:1311.2430 [hep-th]].
}

\lref\orthogpaper{O. Aharony, S. S. Razamat, N.~Seiberg and B.~Willett, 
``3d dualities from 4d dualities for orthogonal groups,''
[arXiv:1307.0511 [hep-th]].
}

\lref\readinglines{
  O.~Aharony, N.~Seiberg and Y.~Tachikawa,
  ``Reading between the lines of four-dimensional gauge theories,''
[arXiv:1305.0318 [hep-th]].
}

\lref\WittenNV{
  E.~Witten,
  ``Supersymmetric index in four-dimensional gauge theories,''
Adv.\ Theor.\ Math.\ Phys.\  {\bf 5}, 841 (2002).
[hep-th/0006010].
}

\lref\GaddeUV{
  A.~Gadde, L.~Rastelli, S.~S.~Razamat and W.~Yan,
  ``Gauge Theories and Macdonald Polynomials,''
Commun.\ Math.\ Phys.\  {\bf 319}, 147 (2013).
[arXiv:1110.3740 [hep-th]].
}

\lref\GaddeIK{
  A.~Gadde, L.~Rastelli, S.~S.~Razamat and W.~Yan,
  ``The 4d Superconformal Index from q-deformed 2d Yang-Mills,''
Phys.\ Rev.\ Lett.\  {\bf 106}, 241602 (2011).
[arXiv:1104.3850 [hep-th]].
}

\lref\GaiottoXA{
  D.~Gaiotto, L.~Rastelli and S.~S.~Razamat,
  ``Bootstrapping the superconformal index with surface defects,''
JHEP {\bf 1301}, 022 (2013).
[arXiv:1207.3577 [hep-th]].
}

\lref\GaiottoUQ{
  D.~Gaiotto and S.~S.~Razamat,
  ``Exceptional Indices,''
JHEP {\bf 1205}, 145 (2012).
[arXiv:1203.5517 [hep-th]].
}

\lref\RazamatUV{
  S.~S.~Razamat,
  ``On a modular property of N=2 superconformal theories in four dimensions,''
JHEP {\bf 1210}, 191 (2012).
[arXiv:1208.5056 [hep-th]].
}

\lref\noumi{
  Y.~Komori, M.~Noumi, J.~Shiraishi,
  ``Kernel Functions for Difference Operators of Ruijsenaars Type and Their Applications,''
SIGMA 5 (2009), 054.
[arXiv:0812.0279 [math.QA]].
}

\lref\SpirWarnaar{
  V.~P.~Spiridonov and S.~O.~Warnaar,
  ``Inversions of integral operators and elliptic beta integrals on root systems,''
Adv. Math. 207 (2006), 91-132
[arXiv:math/0411044].
}

\lref\RazamatJXA{
  S.~S.~Razamat and M.~Yamazaki,
  ``S-duality and the N=2 Lens Space Index,''
[arXiv:1306.1543 [hep-th]].
}

\lref\RazamatOPA{
  S.~S.~Razamat and B.~Willett,
  ``Global Properties of Supersymmetric Theories and the Lens Space,''
[arXiv:1307.4381 [hep-th]].
}

\lref\GaddeTE{
  A.~Gadde, L.~Rastelli, S.~S.~Razamat and W.~Yan,
  ``The Superconformal Index of the $E_6$ SCFT,''
JHEP {\bf 1008}, 107 (2010).
[arXiv:1003.4244 [hep-th]].
}

\lref\deBult{
  F.~J.~van~de~Bult,
  ``An elliptic hypergeometric integral with $W(F_4)$ symmetry,''
The Ramanujan Journal, Volume 25, Issue 1 (2011)
[arXiv:0909.4793[math.CA]].
}

\lref\MaruyoshiCAF{
  K.~Maruyoshi and J.~Yagi,
  ``Surface defects as transfer matrices,''
[arXiv:1606.01041 [hep-th]].
}

\lref\GaddeKB{
  A.~Gadde, E.~Pomoni, L.~Rastelli and S.~S.~Razamat,
  ``S-duality and 2d Topological QFT,''
JHEP {\bf 1003}, 032 (2010).
[arXiv:0910.2225 [hep-th]].
}

\lref\ArgyresCN{
  P.~C.~Argyres and N.~Seiberg,
  ``S-duality in N=2 supersymmetric gauge theories,''
JHEP {\bf 0712}, 088 (2007).
[arXiv:0711.0054 [hep-th]].
}

\lref\SpirWarnaar{
  V.~P.~Spiridonov and S.~O.~Warnaar,
  ``Inversions of integral operators and elliptic beta integrals on root systems,''
Adv. Math. 207 (2006), 91-132
[arXiv:math/0411044].
}

\lref\GaiottoHG{
  D.~Gaiotto, G.~W.~Moore and A.~Neitzke,
  ``Wall-crossing, Hitchin Systems, and the WKB Approximation,''
[arXiv:0907.3987 [hep-th]].
}

\lref\RuijsenaarsVQ{
  S.~N.~M.~Ruijsenaars and H.~Schneider,
  ``A New Class Of Integrable Systems And Its Relation To Solitons,''
Annals Phys.\  {\bf 170}, 370 (1986).
}

\lref\RuijsenaarsPP{
  S.~N.~M.~Ruijsenaars,
  ``Complete Integrability Of Relativistic Calogero-moser Systems And Elliptic Function Identities,''
Commun.\ Math.\ Phys.\  {\bf 110}, 191 (1987).
}

\lref\HallnasNB{
  M.~Hallnas and S.~Ruijsenaars,
  ``Kernel functions and Baecklund transformations for relativistic Calogero-Moser and Toda systems,''
J.\ Math.\ Phys.\  {\bf 53}, 123512 (2012).
}

\lref\kernelA{
S.~Ruijsenaars,
  ``Elliptic integrable systems of Calogero-Moser type: Some new results on joint eigenfunctions'', in Proceedings of the 2004 Kyoto Workshop on "Elliptic integrable systems", (M. Noumi, K. Takasaki, Eds.), Rokko Lectures in Math., no. 18, Dept. of Math., Kobe Univ.
}

\lref\ellRSreview{
Y.~Komori and S.~Ruijsenaars,
  ``Elliptic integrable systems of Calogero-Moser type: A survey'', in Proceedings of the 2004 Kyoto Workshop on "Elliptic integrable systems", (M. Noumi, K. Takasaki, Eds.), Rokko Lectures in Math., no. 18, Dept. of Math., Kobe Univ.
}

\lref\langmann{
E.~Langmann,
  ``An explicit solution of the (quantum) elliptic Calogero-Sutherland model'', [arXiv:math-ph/0407050].
}

\lref\TachikawaWI{
  Y.~Tachikawa,
  ``4d partition function on $S^1 \times S^3$ and 2d Yang-Mills with nonzero area,''
PTEP {\bf 2013}, 013B01 (2013).
[arXiv:1207.3497 [hep-th]].
}

\lref\MinahanFG{
  J.~A.~Minahan and D.~Nemeschansky,
  ``An N=2 superconformal fixed point with E(6) global symmetry,''
Nucl.\ Phys.\ B {\bf 482}, 142 (1996).
[hep-th/9608047].
}

\lref\AldayKDA{
  L.~F.~Alday, M.~Bullimore, M.~Fluder and L.~Hollands,
  ``Surface defects, the superconformal index and q-deformed Yang-Mills,''
[arXiv:1303.4460 [hep-th]].
}

\lref\FukudaJR{
  Y.~Fukuda, T.~Kawano and N.~Matsumiya,
  ``5D SYM and 2D q-Deformed YM,''
Nucl.\ Phys.\ B {\bf 869}, 493 (2013).
[arXiv:1210.2855 [hep-th]].
}

\lref\XieHS{
  D.~Xie,
  ``General Argyres-Douglas Theory,''
JHEP {\bf 1301}, 100 (2013).
[arXiv:1204.2270 [hep-th]].
}

\lref\DrukkerSR{
  N.~Drukker, T.~Okuda and F.~Passerini,
  ``Exact results for vortex loop operators in 3d supersymmetric theories,''
[arXiv:1211.3409 [hep-th]].
}
\lref\XieGMA{
  D.~Xie,
  ``M5 brane and four dimensional N = 1 theories I,''
JHEP {\bf 1404}, 154 (2014).
[arXiv:1307.5877 [hep-th]].
}

\lref\qinteg{
  M.~Rahman, A.~Verma,
  ``A q-integral representation of Rogers' q-ultraspherical polynomials and some applications,''
Constructive Approximation
1986, Volume 2, Issue 1.
}

\lref\qintegOK{
  A.~Okounkov,
  ``(Shifted) Macdonald Polynomials: q-Integral Representation and Combinatorial Formula,''
Compositio Mathematica
June 1998, Volume 112, Issue 2. 
[arXiv:q-alg/9605013].
}

\lref\macNest{
 H.~Awata, S.~Odake, J.~Shiraishi,
  ``Integral Representations of the Macdonald Symmetric Functions,''
Commun. Math. Phys. 179 (1996) 647.
[arXiv:q-alg/9506006].
}

\lref\deBult{
  F.~J.~van~de~Bult,
  ``An elliptic hypergeometric integral with $W(F_4)$ symmetry,''
The Ramanujan Journal, Volume 25, Issue 1 (2011)
[arXiv:0909.4793[math.CA]].
}

\lref\Rains{
E.~M.~Rains,
  ``Transformations of elliptic hypergometric integrals,''
Annals of Mathematics, Volume  171, Issue 1 (2010)
[arXiv:math/0309252].
}

\lref\ItoFPL{
  Y.~Ito and Y.~Yoshida,
  ``Superconformal index with surface defects for class ${\cal S}_k$,''
[arXiv:1606.01653 [hep-th]].
}

\lref\BeniniMZ{
  F.~Benini, Y.~Tachikawa and B.~Wecht,
  ``Sicilian gauge theories and N=1 dualities,''
JHEP {\bf 1001}, 088 (2010).
[arXiv:0909.1327 [hep-th]].
}

\lref\DimoftePD{
  T.~Dimofte and D.~Gaiotto,
  ``An E7 Surprise,''
JHEP {\bf 1210}, 129 (2012).
[arXiv:1209.1404 [hep-th]].
}

\lref\GaddeFMA{
  A.~Gadde, K.~Maruyoshi, Y.~Tachikawa and W.~Yan,
  ``New N=1 Dualities,''
JHEP {\bf 1306}, 056 (2013).
[arXiv:1303.0836 [hep-th]].
}

\lref\AgarwalVLA{
  P.~Agarwal, K.~Intriligator and J.~Song,
  ``Infinitely many $\cal{N}=1 $ dualities from m + 1 - m = 1,''
JHEP {\bf 1510}, 035 (2015).
[arXiv:1505.00255 [hep-th]].
}

\lref\GorskyTN{
  A.~Gorsky,
  ``Dualities in integrable systems and N=2 SUSY theories,''
J.\ Phys.\ A {\bf 34}, 2389 (2001).
[hep-th/9911037].
}

\lref\FockAE{
  V.~Fock, A.~Gorsky, N.~Nekrasov and V.~Rubtsov,
  ``Duality in integrable systems and gauge theories,''
JHEP {\bf 0007}, 028 (2000).
[hep-th/9906235].
}

\lref\RazamatPTA{
  S.~S.~Razamat and B.~Willett,
  ``Down the rabbit hole with theories of class $ \cal S $,''
JHEP {\bf 1410}, 99 (2014).
[arXiv:1403.6107 [hep-th]].
}

\lref\RazamatL{ A.~Gadde, S.~S.~Razamat, and  B.~Willett,
  ``A ``Lagrangian'' for a non-Lagrangian theory,''
  Phys. Rev. Lett. {\bf 115}, 171604 (2015).
  [arXiv:1505.05834 [hep-th]].
  }
  
  \lref\OhmoriPUA{
  K.~Ohmori, H.~Shimizu, Y.~Tachikawa and K.~Yonekura,
  ``6d ${\cal N}=(1,0)$ theories on $T^2$ and class S theories: Part I,''
JHEP {\bf 1507}, 014 (2015).
[arXiv:1503.06217 [hep-th]].
}

\lref\OhmoriPIA{
  K.~Ohmori, H.~Shimizu, Y.~Tachikawa and K.~Yonekura,
  ``6d ${\cal N}=(1,\;0) $ theories on S$^{1}$ /T$^{2}$ and class S theories: part II,''
JHEP {\bf 1512}, 131 (2015).
[arXiv:1508.00915 [hep-th]].
}

\lref\DelZottoRCA{
  M.~Del Zotto, C.~Vafa and D.~Xie,
  ``Geometric engineering, mirror symmetry and $ 6{{d}}_{\left(1,0\right)}\to 4{{d}}_{\left({\cal N}=2\right)} $,''
JHEP {\bf 1511}, 123 (2015).
[arXiv:1504.08348 [hep-th]].
}

\lref\HeckmanBFA{
  J.~J.~Heckman, D.~R.~Morrison, T.~Rudelius and C.~Vafa,
  ``Atomic Classification of 6D SCFTs,''
Fortsch.\ Phys.\  {\bf 63}, 468 (2015).
[arXiv:1502.05405 [hep-th]].
}

\lref\BhardwajXXA{
  L.~Bhardwaj,
  ``Classification of 6d  N=(1,0) gauge theories,''
JHEP {\bf 1511}, 002 (2015).
[arXiv:1502.06594 [hep-th]].
}

\lref\HananyPFA{
  A.~Hanany and K.~Maruyoshi,
  ``Chiral theories of class $ {\cal S} $,''
JHEP {\bf 1512}, 080 (2015).
[arXiv:1505.05053 [hep-th]].
}

\lref\GaiottoLCA{
  D.~Gaiotto and A.~Tomasiello,
  ``Holography for (1,0) theories in six dimensions,''
JHEP {\bf 1412}, 003 (2014).
[arXiv:1404.0711 [hep-th]].
}

\lref\WittenSC{
  E.~Witten,
  ``Solutions of four-dimensional field theories via M theory,''
Nucl.\ Phys.\ B {\bf 500}, 3 (1997).
[hep-th/9703166].
}

\lref\ApruzziZNA{
  F.~Apruzzi, M.~Fazzi, A.~Passias and A.~Tomasiello,
  ``Supersymmetric AdS$_{5}$ solutions of massive IIA supergravity,''
JHEP {\bf 1506}, 195 (2015).
[arXiv:1502.06620 [hep-th]].
}

\lref\ComanBQQ{
  I.~Coman, E.~Pomoni, M.~Taki and F.~Yagi,
  ``Spectral curves of ${\cal N}=1$ theories of class ${\cal S}_k$,''
[arXiv:1512.06079 [hep-th]].
}

\lref\FrancoJNA{
  S.~Franco, H.~Hayashi and A.~Uranga,
  ``Charting Class ${\cal S}_k$ Territory,''
Phys.\ Rev.\ D {\bf 92}, no. 4, 045004 (2015).
[arXiv:1504.05988 [hep-th]].
}

\lref\SpiridonovZR{
  V.~P.~Spiridonov and G.~S.~Vartanov,
 ``Superconformal indices for N = 1 theories with multiple duals,''
Nucl.\ Phys.\ B {\bf 824}, 192 (2010).
[arXiv:0811.1909 [hep-th]].
}

\lref\MorrisonNRT{
  D.~R.~Morrison and C.~Vafa,
  ``F-theory and N = 1 SCFTs in four dimensions,''
JHEP {\bf 1608}, 070 (2016).
[arXiv:1604.03560 [hep-th]].
}

\lref\GukovJK{
  S.~Gukov and E.~Witten,
  ``Gauge Theory, Ramification, And The Geometric Langlands Program,''
[hep-th/0612073].
}

\lref\AharonyDHA{
  O.~Aharony, S.~S.~Razamat, N.~Seiberg and B.~Willett,
  ``3d dualities from 4d dualities,''
JHEP {\bf 1307}, 149 (2013).
[arXiv:1305.3924 [hep-th]].}

\lref\HeckmanPVA{
  J.~J.~Heckman, D.~R.~Morrison and C.~Vafa,
  ``On the Classification of 6D SCFTs and Generalized ADE Orbifolds,''
JHEP {\bf 1405}, 028 (2014), Erratum: [JHEP {\bf 1506}, 017 (2015)].
[arXiv:1312.5746 [hep-th]].
}

\lref\CsakiCU{
  C.~Csaki, M.~Schmaltz, W.~Skiba and J.~Terning,
  ``Selfdual N=1 SUSY gauge theories,''
Phys.\ Rev.\ D {\bf 56}, 1228 (1997).
[hep-th/9701191].
}

\lref\DelZottoHPA{
  M.~Del Zotto, J.~J.~Heckman, A.~Tomasiello and C.~Vafa,
  ``6d Conformal Matter,''
JHEP {\bf 1502}, 054 (2015).
[arXiv:1407.6359 [hep-th]].
}

\lref\GaiottoUSA{
  D.~Gaiotto and S.~S.~Razamat,
  ``${\cal N}=1 $ theories of class $ {\cal S}_k $,''
JHEP {\bf 1507}, 073 (2015).
[arXiv:1503.05159 [hep-th]].
}

\Title{\vbox{\baselineskip12pt
}}
{\vbox{
\centerline{4d ${\cal N}=1$ from 6d $(1,0)$}
\vskip7pt 
\centerline{}
}
}

\centerline{Shlomo S. Razamat,$^a$ Cumrun Vafa,$^b$ and Gabi Zafrir$^{a,c}$}
\bigskip
\centerline{{\it ${}^a$ Physics Department, Technion, Haifa, Israel 32000}}
\centerline{{\it ${}^b$ Jefferson Physical Laboratory, Harvard University, Cambridge, MA 02138, USA}}
\centerline{{\it ${}^c$ Kavli IPMU (WPI), UTIAS, the University of Tokyo, Kashiwa, Chiba 277-8583, Japan}}
\bigskip
\centerline{{\it }}
\vskip.1in \vskip.2in \centerline{\bf Abstract}

\vskip.2in

\noindent We study the geometry of 4d ${\cal N}=1$ SCFT's arising from compactification of 6d $(1,0)$ SCFT's on a Riemann surface.  We show that the conformal manifold of the resulting theory is characterized, in addition to moduli of complex structure of the Riemann surface, by the choice of a connection for a vector bundle on the surface arising from flavor symmetries in 6d.  We exemplify this by considering the case of 4d ${\cal N}=1$ SCFT's arising from M5 branes probing $\Z_k$ singularity compactified on a Riemann surface.   In particular, we study in detail the four dimensional theories arising in the case of two M5 branes on $\Z_2$ singularity. We compute the conformal anomalies and indices of such theories in 4d and find that they are consistent with expectations based on anomaly and the moduli structure derived  from the 6 dimensional perspective.

\vfill

\Date{October 2016}

\newsec{Introduction}

Quantum field theories in a given number of dimensions can exhibit rather surprising properties. Important examples of such properties are the dualities relating seemingly different quantum field theories in four and lower dimensions. In many cases such surprising properties can be given a geometric interpretation once the theories of interest are realized as a dimensional reduction on a compact manifold of a higher dimensional theory. Different duality frames correspond to different ways to construct the same  compact manifold. A paradigmatic example 
of this is the compactifications of $(2,0)$ theories on a Riemann surface giving a wide variety of ${\cal N}=2$ (and ${\cal N}=1$) theories in four dimensions \GaiottoWE. An important by product of this construction is the derivation of existence of new strongly coupled SCFT's with no tunable parameters.  A generic SCFT in four dimensions has then a description in terms of such SCFT's and only in very special cases a weakly coupled Lagrangian can be constructed.

A natural question to ask is whether by starting from more generic six dimensional theories, in particular considering less supersymmetric starting points, new four dimensional phenomena can be derived.
  In particular a variety of six dimensional setups can be considered having $(1,0)$ supersymmetry \DelZottoHPA\ and a classification of them based on F-theory constructions has been proposed \refs{\HeckmanPVA,\HeckmanBFA,\BhardwajXXA}, leading to a vast number of 6d SCFT's. Taking such theories on a Riemann surface would give ${\cal N}=1$ models in four dimensions. These models will be labeled by the choice of the six dimensional theory and by the choices made during the compactification. The latter include the choice of the Riemann surface and choices of bundles for different background vector fields associated to flavor symmetries of the six dimensional theory. Different compactifications, thus different labeling, might produce same theories in four dimensions \refs{\OhmoriPUA,\OhmoriPIA,\DelZottoRCA}, but in a generic scenario the four dimensional theories coming from different compactifications are distinct. This then possibly produces a very wide variety of  theories and relations between those.

The purpose of this paper is to establish a direct link between choices made in six dimensions  and theories conjecturally obtained in four dimensions~\GaiottoUSA (see also \refs{\HananyPFA,\FrancoJNA,\ComanBQQ}). In particular the choices made during compactification determine  in a rather tractable way the dimension of the conformal manifold and the anomalies of the four dimensional theories.  The conformal manifold is determined by the moduli space of complex structures of the Riemann surface and by the moduli space of gauge connections for the symmetries of the six dimensional setup.   These can lead to distinct 4d theories coming from choices of in general non-abelian flat bundles\foot{This can be viewed as the choice of a connection for a holomorphic vector bundle.} leading to a large number of moduli \MorrisonNRT.  More  precisely, let $G$ denote the 6d flavor symmetry.  We pick an abelian subgroup of it $L\subset G$ and denote the commutant by $G'$ which we take to be non-abelian.  In  particular we have
$$L\times G'\subset G$$
and define $G^{max}=L\times G'$. Then we pick a flat gauge field for $G^{max}$ on the Riemann surface modulo the possibility of picking a non-trivial flux for the abelian part $c_1(L)\in {\Z}^{dim(L)}$.
The dimension of the conformal manifold for compactifications with no punctures for each non-trivial choice of $c_1(L)$ is given by,

\eqn\dimco{
dim {\cal M}_{g,0} = 3g-3+(g-1) dim \; G' +g\cdot dim \; L =3g-3+(g-1)dim \; G^{max} + dim \; L .
}    Note that the $-1$ in the $(g-1)$ term is there only for the non-abelian part of $G^{max}$, which comes from the fundamental group relation of the Riemann surface as well as gauge transformations which only impact the non-abelian part of $G^{max}$. The anomalies of the four dimensional theories can be obtained on the other hand by integrating the anomaly polynomial of the six dimensional theory on the Riemann surface taking into account the particular  choice made for the fluxes, $c_1(L)$. 

\

After making some general remarks and predictions we will discuss in detail the case of two M5 branes probing $\Z_2$ singularity. Following \GaiottoUSA\ we will derive a set of theories in four dimensions which we will then map to the six dimensional compactifications on Riemann surfaces with genus $g>1$ and with different choices of fluxes for the flavor symmetry. In particular we will compare the anomalies of the four dimensional construction with those obtained from six dimensions and the dimensions of the conformal manifolds. The theories in four dimensions will have a description in terms of {\it ``strongly coupled''} Lagrangians. The construction will start from weakly coupled Lagrangians and then take us on a parameter space of these models to special loci where certain symmetries are enhanced. Coupling these symmetries to dynamical gauge field the theories which correspond to the different compactifications can then be constructed. Since the enhancement of symmetry used for the construction happens only at strong coupling and we do not have a precise road map to the enhancement locus, we refer to these models as having ``strongly coupled'' Lagrangians. Such Lagrangians are nevertheless sufficient to derive a plethora of protected information about the models. In particular all the anomalies and supersymmetric indices are easily extractable. We will compute here the supersymmetric index and the central charges from these Lagrangians.

\

The organization of this paper is as follows. In section 2 we study the general structure of 6d (1,0) SCFT's compactified on a Riemann surface and the resulting 4d theories.  We exemplify it using 6d SCFT's arising from M5 branes on $G$ singularity compactified on a Riemann surface and in particular predict the dimension of the conformal manifold of the four dimensional field theory arising in low energy. In section 3 we review a four dimensional construction conjectured to give the field theories arising in the compactification, and discuss the dimension of the conformal manifold from this perspective. In section 4
we discuss the salient features of the field theoretic construction when $k=2$ and the number of M5 branes is also two. In section 5 we derive a description of some trinions, theories obtained upon compactification on three punctured sphere, of $k=2$ and two M5 branes case in terms of ``strongly coupled'' Lagrangians.
 We also discuss superconformal indices and anomalies of theories constructed from these building blocks. In section 6 we derive a variety of examples of theories corresponding to different choices of fluxes upon compactifications. In section 7 we compute anomalies of the putative four dimensional theories arising in compactifications from six dimensional anomaly polynomial. In section 8 we summarize our results. Several appendices complement the text.

\  

\newsec{Basic Setup}
In this section we discuss the basic setup we have and use that to make predictions about 4d ${\cal N}=1$ SCFT's arising from compactification of
${\cal N}=(1,0)$ SCFT's in 6d.
In order to preserve half the supersymmetry we need to partially twist the theory along the Riemann surface \BershadskyVM .   What this means is that we modify the spin connection by mixing with it a $u(1)$ in the Cartan of $su(2)$ R-symmetry of the $(1,0)$ theory: $u(1)\subset su(2)_R$.  In this way, on a generic Riemann surface, we preserve half the supersymmetry leading to ${\cal N}=1$ in $d=4$.  We expect, as in the case of its $(2,0)$ cousin, that in the infrared,
which corresponds to the area of the Riemann surface going to zero, we obtain an ${\cal N}=1$ superconformal theory in $d=4$.  Just as in the $(2,0)$ theory, we expect that the resulting theory will depend on the choice of the complex structure of the Riemann surface.  In other words, the conformal manifold, or moduli of the 4d ${\cal N}=1$ theory will include the complex structure moduli of the Riemann surface.  However with lower supersymmetry we will have more options as we go down to 4d.  The reason for this is that the $(1,0)$ theories often (but not always) have additional global symmetries. We can turn on backgrounds \DumitrescuLTQ\ corresponding to gauge field configurations for this flavor symmetry along the Riemann surface in a way that guarantees preserving ${\cal N}=1$ supersymmetry. To see this, note that the condition that supersymmetry be preserved is that the supersymmetry variation of the gaugino fields in the multiplet be zero.  This variation of the gauginos represented by $\lambda$ is given by
\eqn\dellamc{\delta \lambda=F_{\mu \nu} \gamma^{\mu\nu}\cdot \epsilon +D \epsilon\,,}
where the D-term in the variation breaks the symmetry to  an abelian subgroup.  For the non-abelian case (where $D=0$) if we set $F=0$ we see that we can  preserve supersymmetry.  Thus for non-abelian flavor symmetry we can turn on flat non-trivial bundles on the Riemann surface and still preserve the supersymmetry.
For a genus $g$ surface this gives rise to an $(g-1)\;dim\, G$ complex dimensional moduli space where $G$ is the non-abelian flavor group.  To see this note that as usual one can open up the Riemann surface given the $A_i,\; B_j$ cycles.  We can assign arbitrary $G$ holonomies for each $A_i,\;B_j$ subject to
\eqn\sugk{\prod_{i=1}^g [A_i,\,B_i]=1\,,}
and the overall gauge transformation.  This gives $(2g-2)\cdot dim\, G$ real dimensions, or a $(g-1)  \cdot dim\, G$ complex dimensional manifold.

For the abelian $u(1)$ case, we have a more general possibility.  Let $\epsilon$ denote the spinor which is covariantly constant after the topological twist.  Then to preserve the variation along the $\epsilon$ we need to satisfy the equation
\eqn\eqfl{F_{z\overline z}\ {\bar \epsilon} \gamma^{z\overline z}\epsilon +D =F_{z\overline z}\  \epsilon^{z\overline z}+D=0.\,\,,}
  This can be solved by 
\eqn\ergjm{F_{z\overline z}=const.\  \epsilon_{z\overline z}\,.}
We can now have $F$ be non-zero.  In such a case the allowed constants (related to the D-term) in the above equations are quantized because 
${1\over 2\pi}\int_{\Sigma} F=c_1(F)\in \Z$. Even after an F representing a given class in $c_1(F)$ has been fixed, we can still solve the above equation by the addition of flat $u(1)$ gauge fields to it.  So for each $u(1)$ flavor symmetry and for any integer $n$ representing its $c_1$ we will get an additional $g$ complex moduli.

However, there is a more general possibility which allows combining the abelian and non-abelian cases.  Let $G$ denote the flavor symmetry group.  We can choose an abelian subgroup of it $L$ whose non-abelian commutant inside $G$ is given by $G'$: $L\times G'\subset G$.  Then we can choose a non-trivial flux in $L$  which can be deformed by addition of flat bundle in $L \times G'$.  In other words for each $c_1(L)\in \Z^{dim(L)}$ and each genus $g$ we get an ${\cal N}=1$ SCFT in 4d whose conformal manifold ${\cal M}_g$ is expected to have complex dimension
\eqn\dimm{
dim{\cal M}_g=(3g-3)+g\cdot dim(L)+(g-1) dim(G')=(3g-3)+(g-1)dim (G^{max})+dim(L)}
where $G^{max}=L\times G'$.  Note that $G^{max}$ is the maximal symmetry group we expect to have for the 4 dimensional theory corresponding to submoduli where the holonomies of $G'$ are turned off, but for which there is a background flux in $L$.

It is useful to consider a familiar example to illustrate these ideas.  Consider the 6d $(2,0)$ theory but view it as a $(1,0)$ theory as we wish to preserve only
${\cal N}=1$ supersymmetry in 4d.  In such a case the R-symmetry group is $so(5)$.  To twist we need to pick a $u(1)$ in it.  Consider the $su(2)_L\times su(2)_R=so(4)\subset so(5)$.  We choose $so(2)$ R-symmetry in the Cartan of $su(2)_R$ (which can be viewed as the sum of the two canonical Cartans $so(2)\times so(2)\subset so(4)$).   After this twist, we still have an extra flavor symmetry group $G=su(2)_L$.  We can choose to turn on a flat $su(2)_L$ bundle
on the Riemann surface as mentioned above.  This will give us a theory whose moduli space has dimension
\eqn\bimbah{dim{\cal M}_g=(3g-3)+(g-1)\cdot 3=6g-6\,\,,}
which has been studied in \BeniniMZ .  This corresponds to the geometry of the normal bundle to the Riemann surface being $\R\times {\cal L}_{g-1}\oplus {\cal L}_{g-1}$,  where ${\cal L}_r$ denotes a line bundles of degree $r$.  Here the $su(2)$ flat bundle mixes the two line bundles of degree $g-1$.
Instead we can also consider choosing an abelian subgroup $L\subset G$ which in this case is simply $u(1)\subset su(2)_L$, and turn on a non-trivial flux characterized by an integer $c_1(L)=n$.  This corresponds to theories studied in \BahDG\ and corresponds to the normal bundle to the Riemann surface being
$\R\times {\cal L}_{g+n-1}\oplus {\cal L}_{g-n-1}$.  The dimension of the conformal manifold for this case is according to our discussion above
\eqn\beemba{dim{\cal M}_g=(3g-3)+g=4g-3\,\,.}

Let us just consider two examples to illustrate the diversity of choices available for 6d $(1,0)$ theories.  Consider the exceptional 6d SCFT, with F-theory base given by $O(-1)$ bundle over $\P^1$.  This theory has $E_8$ global symmetry.  For this theory we can choose $G'=G=E_8$ and we end up getting the moduli space of flat $E_8$ bundles on the Riemann surface as part of the conformal manifold.  Or we can choose any abelian subgroup $L\subset E_8$ whose dimension can vary from $1,...,8$.
For each dimension there are numerous possibilities for how it embeds in $E_8$ leading to different non-abelian commutants $G'$.  For each such choice, each choice of $c_1(L)$, and each $g$ we get an ${\cal N}=1$ SCFT.  Clearly this is an enormous list of new ${\cal N}=1$ theories arising from a simple
6d SCFT!  We can also contrast this with the 6d SCFT corresponding to F-theory geometry $O(-12)$ bundle over $\P^1$.  This theory has no global symmetries and so for each genus $g$ we get a unique choice whose conformal manifold is simply the complex structure moduli space of the Riemann surface.\foot{This is an ${\cal N}=1$ theory whose moduli space is expected to be exactly that of moduli of genus $g$ surfaces with no extra moduli.  All of the well-studied cases have extra moduli (even those coming from $(2,0)$ theory as already discussed) when viewed as an ${\cal N}=1$ theory in 4d.}

\

 \subsec{Adding punctures}
 So far we have discussed compactifications on Riemann surfaces without punctures.  If we add punctures to the Riemann surface, there would be additional moduli for the resulting conformal field theory.  The most obvious has to do with the choice of the position of the punctures.  If we have $s$ punctures this will add $s$ complex moduli.  Moreover, depending on the type of the puncture, the choice of the holonomy of $G$ will have to be restricted to a special one compatible with preserving conformal symmetry.  One way to think about this is that a puncture on the Riemann surface can be viewed as attaching a semi-infinite cylinder to the surface. So we effectively obtain the reduction of the 6d theory on a circle to 5d.  We will have to choose a holonomy for $G$ along the circle.  Such holonomies typically play the role of mass parameters for 5d theories.  Only at special holonomies can we expect the 5d theory to be gapless.  Thus the choice of such holonomies will typically break the $G$ symmetry to a subgroup $P\subset G$, which preserves the element.  The inequivalent choices of $G$ holonomy
 in a particular conjugacy class, which are preserved by $P$ are given by the coset $G/P$.  In this way we find that at each puncture we have more choices for the moduli, whose complex dimension is given by ${1\over 2}dim( G/P)$ (see \GukovJK\ for a nice discussion of such moduli spaces).
 If the holonomy in the bulk is broken to $G^{max}$ due to picking an Abelian subgroup $L$ to turn on flux in, then the allowed holonomies will be smaller given by $P^{max}=G^{max}\cap P$.  In this way we get the following formula for the dimension of the 4d conformal manifold:
 
\eqn\dimpm{\eqalign{&
dim{\cal M}_g=(3g-3+s)+(g-1)dim (G^{max})+dim(L)+{1\over 2}\sum_i dim\bigg({G^{max}\over P_i^{max}}\bigg)=\cr &\ \ \ \ \ (3g-3+s)+(g-1+{s\over 2})dim (G^{max})+dim(L)-{1\over 2}\sum_i dim(P_i^{max})}}

As we have discussed, the choice of punctures of a given type corresponds to a conjugacy class of $G$ which preserves a subgroup $P$ of $G$. There could be various different though physically equivalent ways to embed $P$ in $G$. These are usually related to one another by automorphisms of $G$, which may be both inner or outer automorphisms.

 For different choices related by an inner automorphism, if $G$ is a simple non-abelian group then there is no need to have an extra designation of the actual element of the conjugacy class of the puncture, because one can obtain all choices of the conjugacy element by the action of $G$ which can be absorbed into the moduli of flat connections on the punctured Riemann surface.  However, even if $G$ is simple, when we choose an abliean subgroup $L$ to put flux in, it could happen that not all conjugacy elements compatible with $P$ symmetry can be reached by the action of
$G^{max}=L\times G'$ which is the left-over symmetry.  In such a case, inequivalent choices of embedding of the conjugacy classes of the punctures in $G$ should be viewed as leading to distinct classes of theories. The same is true also for embeddings differing by an outer automorphism.  These $G^{max}$ inequivalent embeddings of the conjugacy element of the puncture in $G$ is an extra designation for the puncture and explains the appearance of the 
 `color' (and `sign') of the puncture in \GaiottoUSA.  We shall call such extra choices for a puncture for a general theory as the `color' of the puncture.

\

\subsec{M5 branes probing ADE compactified on a surface}

The main class of examples studied in this paper involve compactifications of the 6d SCFT obtained by probing ADE singularities with N M5 branes.
In fact we will concentrate on the $A_{k-1}$ case and our main example will involve $2$ M5 branes probing the $A_1$ singularity.  As we will see, already this example is highly non-trivial and interesting.

Consider M-theory in the presence of $K$-type singularity where $K$ can stand for any of the ADE groups.  In other words, we consider the background given by 

\eqn\bg{{\C^2\over \Gamma_K}\,,}
where $\Gamma_K$ is the discrete subgroup of $su(2)$ associated to the group $K$.  As is well known this gives rise to a 7 dimensional singularity on which a K-type gauge symmetry emerges.

Now consider probing this singularity with $N$ parallel and coincident M5 branes.  Since M5 branes have a 6 dimensional worldvolume, {\it i.e.} 1 lower dimension than the singularity locus, they will appear as points on the line of the singularity.  The resulting theory will have a $G=K_L\times K_R$ symmetry \DelZottoRCA. The $K_{L,R}$  arise from the bulk gauge symmetries of M-theory on the left and right part of the M5 branes respectively.  In the $K=su(k)$ case, which we will be mainly interested in, there is also a global $u(1)$ symmetry coming from the fact that $K$ enhances to $u(k)$.  
Generally these theories are non-trivial 6d $(1,0)$  SCFT's.  However in one case, they are rather simple:  When $N=1$ for the $A_{k-1}$ singularity, they simply give rise to a hypermultiplet which transforms in the $(k,\overline k)_{+1}$ representation of $su(k)_L\times su(k)_R\times u(1)$.
For low $k,N$ the flavor symmetry can get accidental enhancements.  For example for $k=N=2$ it turns out that $su(2)_L\times su(2)_R \times u(1)$ enhances to $G=so(7)$.

Now we wish to compactify this 6d theory on a Riemann surface $\Sigma$ and as already discussed we partially twist the theory by adding to the spin connection the Cartan of $su(2)_R$ symmetry.  In the case of $T^2$ we can preserve all the supersymmetries, leading to an ${\cal N}=2$ supersymmetric theory in 4d.

However, as already discussed we can do more:  Since 
the theory in the bulk has $G$ gauge symmetry we can turn on flat G-bundles on the Riemann surface.  In fact, because the M5 branes split the $K$-symmetry in the bulk to two parts, we can turn on independent flat $K$ bundles for $K_L$ and $K_R$.  This still preserves supersymmetry.  A simple example of this can be seen in the context of $N=1$ M5 brane probing $A_{k-1}$ theory.
In that case the supersymmetry charge is given by $Tr(\psi^* \partial \phi)$ where $\psi$ and $\phi$ are bi-fundamental hypermultiplet fermions and bosons transforming as ,

\eqn\gty{(\psi,\phi) \rightarrow g_L (\psi,\phi ) g_R^{-1}\,,}
which leaves the supercharge invariant as the $\psi$ and $\phi$ undergo $G_L\times G_R$ monodromies around non-trivial cycles of the Riemann surface.  In addition we can turn on a flat bundle for the $u(1)$ global symmetry (which can also have a non-trivial first Chern-class).  Moreover we can also choose other abelian subgroups of $su(k)\times su(k)\times u(1)$ as already discussed, and turn on flux in them.  

Similarly we can add punctures.  There are various punctures allowed.  To characterize them we compactify the theory on a circle to 5d, leading to affine $su(N)^k$ quiver theory.  Punctures which preserve the full symmetry or a simple puncture with just a $u(1)$ symmetry have been studied in \GaiottoUSA.  Moreover the structure of allowed punctures is rather intricate and has been studied in \HeckmanXDL. In this paper we will mainly focus on the full and simple punctures.
For full punctures, we have evidence that the conjugacy class corresponding to preserving superconformal symmetry leads to $P=u(1)^{2k-1}$ for the generic $k$ and $N$,
but for special cases this will be different.  We will find that for the $N=k=2$, the group $P$ corresponding to leaving the holonomy of the full puncture invariant is $su(2)\times u(1)\times u(1)$.

\

\subsec{${\cal N}=2$ sub case and the resolution of a puzzle}
It is instructive to consider the sub-case which leads to higher supersymmetry.
To obtain an ${\cal N}=2$ supersymmetric case we have to do two things:  Take the trivial case of $G=A_0$ and turn on an ${\cal N}=2$ preserving flux.  This leads to the
class ${\cal S}$ theories  \GaiottoWE\GaiottoHG\ with a $3g-3$ complex dimensional moduli.  Or we can consider the $A_{k-1}$ singularity case and consider compactifying the theory on $T^2$ with the $u(1)$
flat connection turned off.  This gives rise to an ${\cal N}=2$ supersymmetric theory whose moduli space is given by the moduli space of flat $su(k)_{diagonal}\subset su(k)\times su(k)$ connections on $T^2$.  This is the case where we obtain a quiver gauge theory using the duality between M-theory compactified on $T^2$ and type IIB on $\S^1$.
Wrapped $M5$ branes on $T^2$ become dual to D3 branes and so in the type IIB theory we have N D3 branes probing $A_{k-1}$ singularity which leads to the affine $su(k)$  ${\cal N}=2$ supersymmetric quiver theory with $su(N)$ gauge groups  on each node of the  quiver.  It is known  \WittenSC\ that the moduli of this theory is that of flat $su(k)$ connections on $T^2$  which is $k$ dimensional.  For a discussion of this in the context of 6d theory compactified on $T^2$ see \DelZottoRCA.

Now we come to a puzzle:  From the discussion of the previous section, we expected the moduli space to be given by {\it two copies} of flat $K$ bundles.  However, here we are only getting one copy.
This is because we are in a special situation where the generic argument discussed in the previous section does not apply. In particular naive application of the formulas of the previous section would have given a factor
of $(g-1)=0$ for the multiplicative factor for the dimension of flat bundles at genus 1.  Of course in that case the moduli would have to be abelian because we are on a torus.  Nevertheless we are still getting only one copy of the flat bundles, which corresponds to turning
on the diagonal flat bundle.  The $g_L\not= g_R$ part of flat bundles lead to mass parameters for the bifundamentals of the quiver theory.  One may wonder whether this is a generic feature, and that somehow the off-diagonal part of K-bundlles
give mass terms.  We will now argue that this is not the case, and the case of torus with no punctures is misleading us.

To gain insight into this consider the case of $N=1$ M5 brane probing $A_{k-1}$ singularity.  If we compactify this theory on $T^2$, and turn on flat $su(k)_L\times su(k)_R$ bundle, indeed if the $L$ and $R$ connections are not equal we will mass up the hypermultiplets
because the light modes will come from the zero modes of the theory on $T^2$ and if the left and right connections are not equal there are no zero modes left.  However, if left and right connections are equal, since the connections are abelian on the torus,
this leads for generic connections to $k$ zero modes leading to $k$ massless bifundamentals, as is expected for affine $A_{k-1}$ quiver with rank 1 on each quiver node.  Having confirmed the expectation based on the ${\cal N}=2$ affine quiver case for this simple example, now let us consider the same theory more generically,
for $g>1$ or even $g=1$ but with punctures.  In that case we may have expected again that if the left and right connections are not equal the theory in 4d will become trivial.  But this is not the case, as we will now argue.

To see this, recall that the theory on a curved Riemann surface is twisted, which means that the fields acquire non-traditional spins.  In particular let us consider the fermions in the hypermultiplet.  They will transform as 1-forms as well as 0-forms, coupled to the $K_L\times K_R$ bundle.  To find if there are any low energy 4d modes left, we need to count the zero modes of the internal theory.  By Hirzebruch-Riemann-Roch theorem we deduce that 

\eqn\rhro{n^1-n^0=(dim V)(2g-2+s)\,,}
where $n^1$ denotes the number of 1-form zero modes and $n^0$ the number of 0-form zero modes leading in the 4d ${\cal N}=1$ theory to chiral and anti-chiral fields respectively.  Here $dim V=k^2$.  We thus learn that for $2g-2+s>0$
we will typically get 1-form zero modes and no 0-forms.  {\it I.e.} we have chiral fields.  For $g=1,s=0$ generically we get no zero modes, which is what we already observed.  For the special case when $g=1,s=0$ and if we set left and right $K$-connections equal, then we get equal number of $n^1$ and $n^0$ modes
($k$ of each), leading to enhancement of supersymmetry to ${\cal N}=2$.  This resolves the puzzle and strengthens the picture that for arbitrary flat
$K_L\times K_R$ connections on the Riemann surface we expect non-trivial 4d ${\cal N}=1$ SCFT in the IR.

\

\newsec{Four dimensional perspective: preliminaries}

We now turn our attention to field theories in four dimensions conjectured to be obtained once $N$ M5 branes  probing  $\Z_k$ singularity get compactified on a Riemann surface. The case of $k=1$ is well studied and gives rise to ${\cal N}=2$ theories \GaiottoWE\ or ${\cal N}=1$ theories \BeniniGI\BahDG,  while for $k>1$ one in general obtains ${\cal N}=1$ theories. Some of the models obtained in this way are quiver theories with standard descriptions in terms of Lagrangians, but the vast majority are built from strongly coupled ingredients. The theories possessing standard Lagrangians correspond, for general $N$, to Riemann surfaces with genus zero and punctures with low enough symmetry, linear superconformal quivers, or to genus one with a number of $u(1)$ punctures, circular/toric quivers. For higher genus and ($N>2$ when $k=1$) the theories usually have strongly coupled ingredients. In what follows we will first review the essentials of the linear and toric quiver cases 
and then discuss the higher genus generic theories.

\subsec{Linear and circular quivers}

We start with reviewing some aspects of the linear and circular quivers, for details we refer the reader to \GaiottoUSA.
The building block for constructing linear and circular quivers is the free trinion. This is a collection of free chiral fields which we organize into $k$  fields $Q_i$ and $k$  fields $\widetilde Q_i$, Fig. 1.  The chiral fields $Q_l$ are in the bifundamental representation of  $su(N)^a_l\times su(N)^b_l$ , and $\widetilde Q_l$ are in the bifundamental representation of $su(N)^b_{l+1}\times su(N)_l^a$. We also have $2(k-1)+1+1$ abelian symmetries $u(1)_t\times u(1)_\beta^{k-1}\times u(1)^{k-1}_\gamma\times u(1)_\alpha$.  It is convenient to encode the charges of the fields in fugacities. For $Q_l$ we have $t^{\frac12}\alpha^{-1} \beta_l$ and for $\widetilde Q_l$ $t^{\frac12}\alpha\gamma_\ell$. We think of this theory as associated to a three punctured sphere with two maximal punctures with flavor symmetry $su(N)_a^k$ and $su(N)_b^k$, and a minimal puncture with symmetry $u(1)_\alpha$. The other symmetries $(\beta,\gamma, t)$ are not associated to the punctures but have more general geometric origin. We will refer to  the symmetries which are not associated to punctures as internal symmetries. These symmetries are conjectured to come from the Cartan subgroup of $u(1)\times su(k)_L\times su(k)_R$ discussed in the second section.

\

\centerline{\figscale{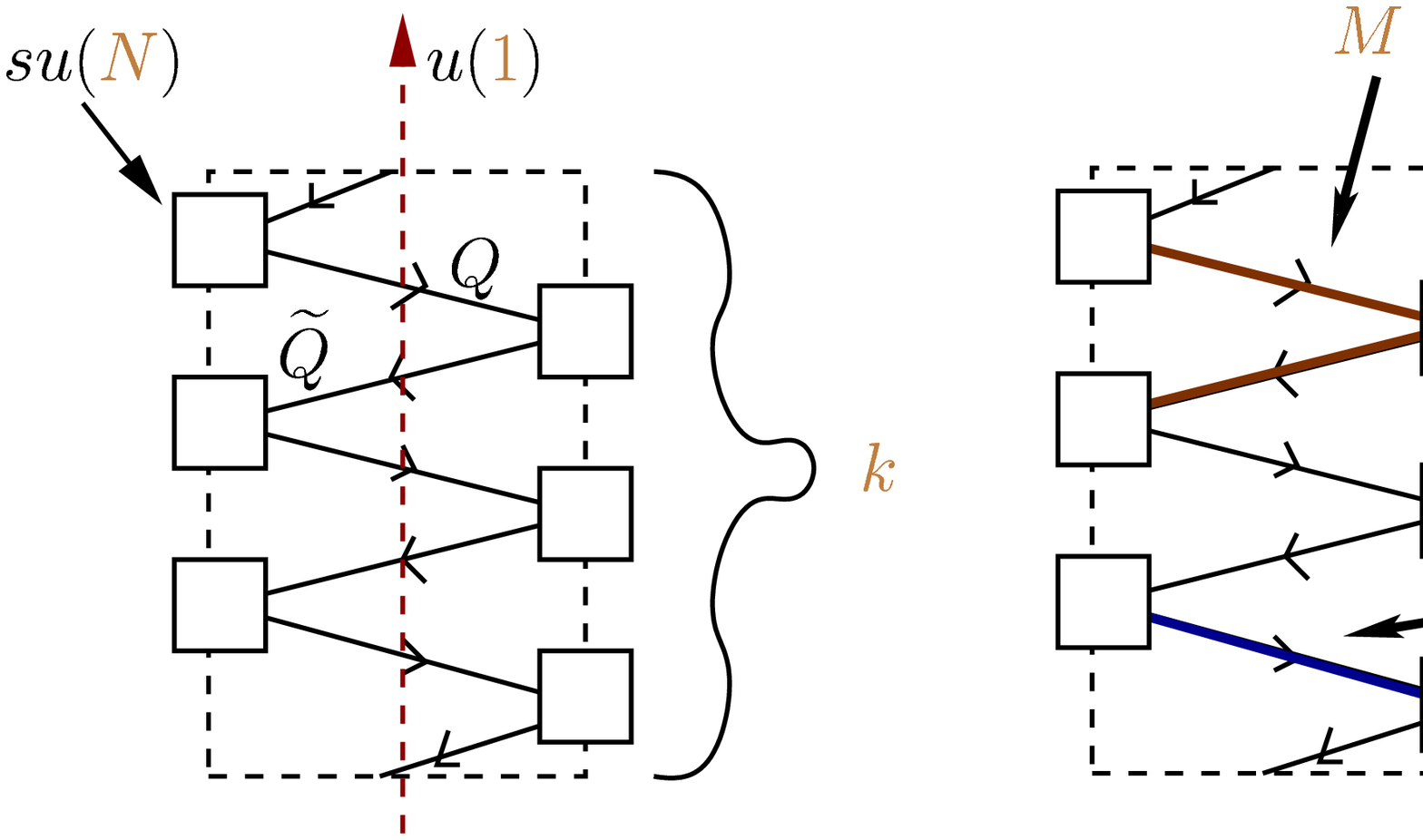}{3.7in}}
\medskip\centerline{\vbox{
\baselineskip12pt\advance\hsize by -1truein
\noindent\footnotefont{\bf Fig.~1:} The free trinion on the left and the mesonic and baryonic operators associated to punctures. The mesons are the bilinear combinations of $Q$ and $\widetilde Q$ denoted in brown in the picture. In the example here the meson is charged under the maximal puncture symmetry corresponding to the groups on the left side of the quiver and singlet under other puncture symmetries. 
Note that the meson operators in the free trinion are ``mesons'' of the maximal symmetry on the right in the usual sense of the word, that is are bilinears and singlets under that symmetry, but are operators we associate to the symmetry on the left under which they are charged in the bifundamental representation.
The baryons are operators of the form $Q^N$ and are singlets of the maximal puncture symmetries while being charge under the minimal puncture abelian symmetry.}}

We have natural operators which are charged only under the symmetry of one of the maximal punctures and internal symmetries, the mesonic operators $M^b_l=Q_l\widetilde Q_l$ and $M^a_l=Q_l\widetilde Q_{l-1}$.  There are also operators which are charged only under the minimal puncture symmetries, the baryons $\epsilon Q_l^N$ and $\epsilon\widetilde Q_l^N$. There are $k$ mesonic operators for every maximal puncture and $2k$ baryonic operators for every minimal puncture. The $k$ mesons for the two maximal punctures of the free trinion have different charges under the $2k-1$ internal symmetries. They have charge $+1$ under $u(1)_t$ and are charged under certain diagonal combination of the $u(1)^{k-1}_\beta$ and $u(1)_\gamma^{k-1}$ and have charge zero under the complementary combination. In fact there is a choice of the diagonal subgroup corresponding to mapping $u(1)^{k-1}_\gamma$ to $u(1)_\beta^{k-1}$. We could have in principle maximal punctures with mesons charged under any one of these choices. We associate a label to the maximal punctures we call color defining the choice of the diagonal abelian group, that is a map between the $u(1)_\beta$ to $u(1)_\gamma$ symmetries. In linear quivers we discuss here only a $\Z_k$ valued index will appear, \GaiottoUSA. We note that in the special case of $k=2$ and also $N=2$ the abelian symmetry under which the mesons are charged enhances to $su(2)\times u(1)$. In this case we have two mesons in the bifundamental representations of the same groups and having opposite charges under a diagonal combination of $u(1)_\beta$ and $u(1)_\gamma$. We can think of the punctures as breaking the $su(k)_L\times su(k)_R\times u(1)$ symmetry to the Cartan in general, and to $su(2)_{diag.}\times u(1)\times u(1)_t$ when $k$ and $N$ are both two. From the point of view of six dimensions the internal group is enhanced to $so(7)$ with the $su(2)_{diag.}$ a particular subgroup, which we will discuss in detail soon.
We  denote the symmetries that the punctures preserve as $P$. This will become important in what follows. We will momentarily discuss a six dimensional interpretation of the color label.

We can combine two theories together by gauging a diagonal combination of symmetries associated  to maximal punctures, one from each theory. There are two different ways to perform the gluing.
The gluing we refer to as $\Phi$-gluing, see Fig. 2,  also introduces $k$ bifundamental chiral fields $\Phi_l$ which couple to the mesonic operators through a superpotential term $\sum_l (M_l\Phi_l-\Phi_l M'_l)$, where $M_l$ and $M'_l$ are the mesons coming from the two different theories.  This gluing is the $\Z_k$ orbifold of gluing with ${\cal N}=2$ vector field in class ${\cal S}$ with the bifundamental originating from the adjoint chiral field in the ${\cal N}=2$ vector. Gluing two trinions together the resulting theory will have two maximal punctures, two minimal punctures, and also have all the additional abelian symmetries discussed above. It is important to note that the maximal punctures have a natural cyclic ordering of the $k$ $su(N)$ groups and when gluing we keep that ordering. The gluing together of free trinions into a linear quiver theory breaks no global symmetries specified above.

\

\centerline{\figscale{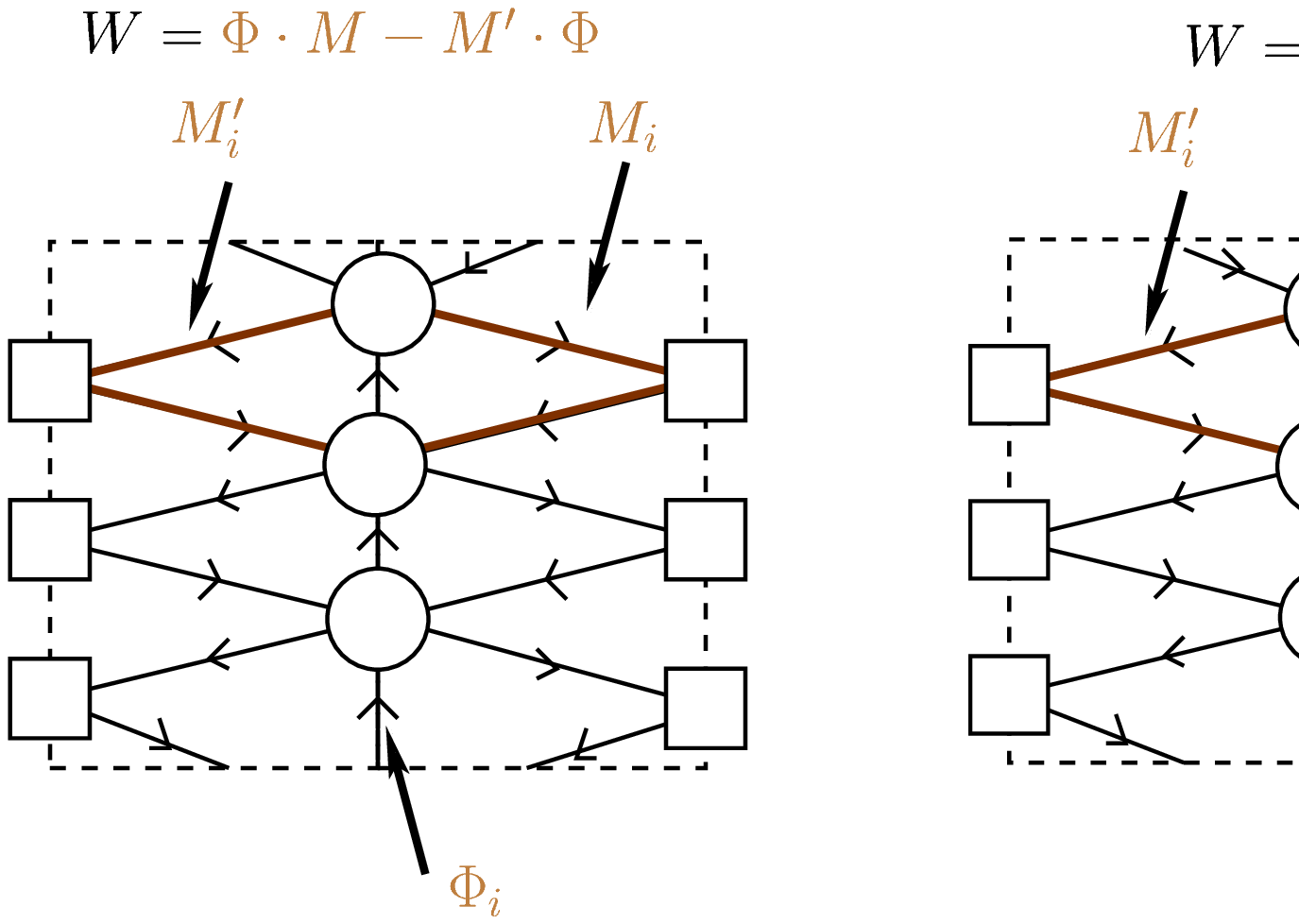}{3.5in}}
\medskip\centerline{\vbox{
\baselineskip12pt\advance\hsize by -1truein
\noindent\footnotefont{\bf Fig.~2:} Gluing two theories together. We have $\Phi$-gluing on the left and $S$-gluing on the right.}} 

The linear quivers have a conformal manifold. A sub-manifold of this preserves all the symmetries and has dimension equal to the number of minimal punctures minus one.
The linear quivers enjoy dualities which geometrically are associated to exchanging the minimal punctures. For $k=1$ the duality is just ${\cal N}=2$ S-duality of $su(N)$ theory with $2N$ flavors. For $k>1$ it is a combination of the ${\cal N}=2$ duality and Seiberg duality of ${\cal N}=1$ $su(N)$ theories with $3N$ flavors \GaiottoUSA\ .

We can also glue two maximal punctures of a linear quiver together. The theory so obtained is associated to torus with a number of minimal punctures. If the number of minimal punctures is a multiple of $k$ no symmetries are broken by the gauging, however if the number of minimal punctures is not a multiple of $k$, a $u(1)^{k-1}$ subgroup of the $u(1)^{2k-1}$ internal symmetry group is broken.
As a basic example, gluing  two maximal punctures of the free trinion one obtains the affine ${\cal N}=2$ quiver with $k$ nodes coupled to additional $k$ singlet chiral fields. This theory corresponds to torus with one minimal puncture. Removing the singlet chiral fields one interprets the affine quiver itself as associated to torus with no punctures \DelZottoRCA. This theory has $u(1)_\alpha\times u(1)_t \times  \,u(1)^{k-1}_\beta$ symmetry. It has $k+1$ dimensional manifold of conformal deformations. This manifold can be thought of as $1$ for complex structure moduli, $k-1$ for $su(k)$ holonomies, $1$ for $u(1)_t$ holonomy. In ${\cal N}=2$ language the affine quiver is a torus with $k$ punctures. We have $k$ complex structure moduli and one ${\cal N}=1$ deformation. 
 
 The breaking of the symmetry when one glues a torus can be understood by considering the $\Z_k$ valued label, color, of maximal punctures introduced above. The basic trinion has two maximal punctures with color differing by one unit. 
Cyclic shift of the color label has no physical meaning so gluing several free trinions we can always think of them as glued along maximal punctures of same color. The colors of two maximal punctures of a sphere with some number of minimal punctures differ by that number. When we glue two ends together however if the number of minimal punctures is not a multiple of $k$ we glue two punctures of different colors together and this breaks some symmetries. 

We can obtain different theories by starting from one of the theories above and closing different punctures. This corresponds to vacuum expectation values for baryonic operators, which closes minimal punctures, or mesonic ones which closes maximal punctures down to minimal ones. In general closing minimal punctures does not correspond to a theory with one minimal puncture less. This leads to the need to introduce discrete charges for the symmetries. 
 We will discuss this in more detail for the $k=2$ and $N=2$ M5 branes in what follows. 

 We can also discuss a choice of such a bundle for the additional $u(1)$ symmetry \refs{\BahDG,\GaiottoUSA,\FrancoJNA,\HananyPFA}. This corresponds to another type of gluing which we call $S$-gluing (Fig. 2) .  This is the $\Z_k$ orbifold of gluing with an ${\cal N}=1$ vector in class ${\cal S}$. Since there are no adjoint chiral fields, no additional fields are introduced by orbifolding. Therefore in $S$-gluing, we do not introduce the bi-fundamental fields when two models are glued but only turn on a superpotential coupling the mesons of the two maximal punctures. For this gauging not to break the symmetries we need to map appropriately the internal symmetries of the glued components. In particular, we couple the mesons of the two copies and these are charged under $u(1)_t$. For this symmetry to be present after the gluing the $u(1)_t$ symmetry of one theory is identified with $u(1)_{\frac1t}$ of the other theory. This leads to the notion of a {\it sign} of maximal punctures depending on the pattern of charges under $u(1)_t$. This idea was introduced in \BahDG\ and further developed in \refs{\BeemYN,\XieGMA,\AgarwalVLA,\GaiottoUSA,\HananyPFA,\FrancoJNA}.
 
 One can dwell on the question of what happens when a maximal puncture is closed. In field theory closing a puncture corresponds to turning on vacuum expectation values to operators which are charged under the puncture symmetries. For general $N$ and $k$ giving a vacuum expectation value to a meson operator for example will change the maximal puncture having $su(N)^k$ symmetry to a non maximal one having a smaller symmetry. Closing a puncture, one needs to give vacuum expectation values to several mesons completely breaking the symmetry. It was observed in \GaiottoUSA\  that doing so one does not just obtain a theory without a puncture but rather  one has a theory corresponding to a Riemann surface with one less puncture and with certain additional fluxes/discrete charges turned on. 

\ 

To summarize, the theories with Lagrangians, corresponding naturally to certain sphere and torus compactifications of M5 branes on $A_{k-1}$ singularity, possess $2k-1$ abelian symmetries which are not associated to punctures. These symmetries can be thought of as corresponding to the Cartan of $su(k)_L\times su(k)_R\times u(1)$ of the basic geometric setup of section 2. The discrete charges are possible choices of bundles/fluxes for these abelian symmetries.

\

\subsec{Higher genus theories}

We wish to construct theories associated to general surfaces. This requires introducing theories associated to a genus zero surface and having three maximal punctures. Such theories are strongly interacting SCFT's except for the $su(2)$ $k=1$ case. We thus do not have much control over them. In the $k=1$ case a lot of properties of such theories are known mostly by indirect arguments involving dualities and cross dimensional relations. 

Much more is understood for $k=1$ than for higher $k$ . The interacting three punctured sphere, trinion, here is denoted by $T_N$. These theories for example can be seen at strong coupling cusps of a linear quiver with $N-1$ minimal punctures. For $su(2)$ this quiver itself is the $T_2$ theory, but for higher $N$ the resulting theory is the $T_N$ theory with certain superconformal tail attached to it by gauging a subgroup of it's symmetry \GaiottoWE.  The construction of the interacting trinions in the $k>1$ case is more involved \GaiottoUSA. We will discuss in this paper in detail the trinions for two M5 branes and $k=2$. 
 
The addition of  handles  brings several complications. Gluing two punctures of the same theory together might break some symmetries as we mentioned. We will not consider such situations in general. 
We can also have additional moduli associated to cycles of the surface and not just to the number of tubes as we will soon discuss at length. Here we just mention that already in the ${\cal N}=2$ case the number of moduli is found to be $3g+s-3+g$ where the extra $g$ break supersymmetry down to minimal one and come from holonomies for the Cartan of the enhanced non abelian R symmetry \refs{\BahDG,\BeemYN}. Tuning the degrees of the line bundles this number can be enhanced to $3g+s-3+3g-3+s$ where the term coming in addition to the complex structure is given by the number of flat $su(2)$ holonomies \BeniniGI.

\

\

\subsec{Marginal directions} 

Let us constrain the number of marginal deformations of a theory corresponding to $M5$ branes on $\Z_k$ singularity.
To derive a constraint we will have to assume something about the theories. The assumptions we choose to make are
motivated by the linear quivers we have discussed. We assume the following,

\

\noindent$\qquad\star$ -- For each puncture we have  $k$ relevant operators $M_i$  in a bifundamental representation of an $su(N)_i\times su(N)_{i+1}$ subgroup of the $su(N)^k$ flavor group associated to the maximal puncture.

\noindent$\qquad\star$ -- Gluing two maximal punctures together can be achieved in two different ways, where in both we gauge a diagonal combination of the $su(N)^k$ symmetry of the punctures. In the first method of gluing, the $\Phi$-gluing,  we introduce $k$ fields $\Phi_i$  charged $-1$ under $u(1)_t$ and in bifundamental representation of of $su(N)_{i+1}\times su(N)_i$. We also turn on the superpotential,

\eqn\supe{
M_i\Phi_i-\Phi_i M'_i
\,.} In the second way of gluing,the $S$-gluing, no bifundamental fields are introduced and we turn on a superpotential coupling the meson fields of the punctures,

\eqn\supo{
M_iM'_i\,.
}

\noindent$\qquad \star$ -- The only new marginal operators after we perform the gluing are the ones appearing in the superpotential and the ones one can build from $\Phi_i$ in the $\Phi$-gluing, and the field strength $W^{(i)}_\alpha$ corresponding to the gauged vector fields. We also assume that operators which are marginal before gluing remain marginal.

\noindent$\qquad\star$ -- The number of marginal operators which are singlets of the symmetries associated to punctures minus the conserved currents for the internal symmetries will be assumed to be given by a linear expression,

\eqn\opsf{
m(g,s) = A g+B s+C\,.
} Here $g$ denotes the genus of the surface and $s$ the number of maximal punctures. We do not consider here more general types of punctures.  We are considering marginal directions minus the currents as this is a robust and easily computable quantity parametrizing a model. For example, it is an invariant of the Higgs mechanism for flavor symmetry \GreenDA.

\

Using these assumptions we can quickly derive a constraint on $m(g,s)$. When we glue two theories together we combine the genera and the number of punctures is the sum of punctures minus two. The number of marginal operators minus currents is the sum of these numbers of the two components plus the following new operators. 
We analyze first the $\Phi$-gluing. We have $k$ operators $M_i\Phi_i$, $k$ $M'_i\Phi_i$, $k$ gaugino bilinears $W^{(i)} W^{(i)}$, and we have $k$ fermionic opertaors $\bar \psi_i \phi_i$. Here $\psi_i$ and $\phi_i$ are the fermionic and bosonic components of $\Phi_i$. The latter operators correspond to the $k$ $u(1)$ symmetries rotating the $\Phi_i$ which are broken by the superpotential/anomalies. All in all we deduce,

\eqn\gll{
m(g_a+g_b,s_a+s_b-2) =m(g_a,s_a)+m(g_b,s_b) + 2k\,.
} The same computation applied to increasing the genus by gluing two punctures of the same theory gives,

\eqn\genb{
m(g+1,s-2)=m(g,s)+2k\,.
} In the $S$-gluing we have $k$  gaugino bilinears $W^{(i)} W^{(i)}$, and we have $k$  operators $M_iM'_i$.  Thus again every $S$-gauging introduces $2k$ marginal operators and the relations \gll\ and \genb\ derived for $\Phi$-gauging still hold. 

\

From relations \genb\ and \gll\ we deduce,

\eqn\reabc{
C=-A\,,\qquad B=\frac12 A-k\,.
} We deduce two observations. First, if we do not have any punctures the number of marginal deformations minus the currents is proportional to $g-1$. We also can write a general solution to the above constraints in a informative way,

\eqn\martyu{
m(g,s)=3g-3+s+{\Upsilon} (g-1+\frac{s}2)-\frac{s}2 \, (2k-1)\,.
} Here $\Upsilon$ is a parameter which should be integer.  Note that from our discussion in six dimensions both $\Upsilon$ and $2k-1$ should be dimensions of groups. The former is the dimension of the group preserved on the Riemann surface, $\Upsilon =dim\, G^{max}$, whence the latter is the dimension of the intersection of the group fixed by the puncture and the group preserved by the surface, $2k-1=dim(G^{max}\cap P)$. The rank of $G^{max}$ is $2k-1$ and thus we see that in a generic compactification our result is consistent with  fixing the conjugacy classes of the holonomies around punctures. There are a variety of values $\Upsilon$ can have depending on the choice of R symmetry. 
Taking $\Upsilon=2k-1$ we get that there are $u(1)^{2k-1}$ abelian symmetries. This is the symmetry we obtained from our Lagrangian constructions of linear and toric quivers, and we will argue that in a general compactification  it gives the correct dimension (minus the currents) of the conformal manifold. Another natural choice we could consider would be $\Upsilon=2k^2-1$. This gives $m(g,s)=3g+s-3+g-1+(k^2-1)(2g-2)+(k^2-k)s$ which implies that the $G^{max}$ symmetry is  $su(k)\times su(k)\times u(1)$  and the expression gives the number of complex structure moduli with flat connections for these groups.

\

The general arguments  presented here should be modified in certain cases. We give several examples. Let us consider first the $S$-gauging. The simplest special case is $k=1$. In excess to the operators we counted also $TrM^2$ and $TrM'^2$ here are gauge invariant since the mesons here are the moment map operators in the adjoint representation of the gauge group.
These operators can be also marginal in certain cases and we will assume they are in what follows. The mesons squared are associated to the each glued component and thus do not change the number of marginal directions associated to gluing, but do change the values of $\Upsilon $ we should consider. Each gauging adds then $2$ operators. Number of marginal operators minus the conserved currents is,

\eqn\kko{
m(g,s)=3g-3+s+\Upsilon(g-1+\frac{s}2)-\frac{s}2\,.
} From six dimensions the group preserved by the compactification for which the additional operators are marginal  is $so(3)$. We choose a $u(1)$ sub-group of the $so(5)$ R symmetry to perform the compactification and the commutant is the global symmetry. We thus expect $\Upsilon$ to be equal to $dim \, so(3)$ and we can write the above as,

\eqn\kkio{
m(g,s)=3g-3+s+(g-1+\frac{s}2)dim \, so(3)-\frac{s}2\, rank \, so(3)\,.
} This is the known result of~\BeniniMZ\BahDG.

Other interesting cases are when $k=2$ and/or $N=2$, where some operators which are otherwise not gauge invariant and/or marginal might become such. Take $k=2$ and general $N$. The operators $M_1M_2$ and $M'_1M'_2$ in this particular case become  gauge invariant. In some situations these become also marginal. Since these operators are from only one of the glued copies they do not change the counting of operators one has to add but this does affect the number of operators. From the six dimensional analysis the symmetry here enhances to $su(2)^3$. We thus would expect there to be marginal directions for flat holonomies of this symmetry.
In this case then we can deduce from six dimensions that $A=3+dim\, su(2)^3=12$ giving $B=4$. The number of marginal operators minus currents then can be written as,

\eqn\gauters{
m(g,s)=3g-3+s+(g-1+\frac{s}2)dim \, su(2)^3-\frac{s}2 \, rank \,su(2)^3\,.
} In the case of $N=2$ we again have more gauge invariant operators, $M_iM_i$ and $M_j'M_j'$, which come from one of the two glued theories. Also here in certain situations these are marginal. Here the dimension of the symmetry group enhances to $su(2k)$. This gives $A=3+dim \, su(2k)=4k^2+2$ which sets $B=2k^2-k+1$.  Then the dimension of the conformal manifold minus the conserved currents becomes,

\eqn\gautyy{
m(g,s)=3g-3+s+  (g-1+\frac{s}2)dim\,su(2k)-\frac{s}2 rank\,su(2k)\,.
} Finally in case both $k$ and the number of branes are equal to two we have the gauge invariant deformations, which are also marginal in some cases, derived above and two additional deformations constructed from ingredients of the two copies, $M'_1M_2$ and $M_1M_2'$. The latter operators are added when we glue theories together and thus they change the number of operators added for each gluing to six. The group here enhances to $so(7)$ which determines $A=3+dim \, so(7)=24$ and $B=9$ leading to  the marginal operators minus currents,

\eqn\dgtyp{\eqalign{&
m(g,s)=3g-3+s+(g-1+\frac{s}2) \, dim \, so(7)-\frac{s}2\,5= \cr &
\;\;\;\;\; 3g-3+s+(g-1+\frac{s}2 ) \, dim \, so(7)-\frac{s}2\,(dim \, su(2)+1+1)\,.}}
We will analyze this case in detail in what follows and will see this result emerging from computations.

\

Next we consider the $\Phi$-gauging.  When both $k$ and number of M5 branes are two  we  have additional gauge invariant  deformations becoming marginal in certain compactifications. These are $M_1\Phi_2$, $M_2\Phi_1$, $M_2'\Phi_1$, $M_1'\Phi_2$. We also have more fermionic operators, $\phi_1\bar\psi_2$ and $\bar\psi_1\phi_2$. Gluing thus changes the number of marginal directions minus currents. The six dimensional symmetries here are broken to $so(5)\times u(1)$. Thus we expect to have $A=3+dim \, so(5)u(1)=14$ while $B=4$. Number of marginal operators minus the currents is,

\eqn\martyuro{
m(g,s)=3g-3+s+(g-1+\frac{s}2)dim \, so(5)-\frac{s}2 (dim \, su(2)+1)+g-1\,.}
We will recover this expression in explicit computations in what follows. Here too in general compactification the additional marginal operators are not marginal and only for special choices of R symmetry they are exactly marginal.

\

In all the cases above we assumed that the additional operators we add are exactly marginal. Although they are marginal for any setup they are only exactly marginal for compactifications with particular choices of R symmetry. Some of the operators which are not marginal before gluing can become marginal in certain cases which will violate our assumptions. We will see how this comes about in the $k=2$ and $N=2$ case in detail in the rest of the  sections of this paper.
There can be additional cases when  operators become marginal, for example operators  made from $\Phi$s,  for some special choices of the R symmetry. It will be interesting to explore this further.

We have derived here constraints on the structure of the conformal manifolds of the four dimensional theories based on simple assumptions of what the punctures are and how you glue them together. This structure fits well with the geometric expectation of section 2 and we will see in what follows that even finer details, like values of $\Upsilon$, agree with the expectations.

\

\newsec{Two M5 branes on $A_1$ singularity: preliminaries and summary}

We will next study in detail the case of two M5 branes probing $\Z_2$ singularity. The symmetry of the six dimensional setup, which for general $\Z_k$ is $su(k)\times su(k)\times u(1)_t$, here enhances to $so(7)$~\refs{\DelZottoRCA,\OhmoriPIA}. This fact makes this case in certain respects richer than the general setup. However, we will be able to completely determine supersymmetric properties of the four dimensional theories corresponding to such compactifications. 
In this section we discuss the general properties of this case and state the main results to be derived more rigorously in the following sections. 
To avoid confusion since both $k$, the order of the orbifold, and $N$, the number of branes, are two we will try to keep the variable $N$ where the confusion between the two might arise. Everywhere in the following sections unless explicitly said otherwise $N$ stands for the number two.

\

\subsec{Symmetries and group theory}

When we compactify the six dimensional model, we can turn on a flux on the Riemann surface for an abelian subgroup $L=u(1)^r$ of the flavor symmetry, $so(7)$ in this case.\foot{In most parts of this paper we are not cautious with the global structure of the groups. In particular by quoting  groups we are referring to their Lie algebras.} The four dimensional theories obtained when compactified on a Riemann surface with no punctures will have a conformal manifold with maximal symmetry on a sub locus of it being the commutant of the chosen $u(1)^r$ in $so(7)$. We denote this symmetry by $G^{max}$. On a general point of the conformal manifold the symmetry will be given by the abelian sub-group of $G^{max}$ which is $L$. The 8 possible values of $G^{max}$ and $L$ for the $k=N=2$ theory are summarized in the table below and in table (4.7).

\eqn\theor{
\vbox{\offinterlineskip\tabskip=0pt
\halign{\strut\vrule#
&~$#$~\hfil\vrule
&~$#$~\hfil\vrule
&~$#$~\hfil\vrule
&~$#$~\hfil\vrule
&~$#$\hfil\vrule
&\strut\vrule#
\cr
\noalign{\hrule}
& G^{max}  &u(1)^3&su(2)u(1)^2   &su(2)_{diag}u(1)^2&su(2)su(2)u(1)&  \cr
\noalign{\hrule}
&  \;L       &u(1)^3  &\;\;\;\;\;u(1)^2 &\;\;\;\;\;u(1)^2&\;\;\;\;u(1) &\cr\noalign{\hrule}
& {\cal F}  &\,(a,b,c)&(a,0,b)/(0,a,b)   &\;\;\;(a,\pm a,b)&\; (a,0,0)/(0,a,0)&  \cr
\noalign{\hrule}\cr
\noalign{\hrule}
& G^{max}  &\widetilde {so(5)}u(1)& so(5)u(1)& su(3) u(1) & so(7)&  \cr
\noalign{\hrule}
&  \;L       &\;\;\;\;u(1) &\;\;\;u(1) & \;\;\;u(1) &\;\; \; \,\emptyset&\cr\noalign{\hrule}
& {\cal F}  &(a,\pm a,\;0)&\; (0,0,a)&\; (a,0,\pm a)/(0,a,\pm a)& (0,0,0)&  \cr
}\hrule}}  We will refer to compactification leading to $G^{max}$ maximal symmetry on a locus of conformal manifold as being of type $G^{max}$.  In the table  ${\cal F}$ denotes a triplet of fluxes for the Cartan of $so(7)$ one turns on. The three Cartans we  will denote as $u(1)_\beta\times u(1)_\gamma\times u(1)_t$ and will now define.
It is convenient to parametrize the symmetries with fugacities. The character of the adjoint representation of $so(7)$ we will parametrize as,

\eqn\csos{
{\bf 21}_{so(7)}= 1+{\bf 10}_{so(5)}+(t^2+\frac1{t^2}){\bf 5}_{so(5)}\,,
} where we have defined $so(5)$ characters,

\eqn\csof{\eqalign{
&{\bf 10}_{so(5)}={\bf 3}_{su(2)_1}+{\bf 3}_{su(2)_2}+{\bf 2}_{su(2)_1}{\bf 2}_{su(2)_2}\,,\cr
&{\bf 5}_{so(5)}=1+{\bf 2}_{su(2)_1}{\bf 2}_{su(2)_2}\,,
}} and $su(2)$ characters are defined as,

\eqn\csot{
{\bf 3}_{su(2)_1}=1+\frac1{\beta^4}+\beta^4\,,\quad
{\bf 3}_{su(2)_2}=1+\frac1{\gamma^4}+\gamma^4\,\quad
{\bf 2}_{su(2)_1}=\frac1{\beta^2}+\beta^2\,,\quad {\bf 2}_{su(2)_2}=\frac1{\gamma^2}+\gamma^2\,.}We choose here a somewhat unusual normalization for the $su(2)$ fugaicities which is done to 
be consistent with \GaiottoUSA\ and avoid proliferation of square roots in the equations.
In terms of these fugacities the groups appearing in the table above are as follows,

\eqn\symmfu{\eqalign{
u(1)^3 &= u(1)_t\times u(1)_\beta\times u(1)_\gamma\,,\cr
su(2)u(1)^2 &= u(1)_t\times u(1)_\beta\times su(2)_2\,,\; {or }\;\;   u(1)_t\times u(1)_\gamma\times su(2)_1 \,,\cr
su(2)_{diag}u(1)^2&=u(1)_t\times u(1)_{\beta\gamma}\times su(2)_{\beta/\gamma}\,,\; {or }\; \;u(1)_t \times u(1)_{\beta/\gamma}\times su(2)_{\gamma\beta}\,,\cr
su(2)\times su(2)u(1) & = u(1)_\beta\times su(2)_t\times su(2)_\gamma\,,\; {or }\; \;u(1)_\gamma \times su(2)_t\times su(2)_\beta\,,\cr
so(5)u(1)&=so(5)\times u(1)_t\,,\cr
\widetilde{so(5)} u(1)&=(su(2)_{\beta\gamma^{\mp1}/t} su(2)_{\beta\gamma^{\mp1}t})\times u(1)_{\beta\gamma^{\pm1}}\,,\cr
su(3)u(1)&= (su(2)_1\times u(1)_{\gamma^2t^{\pm2}})\times u(1)_{\gamma^2 t^{\mp2}}\,,\;\;{ or}\;\;\; (su(2)_2\times u(1)_{\beta^2t^{\mp2}})\times u(1)_{\beta^2t^{\pm2}}\,,\cr
so(7)&=so(7)\,.
}} 
It is informative to decompose $so(7)$ into its $so(6)$ maximal subgroup. The $so(6)$ maximal subgroup is given in terms of an $so(3)\times so(3)=su(2)\times su(2)$ decomposition as $su(2)_{\beta/\gamma}\times su(2)_{\beta\gamma}$\foot{One can also decompose $so(6)$ into $su(2)_{t}\times su(2)_{\beta\gamma}$ or $su(2)_{\beta/\gamma}\times su(2)_{t}$ with similar expressions. These are equivalent decompositions of $so(7)$ differing by a Weyl transformation.}. In particular the adjoint of $so(7)$ decomposes as ${\bf 15}+{\bf 6}$ of $so(6)$ where,

\eqn\suytin{\eqalign{&
{\bf 15}=(1+\frac{\gamma^2}{\beta^2}+\frac{\beta^2}{\gamma^2})+(1+\frac{1}{\gamma^2\beta^2}+\beta^2\gamma^2)+(1+\frac{\gamma^2}{\beta^2}+\frac{\beta^2}{\gamma^2})(1+\frac{1}{\gamma^2\beta^2}+\beta^2\gamma^2)\,,\cr&
{\bf 6}=(1+\frac{\gamma^2}{\beta^2}+\frac{\beta^2}{\gamma^2})+(1+\frac{1}{\gamma^2\beta^2}+\beta^2\gamma^2)\,.}} The $su(2)_{diag}$ that we encounter are the $su(2)_{\beta/\gamma}$ and $su(2)_{\beta\gamma}$ we see here.

Note that $so(5)$ and $\widetilde{so(5)}$ from both the six dimensional point of view and group theory wise are equivalent and related by a choice of $u(1)$ in $so(7)$. However, we treat them differently here as in the four dimensional constructions the field theoretic description of the two cases is different. Because the six dimensional origin of the two is the same the theories with $G^{max}=\widetilde{so(5)}u(1)$ and $G^{max}=so(5)u(1)$ should be dual to each other.  There are more choices of fluxes which give same groups as the ones appearing in table \theor\ having different four dimensional constructions. We list these for completeness but will not discuss these in detail in what follows as an interested reader can easily generate field theories corresponding to them from the constructions of other models.

\eqn\wetty{
\vbox{\offinterlineskip\tabskip=0pt
\halign{\strut\vrule#
&~$#$~\hfil\vrule
&~$#$~\hfil\vrule
&~$#$~\hfil\vrule
&~$#$\hfil\vrule
&\strut\vrule#
\cr
\noalign{\hrule}
& G^{max}  &su(2)u(1)^2   &su(2)_{diag}u(1)^2&su(2)su(2)u(1)&  \cr
\noalign{\hrule}
&  \;L        &\;\;\;\;\;u(1)^2 &\;\;\;\;\;u(1)^2&\;\;\;\;u(1) &\cr\noalign{\hrule}
& {\cal F}  &(a,b,\pm a\pm b)   &\;\;\;(a,b,0)&\;(a,\pm a,\pm 2 a)&  \cr
}\hrule}} 
\medskip\centerline{\vbox{
\baselineskip12pt\advance\hsize by -1truein
\noindent\footnotefont{} Other flux choices leading to symmetry enhancement. Note that the signs appearing in the table are not correlated.}}

We will obtain four dimensional ${\cal N}=1$ QFTs corresponding to each one of the choices for $G^{max}$ in \theor. The resulting theories are different but they are related by RG flows triggered by either superpotentials or vacuum expectation values of operators charged under puncture symmetries. We will discuss the latter next and here let us comment on the former.
Let us denote as ${\cal M}_{g,s}^{G^{max}}$ the conformal manifold of a theory corresponding to genus $g$ with $s$ maximal punctures compactification of type $G^{max}$. Then we will find relevant operators which will bring us from the conformal manifold of a theory with a given $G^{max}$ to ones  with larger $G^{max}$. For example,

\eqn\flg{
{\cal M}_{g,s}^{u(1)^3}\;\to\;
{\cal M}_{g,s}^{su(2)_{diag}u(1)^2}\;\to\;
{\cal M}_{g,s}^{so(5)u(1)}\;\to\;
{\cal M}_{g,s}^{so(7)}\,.
} In particular theories of type $so(7)$ with no punctures have no relevant deformations.  The reason for the fact that theories with more fluxes has more degrees of freedom is not apriori clear.  One explanation may be that with flux we get more zero modes on the Riemann surface leading to more degrees of freedom surviving in 4d.\foot{We thank Thomas Dumitrescu for a discussion on this point.} 
We will see how this type of flows arise in what follows. In all the examples listed here we will find that as expected from \dimm

\eqn\dimmp{
dim{\cal M}_{g,0}^{G^{max}} =(g-1)dim \; G^{max}+dim \,L+3g-3\,.
}

\

\subsec{Punctures and gluings}

In each one of the different types of compactification we can introduce maximal and minimal punctures as well as more general punctures (see \refs{\GaiottoUSA,\HeckmanXDL}) which we will not consider here. Let us first discuss the maximal punctures.  These punctures introduce an $su(N)^2$ ($su(2)_a\times su(2)_b$) factor into the four dimensional global symmetry. However, they also break some of the $so(7)$ symmetry of the six dimensional set-up, see for example \BeniniMZ. The maximal punctures are classified according to their color, sign, and orientation. Color and sign are related to embedding of the symmetry preserved by the puncture in $so(7)$. Different embeddings are related by a Weyl transformation of $so(7)$. Though group-theoretically they are equivalent, when two punctures with two different embeddings appear in the same model,  and the element of the Weyl group relating them is not in $G^{max}$, the difference between the embeddings is physically meaningful.  Very concretely the choices of embeddings are classified as follows.
Color determines what is the group un-broken by the punctures. The symmetry preserved by the full punctures turns out to be an $su(2)_{diag}u(1)^2$ subgroup of $so(7)$. One has several choices for the embedding of the puncture symmetry in $so(7)$ which is conveniently parametrized by taking an $so(5)u(1)\subset so(7)$,  for which we have three natural choices, and then also an $su(2)^2$ decomposition of $so(5)$. The preserved $su(2)_{diag}$ is the diagonal combination of $su(2)^2$. We will restrict our investigation in what follows, mostly, to the decomposition $so(5)u(1)_t$ since it can be generalized to higher $k$. 
The two possibilities for the puncture symmetry $P$ are $su(2)_{\gamma/\beta}\times u(1)_{\beta\gamma}\times u(1)_t$ and $su(2)_{\beta\gamma}\times u(1)_{\beta/\gamma}\times u(1)_t$. 
We denote these groups as $P_1$ and $P_2$ respectively. The groups $P_1$ and $P_2$ are not to be confused with the $su(N)^2$ group associated to the puncture, rather these are sub-groups of $so(7)$ preserved by the puncture. The orientation is related to the ordering of the two $su(N)$ factors associated to the puncture $(su(2)_a, su(2)_b)$ or $(su(2)_b,su(2)_a)$. The sign is related to the two choices of embedding of $u(1)_t$ inside $so(7)$ which differ by complex conjugation\foot{Note that for $so(7)$ the operation that inverts the charge under $u(1)_t$ is part of the Weyl group. While also true for the $k=2$ case, this fails in the $k>2$ case. In those cases the sign appears to be related to embeddings differing by an outer automorphism while the color to ones differing by an inner automorphism.}. This is illustrated by considering the mesonic operators associated to the maximal puncture. These operators are in the bifundamental representation of $su(2)_a\times su(2)_b$. We will associate to them R-charge one which in general will not be the superconformal one. They are in the ${\bf 2}$ of $su(2)_{\beta\gamma}$ and have zero charge under $u(1)_{\beta/\gamma}$ for punctures of one color, and in the ${\bf 2}$ of $su(2)_{\beta/\gamma}$ and have zero charge under $u(1)_{\gamma\beta}$ for the other color. The mesons have $u(1)_t$ charge $+1$ for negative punctures, and $-1$ for positive ones. As far as orientation goes the mesonic operators are the same but other operators can and are different.
 
\

When gluing two punctures together we glue punctures of opposite orientation and of the same color.  The curvature triplet ${\cal F}$ of the theory obtained by gluing two surfaces is the sum of the triplets associated with the two summands. This is a non trivial statement and we will check its validity in numerous examples in the next sections.
In field theory the gluing is associated with gauging the diagonal combination of the $su(N)^2$ symmetry associated to the punctures.  When gluing punctures of the same sign we use the $\Phi$ gluing,  introduce dynamical vector fields for $su(N)^2$ and bi-fundamental fields $\Phi$ in a doublet representation of $su(2)_{diag}$ which flip the mesonic operators associated to the punctures~\GaiottoUSA, 

\eqn\suposf{
W=\Phi\cdot M-M'\cdot \Phi\,.
} The different ordering here denotes the way the two factors of $su(N)^2$ are contracted between $\Phi$ and the mesons coming from the oppositely oriented punctures. When gluing two opposite sign punctures together we use the $S$ gluing,  introduce the dynamical vector fields and couple the mesonic operators of the two maximal punctures to each other forming a singlet of $su(2)_{diag}$~\refs{\GaiottoUSA,\HananyPFA,\FrancoJNA},
\eqn\supoopk{
W= M\cdot M'\,.
}
Since the mesonic operators have opposite $u(1)_t$ charges and are not charged under the additional $u(1)$s no symmetries are broken. 

\

The dimension of the conformal manifold of  theory of type $G^{max}$ with $s_2$ punctures preserving $P_2$ and $s_1$ punctures preserving $P_1$ will be found to have the form (consistent with eq. \dimpm),

\eqn\comg{\eqalign{&
dim {\cal M}_{g,(s_1,s_2)}^{G^{max}} =3g-3+s_2+s_1 + \cr
&\;\; (g-1+\frac{s_2+s_1}2) dim\ G^{max}-\frac{s_1}2 dim (G^{max}\cap P_1) -\frac{s_2}2 dim (G^{max}\cap P_2)+dim\,L\,.
} } 

\

We can give vacuum expectation values to one of the two mesonic operators associated to the maximal puncture. This will break the flavor symmetry of the puncture to $u(1)$. This symmetry is associated to a minimal puncture. There are two independent ways to give such a vacuum expectation value for a maximal puncture of a given color~\GaiottoUSA. 
The remaining $u(1)$ symmetry in the two choices is the Cartan of one of the two possible diagonal $su(2)$s built from $su(N)^2$ associated to the maximal puncture. The theory in the IR is associated to compactification on a surface with the maximal puncture traded with minimal one and the triplet of curvatures shifted by ${\cal F}=(\pm\frac14,\pm\frac14,\pm\frac12)$ relative to the original theory with the signs determined by the color and sign of the original puncture, and  the choice of the meson for the vacuum expectation value. We note that the above fluxes are either the minimal or half the minimal possible ones for the corresponding symmetries, depending on the global properties of the flavor symmetry group. We shall discuss the normalization for fluxes in section seven.

We can use this to gain an understanding of the color parameter of the punctures. The minimal punctures only differ by their sign \GaiottoUSA\ and do not carry the color label.  The positive puncture preserving symmetry $P_1$ when closed shifts the curvature triplet with ${\cal F}=(\pm\frac14,\,\pm \frac14,\;-\frac12)$, whence the positive one preserving $P_2$ shifts by ${\cal F}=(\pm\frac14,\;\mp\frac14,\,-\frac12)$. For negative maximal punctures the shifts are minus of the above. We will check this association of fluxes with six dimensional computations through computations of central charges which depend explicitly on the fluxes. We will perform the computation in six dimensions only for the surfaces with no punctures. However, since the construction of those in four dimensions involves combining theories with punctures it will be a strong check of this logic.
  We will give more details in next sections.
  
  We will not deal in detail with minimal punctures in what follows. Let us mention that closing a minimal puncture can be achieved by turning on a vacuum expectation value to baryonic operators and this will shift the fluxes for positive minimal punctures by ${\cal F}=(\pm\frac12,0,-\frac12),\;\,(0,\pm\frac12,\,-\frac12)$ depending on choice of the baryon, and the same with opposite signs for negative punctures.
   Here we also want to mention that the triplet of charges associated to the free trinion, which is sphere with two maximal punctures of opposite color but same signs and orientations, is ${\cal F}=(0,0,\pm\frac12)$ with the sign determined by the sign of the punctures which is the same for all three here.

\

We will next define the trinion theories corresponding to the different types of compactifications. We will glue from these trinions theories which correspond to different $G^{max}$, genera, and numbers of punctures. In the following sub-sections we will  state the basic facts of the $G^{max}=so(7)$ and $G^{max}=so(5)u(1)$ models and postpone the more general cases and rigorous but technical derivations to the next sections.

\

\subsec{The $G^{max}=so(7)$ models} 

\

We begin unfolding the story with constructing the field theories corresponding to compactifications with no flux for $so(7)$ turned on. 
In what follows we just state what is the theory we conjecture is obtained by such a compactification. The derivation of this result follows several steps detailed in the next sections.

Let us consider the theory of Fig. 3. This model is constructed from two copies of $su(2)$ SQCD with four flavors coupled through quartic superpotentials coupling the mesons of the two copies. We can also construct this model by gluing two free trinions using the $S$ gluing. Since here the groups are $su(2)$ we can turn on quartic superpotential coupling all the mesonic operators associated to maximal punctures, or saying this differently, turning on the most generic quartic superpotentials preserving puncture symmetries breaks all the $u(1)_t\times u(1)_\beta\times u(1)_\gamma$ symmetry.
Alternatively we can turn on only a subset of quartic couplings. One such choice is turning on the couplings preserving the diagonal $su(8)$ global symmetry of the two copies, this is the theory discussed in \DimoftePD. Another choice preserving larger rank symmetry~\GaiottoUSA\ is to couple the mesons as in Fig. 2. All these different choices sit on the same conformal manifold as the superconformal R-symmetry determined by these superpotentials, and thus the conformal anomalies, are the same.  The number of marginal operators (minus the currents) is large and equals $1330$. It was claimed in \DimoftePD\ that on some locus of this manifold the flavor symmetry enhances to $E_7$.

\

 \centerline{\figscale{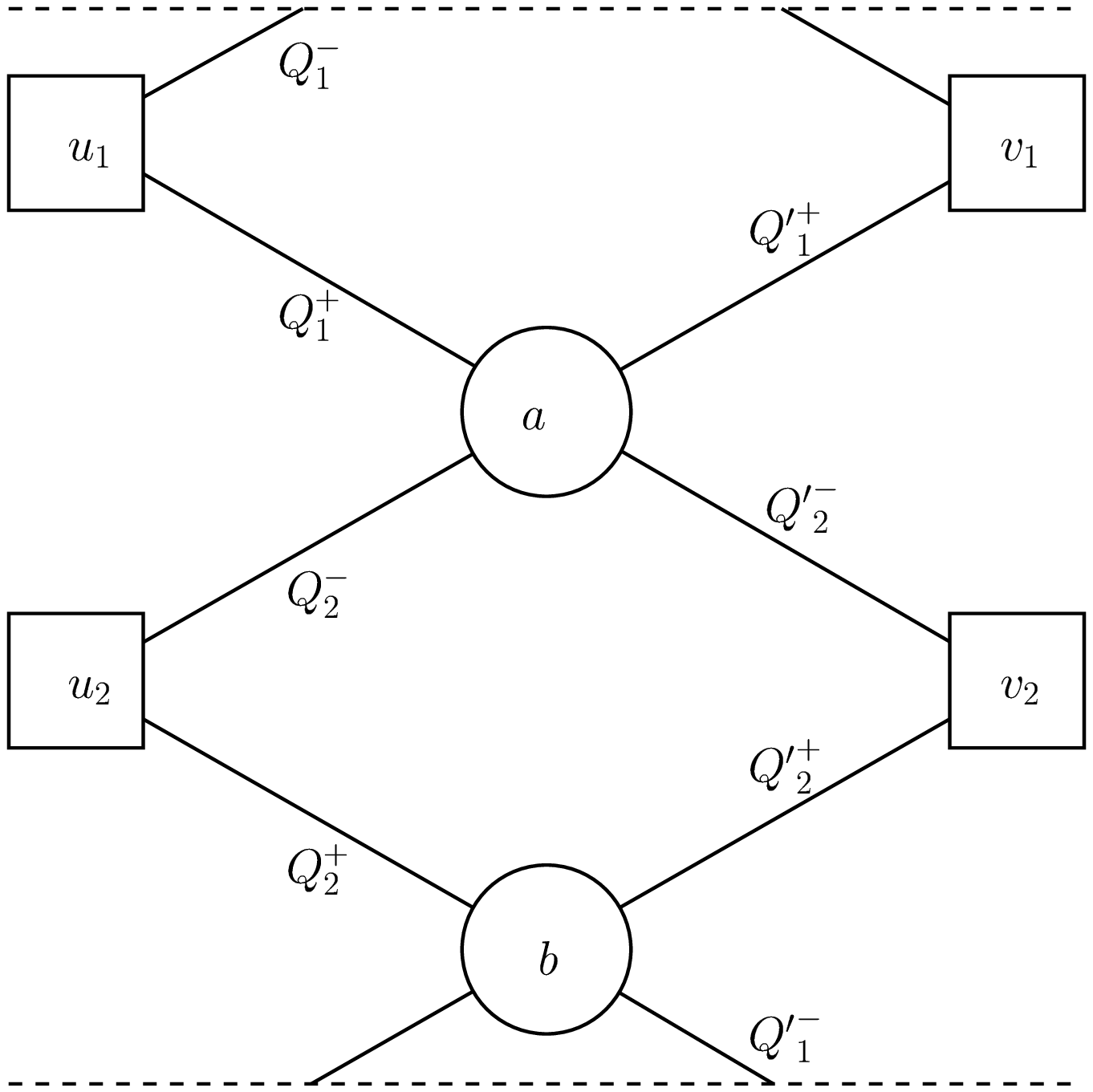}{2.1in}}
\medskip\centerline{\vbox{
\baselineskip12pt\advance\hsize by -1truein
\noindent\footnotefont{\bf Fig.~3:} The $E_7$ surprise theory. The boxes are $su(2)$ flavor groups, circles $su(2)$ gauge groups. Fields $Q_i^\pm$ have charge $\pm1$ under $u(1)_\alpha$, and fields ${Q'}^\pm_i$ have charge $\pm1$ under $u(1)_\delta$. The quiver is drawn on a cylinder and has quartic superpotential interactions preserving the four $su(2)$s and the two $u(1)$s. All the fields have R-charge half.}} 

\

We want to understand this theory starting from the Lagrangian of Fig. 3. In this language we have two copies of $su(N)^2$ factors in the flavor symmetry along with five abelian factors.
  As we discussed in previous sections two of the abelian factors correspond to minimal punctures, which we denote by $u(1)_\alpha$ and $u(1)_\delta$, and three to the Cartan of the $so(7)$ symmetry of the six dimensional theory, which is denoted as $u(1)_\beta\times u(1)_\gamma\times u(1)_t$.  In the caption of Fig. 3 we identified an $su(N)^2_u\times su(N)^2_v\times u(1)_\delta\times u(1)_\alpha$ subgroup of the global symmetry.

We start exploring the conformal manifold by turning on exactly marginal operators. There are such operators which preserve the $u(1)_\alpha\times u(1)_\delta\times u(1)_{\gamma/\beta}$ and the Cartan of the symmetry associated to the maximal punctures. That is we break the $u(1)_{\beta\gamma}\times u(1)_t$ symmetry and the non-abelian structure of the puncture symmetries.
At the point of $E_7$ enhancement \DimoftePD\ the $u(1)_\alpha\times u(1)_\delta$ symmetry in particular enhances to $su(2)_{\alpha/\delta}\times su(2)_{\delta\alpha}$ and $u(1)_{\gamma/\beta}$ enhances to $su(2)_{\beta/\gamma}$.
We have seven factors of $su(2)$ groups which enhance to $E_7$. Denoting $(\alpha/\delta,\delta\alpha)$ as $(w_1,w_2)$ the factors $su(N)^2_u$, $su(N)^2_v$, $su(N)^2_w$, appear completely symmetrically. We claim that this theory resides on the conformal manifold of  the trinion theory for the $so(7)$ compactifications with the $su(N)^2$ symmetries associated to the maximal punctures. Note that this theory is {\it almost} Lagrangian. The only subtlety here is that we have to tune the couplings of the model in Fig. 3 to the symmetry enhanced point. The vector of fluxes of this theory is ${\cal F}=(0,0,0)$.
We state this here and will verify this statement by comparing the conformal central charges later in the paper.

\

We can glue such theories together to form arbitrary Riemann surfaces by gauging the symmetries associated to the punctures. When doing so we will in general turn on  superpotentials coupling the mesons of the glued theories.

 We can readily compute the indices of such models. Since the indices are the same independent of the locus of the conformal manifold we can compute them from the Lagrangian of Fig. 3. In particular one observes the enhancement of symmetry in the index. The index of general genus $g$ theory with $s$ maximal punctures is then (suppressing fugacities for $u(1)_\beta\times u(1)_\gamma\times u(1)_t$), \foot{In appendix D the interested reader can find all the details of the index and anomaly computations reported in the rest of the paper.}
 
\eqn\sosi{\eqalign{
&{\cal I}^{(so(7))}_{g,s}= 1+ 2\sum_{j=1}^s{\bf 2}^{(j)}_1{\bf 2}^{(j)}_2\,(p q)^{\frac12} + \biggl\{(g-1+\frac{s}2)dim(so(7))-\frac{s}2dim(su(2)_{diag}u(1)^2) +\cr &\;\;\;\;\;\;\;3g-3+s+ 3 \sum_{j=1}^s{\bf 3}^{(j)}_1{\bf 3}^{(j)}_2+\sum_{j=1}^s({\bf 3}^{(j)}_1+{\bf 3}^{(j)}_2)+
4\sum_{j\neq l}{\bf 2}^{(j)}_1{\bf 2}^{(j)}_2{\bf 2}^{(l)}_1{\bf 2}^{(l)}_2\biggr\} pq  +  \cr
&+2\sum_{i=1}^s{\bf 2}^{(i)}_1{\bf 2}^{(i)}_2 (p q)^{\frac12}(p+q)+\cdots .}
} Note that the superconformal R-charges are here completely fixed by the interactions since on a generic point of the conformal manifold there are no abelian symmetries left. We readily see from this expression that there is a conformal manifold on which the puncture symmetries are not broken, dimension of which is,

\eqn\sisos{
dim{\cal M}^{so(7)}_{g,s} = (g-1+\frac{s}2)dim(so(7))-\frac{s}2dim(su(2)_{diag}u(1)^2)+3g-3+s\,.
} Since here $so(7)\cap su(2)_{diag}u(1)^2=su(2)_{diag}u(1)^2$ and $L=\emptyset$ this is exactly what we would expect from eq. \dimpm.  We remind the reader that the marginal operators reside in order $p q$ of the index together with the conserved currents which appear with a negative sign \BeemYN. Computing the dimension for three punctured sphere the above implies that $dim\, {\cal M}_{0,3}=3$. We can compute  this dimension using the Lagrangian. At the point of enhancement to $E_7$ the marginal operators form  ${\bf 1463}$ representation of $E_7$. There are thus no exactly marginal directions preserving the full $E_7$ symmetry. However we have exactly marginal deformations preserving the puncture symmetries, three copies of $su(N)^2$. The group $E_7$ decomposes into $so(12)\times su(2)_{\beta/\gamma}$, and we can further decompose $so(12)$ to $su(N)^2\times su(N)^2\times su(N)^2$ (remember that $N$ stands for $2$).  In particular we obtain two singlets of $su(N)^2\times su(N)^2\times su(N)^2$ which are in the adjoint of $su(2)_{\beta/\gamma}$. Turning on this operators we break $su(2)_{\beta/\gamma}$ and thus three of the six marginal operators combine with the current of $su(2)_{\beta/\gamma}$ to form long multiplet leaving behind three exactly marginal directions as \sisos\ tells us (see appendix E for details). Note that at the $E_7$ enhanced point, the $u(1)_t$ and $u(1)_{\beta\gamma}$ symmetries are broken. This implies that for the symmetry to enhance to $E_7$, flat connection for the two symmetries needs to be turned on, and where
the puncture symmetries are not aligned within the $so(7)$ group.

The central charges of this models are obtained readily through the Lagrangians with the fixed R symmetry,

\eqn\anl{
a=\frac38((g-1)17+7s)\,,\qquad\quad 
  c=\frac18((g-1)52+23 s)\,.} 
Again, these quantities are the same for all loci of the conformal manifold and thus can be read off from the Lagrangian of Fig. 3 for the three    punctured  sphere,  even though it describes a corner of the moduli space at which enhancement of symmetry does not occur. For generic genus and number of punctures the anomalies of the trinions and the vectors just combine as the superconformal R symmetry is fixed, and no abelian symmetries survive at a generic point on the conformal manifold.

We will rederive these models a bit more constructively in section 6 and match the anomalies \anl\ with the six dimensional computation in section 7. In particular we will see in section 6 that with no punctures there is a locus on the conformal manifold on which we have the full $so(7)$ symmetry. We note that the $so(7)$ theory is the analogue of the Sicilian theories of \BeniniMZ\ in the $k=1$ case.
 
\

In the language of the previous section and from \sisos\ we have $A=24$ and $B=10$, which implies that each gluing add $4$ marginals minus currents. These are exactly the $2k$ we obtained on general grounds. 

\

\

\subsec{The $G^{max}=so(5)u(1)$ models}

We now turn to theories which are associated to compactifications with some specific  $u(1)_t$ flux turned on.
The theory discussed here is the closest analogue of the ${\cal N}=2$ theories in the $k=1$ case.
 These theories, when punctures are not present, will have $so(5)\times u(1)_t$ symmetry on some locus of their conformal manifold.
We start from the theory which was associated to a sphere with two maximal and two minimal punctures which is just the orbifold of ${\cal N}=2$ $su(2)$ SQCD with four flavors \GaiottoUSA. We detail the fields in Fig. 4.

\

\centerline{\figscale{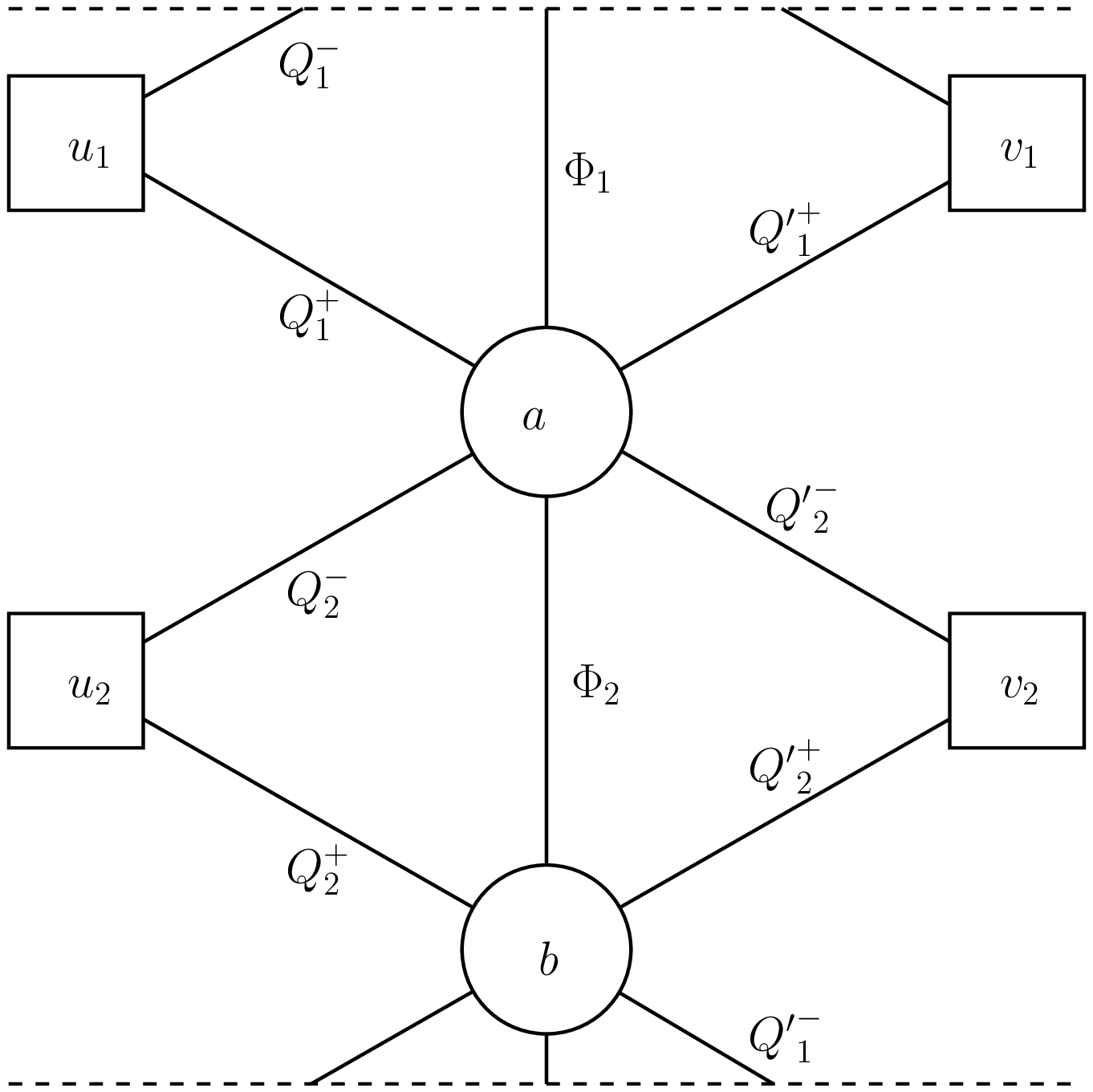}{2.1in}}
\medskip\centerline{\vbox{
\baselineskip12pt\advance\hsize by -1truein
\noindent\footnotefont{\bf Fig.~4:} The $\Z_2$ orbifold of ${\cal N}=2$ $su(2)$ SQCD with four flavors. The quiver is drawn on a cylinder. There is a superpotential term for each triangle in the graph.}} 

\

Throughout this paper we will refer to  this theory as {\it the orbifold theory}. The theory has various symmetries charges under which are summarized in the following table.

\eqn\theor{
\vbox{\offinterlineskip\tabskip=0pt
\halign{\strut\vrule#
&~$#$~\hfil\vrule
&~$#$~\hfil
&~$#$~\hfil\vrule
&~$#$~\hfil
&~$#$\hfil\vrule
&~$#$\hfil
&~$#$\hfil\vrule
&~$#$\hfil
&~$#$\hfil\vrule
&~$#$\hfil\vrule
&~$#$\hfil\vrule
&~$#$\hfil\vrule
&~$#$\hfil\vrule
&\vrule#
\cr
\noalign{\hrule}
&  &  su(2)_a & su(2)_b & u(1)_\delta   &u(1)_\alpha& su(2)^u_1 & su(2)^u_2 & su(2)^v_1 & su(2)^v_2& u(1)_\beta & u(1)_\gamma&u(1)_t&u(1)_R \cr
\noalign{\hrule}
&  Q^+_1         & \; \quad {\bf  2}     & \quad {\bf 1} &\; -1&\quad 0   &\quad {\bf 2} & \quad {\bf 1}  &    \quad  {\bf 1}  & \quad {\bf 1} & \quad 1&\quad 0&\quad\frac12&\quad\frac23&\cr
&  Q^-_1   & \; \quad {\bf  1}     & \quad {\bf 2}    &\quad 1&\quad 0   &\quad {\bf 2} & \quad {\bf 1}  &    \quad  {\bf 1}  & \quad {\bf 1} & \quad 0&\quad 1&\quad\frac12&\quad\frac23&\cr
&  Q^+_2         & \; \quad {\bf  1}     & \quad {\bf 2} &\; -1&\quad 0   &\quad {\bf 1} & \quad {\bf 2}  &    \quad  {\bf 1}  & \quad {\bf 1} & \; -1&\quad 0&\quad\frac12&\quad\frac23&\cr
&  Q^-_2   & \; \quad {\bf  2}     & \quad {\bf 1}    &\quad 1&\quad 0   &\quad {\bf 1} & \quad {\bf 2}  &    \quad  {\bf 1}  & \quad {\bf 1} & \quad 0&\; -1&\quad\frac12&\quad\frac23&\cr
&  {Q'}^+_1         & \; \quad {\bf  2}     & \quad {\bf 1} &\quad 0&\quad    1&\quad {\bf 1} & \quad {\bf 1}  &    \quad  {\bf 2}  & \quad {\bf 1} & \quad 0&\quad 1&\quad\frac12&\quad\frac23&\cr
&  {Q'}^-_1   & \; \quad {\bf  1}     & \quad {\bf 2}    &\quad 0&\; -1   &\quad {\bf 1} & \quad {\bf 1}  &    \quad  {\bf 2}  & \quad {\bf 1} & \quad 1&\quad 0&\quad\frac12&\quad\frac23&\cr
&  {Q'}^+_2         & \; \quad {\bf  1}     & \quad {\bf 2} &\quad 0 &\quad 1   &\quad {\bf 1} & \quad {\bf 1}  &    \quad  {\bf 1}  & \quad {\bf 2} & \quad 0&\; -1&\quad\frac12&\quad\frac23&\cr
&  {Q'}^-_2   & \; \quad {\bf  2}     & \quad {\bf 1}    &\quad 0&\; -1   &\quad {\bf 1} & \quad {\bf 1}  &    \quad  {\bf 1}  & \quad {\bf 2} & \; -1&\quad 0&\quad\frac12&\quad\frac23&\cr
\noalign{\hrule}
&  \Phi_1      & \; \quad {\bf  2}     & \quad {\bf 2} &\quad 0 &\quad 0   &\quad {\bf 1} & \quad {\bf 1}  &    \quad  {\bf 1}  & \quad {\bf 1} & \; -1&\; -1&\; -1&\quad\frac23&\cr
&  \Phi_2       & \; \quad {\bf  2}     & \quad {\bf 2} &\quad 0 &\quad 0   &\quad {\bf 1} & \quad {\bf 1}  &    \quad  {\bf 1}  & \quad {\bf 1} & \quad 1&\quad 1&\; -1&\quad\frac23&\cr
}\hrule}}  The theory has a superpotential,

\eqn\spro{
W=Q^+_1Q^-_1 \Phi_1 - Q^+_2 Q^-_2 \Phi_2 + {Q'}^+_1{Q'}^-_1 \Phi_1-{Q'}^+_2{Q'}^-_2\Phi_2\,.
} The symmetry of this theory actually enhances to $su(4)\times su(4)\times u(1)_t\times u(1)_{\alpha\delta}\times u(1)_{\gamma\beta}$ but it will be convenient for us to discuss it in terms of the  sub-group appearing in the table.
The two factors of $su(4)$ in the $su(2)\times su(2)\times u(1)$ decomposition are,

\eqn\sfs{\eqalign{&
su(4)_1 \to su(2)_{u_1}\times su(2)_{v_1}\times u(1)_{\sqrt{\alpha\beta\delta/\gamma}}\,,\cr&
su(4)_2 \to su(2)_{u_2}\times su(2)_{v_2}\times u(1)_{\sqrt{\alpha\gamma\delta/\beta}}\,.}}

The orbifold theory has a six dimensional conformal manifold, see appendix E. On a one dimensional locus of the manifold no symmetries are broken. In \GaiottoUSA\ this locus was identified as corresponding to the complex structure modulus of the four-punctures sphere. On a general point of the conformal manifold the symmetry $u(1)_{\beta\gamma}$ is broken as well as all the $su(2)$ symmetries being broken to their Cartan. In particular we now have six abelian symmetries which come from the punctures. It is easy to see, for example from index computations, that these symmetries come on the same footing. To be more precise, the diagonal and anti-diagonal combinations of the Cartans of $su(N)^2$ symmetries associated to the punctures are interchangeable with $u(1)_\alpha$ and $u(1)_\delta$ symmetries associated to the minimal punctures. 

The conformal manifold thus has an interesting structure. At a generic point it looks as if it should correspond to a sphere with six minimal punctures. At a special one dimensional loci it is natural to associate it to a sphere with two minimal and two maximal punctures. There should be loci where it should correspond to one maximal and four minimal punctures. It is also plausible that at some locus the minimal punctures combine in pairs to form three maximal punctures. We interpret the theory at that locus as the trinion of the $so(5)$ models. Again, we will give a more constructive derivation of this claim in section six. As in previous subsection this trinion is almost Lagrangian as it involves going to  a special point of a conformal manifold of  a theory with completely standard Lagrangian.  The vector of fluxes of this theory is the sum of the ones for free trinions and is ${\cal F}=(0,0,1)$. This is obtained as twice the flux of free trinion and we will test this statement against six dimensional computation of anomalies in section seven.

We can glue trinions together to form general Riemann surfaces. The gluing here corresponds to gauging a diagonal combination of the $su(N)^2$ symmetry as well as introducing bi-fundamental fields $\Phi_i$ coupled to the mesonic operators associated to the punctures. The gauging in this theory is conformal, a fact that we want to stress here.

Using this Lagrangian we can compute the indices and the conformal anomalies of the models. The former is given by,

\eqn\incvuppe{\eqalign{
&{\cal I}^{so(5)}_{g,s}({\bf u}_i) =
\cr&\;\;\;\; \; 1+
\left((3g-3+s)5+s\right)\frac{(p q)^{\frac23}}{t^2}+
\left(2\;\,t\sum_{j=1}^s (u^{(1)}_j)^{\pm1}(u^{(2)}_j)^{\pm1}\right)
(p q)^{\frac23}+\cr
&\;\;\;\biggl\{(g-1+\frac{s}2)10-2s+3g-3+s+g-1\biggr\}p q+\cdots\,.
}}  We us here the R-symmetry of table \theor .
 The conformal manifold here has directions on which the puncture symmetries are broken and we will discuss this in detail in the next sections. Focusing on directions on which the puncture symmetries are not broken
we note that the parameter $A$  is $14$ and $B=4$. From here the number of deformations which are added when we glue is $6$ which corresponds to the $M_i\Phi_j$, $M_i'\Phi_j$, gaugino bilinear minus the four operators $\bar \psi_i\phi_j$.  

The conformal manifold when we turn on only maximal punctures and preserve their symmetry has dimension,

\eqn\dimcsop{
dim{\cal M}^{so(5)}_{g,s}=(g-1+\frac{s}2)dimso(5)u(1)-\frac{s}{2} dim (su(2)_{diag}u(1)^2) +3g-3+s+1\,.
} Here we have the symmetry preserved by the punctures is a subgroup of $so(5)u(1)$ and this is equal to the intersection of the two. We have $L=u(1)$ and thus the last $1$ in the equation.

The conformal anomalies here can be easily computed using the Lagrangian. The only symmetry which can be admixed to the $u(1)_R$ symmetry is $u(1)_t$. Performing a maximization we find that the R-charges of \theor\ are actually the superconformal ones. In particular this implies that the gluings in the $so(5)$ theories are superconformal with the gauge couplings being exactly marginal when correlated with suitable superpotentials. The anomalies here are given by,

\eqn\angst{
a=\frac1{24}((g-1)187+78 s)\,,\qquad\;\;\; c=\frac1{12}((g-1)97+42 s)\,.
} We will derive these models in a more systematic way in section 6 and in section 7 we will see by matching central charges that they correspond, when no punctures appear, to compactifications with $2g-2$ units of $u(1)_t$ flux. These theories are the analogues of the ${\cal N}=2$ twist in the $k=1$ case.  We will construct $so(5)$ theories with other values for this flux in the next sections.

\

\

\newsec{The $G^{max}=su(2)_{diag}u(1)^2$ models: derivation from dualities}

We now turn our attention to systematic derivation of $su(2)_{diag}u(1)^2$ theories. This derivation is rather technical and relies heavily on the results of \GaiottoUSA. However, we will be able to use the construction of this section to derive and extend the results of the previous sections as well as to deduce the theories which should correspond to compactifications with arbitrary fluxes.

\

This section involves many  technical considerations so let us first outline the general procedure. We will start by constructing certain theories which are naturally associated to some compactification on a three punctured sphere, {\it i.e.} trinions with maximal punctures. The construction appears  in \GaiottoUSA\ and we
will make it more explicit here. This involves harnessing dualities of linear quivers and constructing ``strongly coupled'' Lagrangians for certain trinions. 
The Lagrangians will have some of the couplings tuned to infinite values and some of the gauged symmetries not
manifest in the description. It is analogous to the construction of a ``strongly coupled'' Lagrangian for the $E_6$ SCFT in \RazamatL.\foot{For somewhat related ideas for constructing Lagrangians by giving up on some symmetries of the models see \MaruyoshiTQK.} 

\

\centerline{\figscale{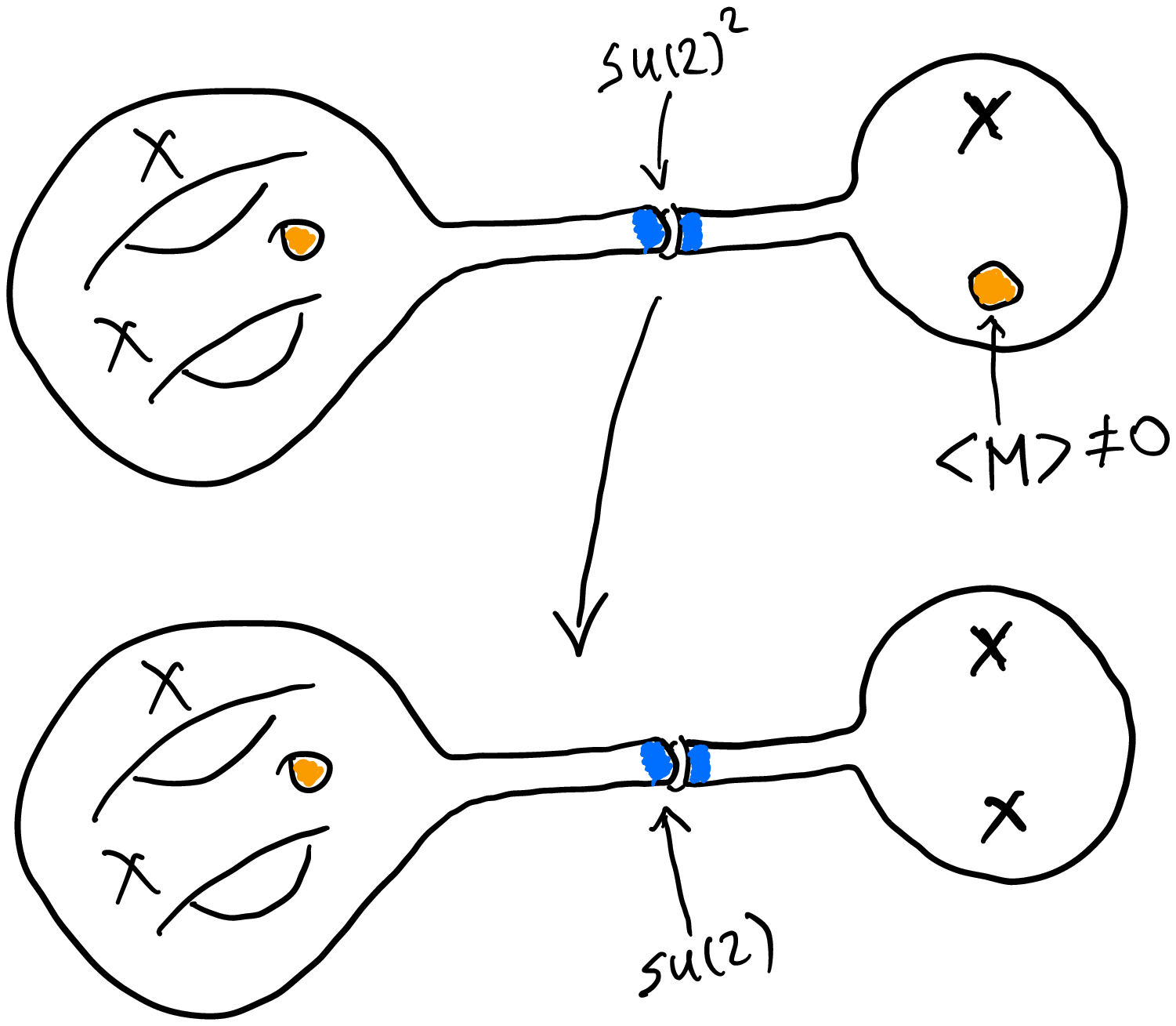}{1.7in}
$\;\;\;$ \figscale{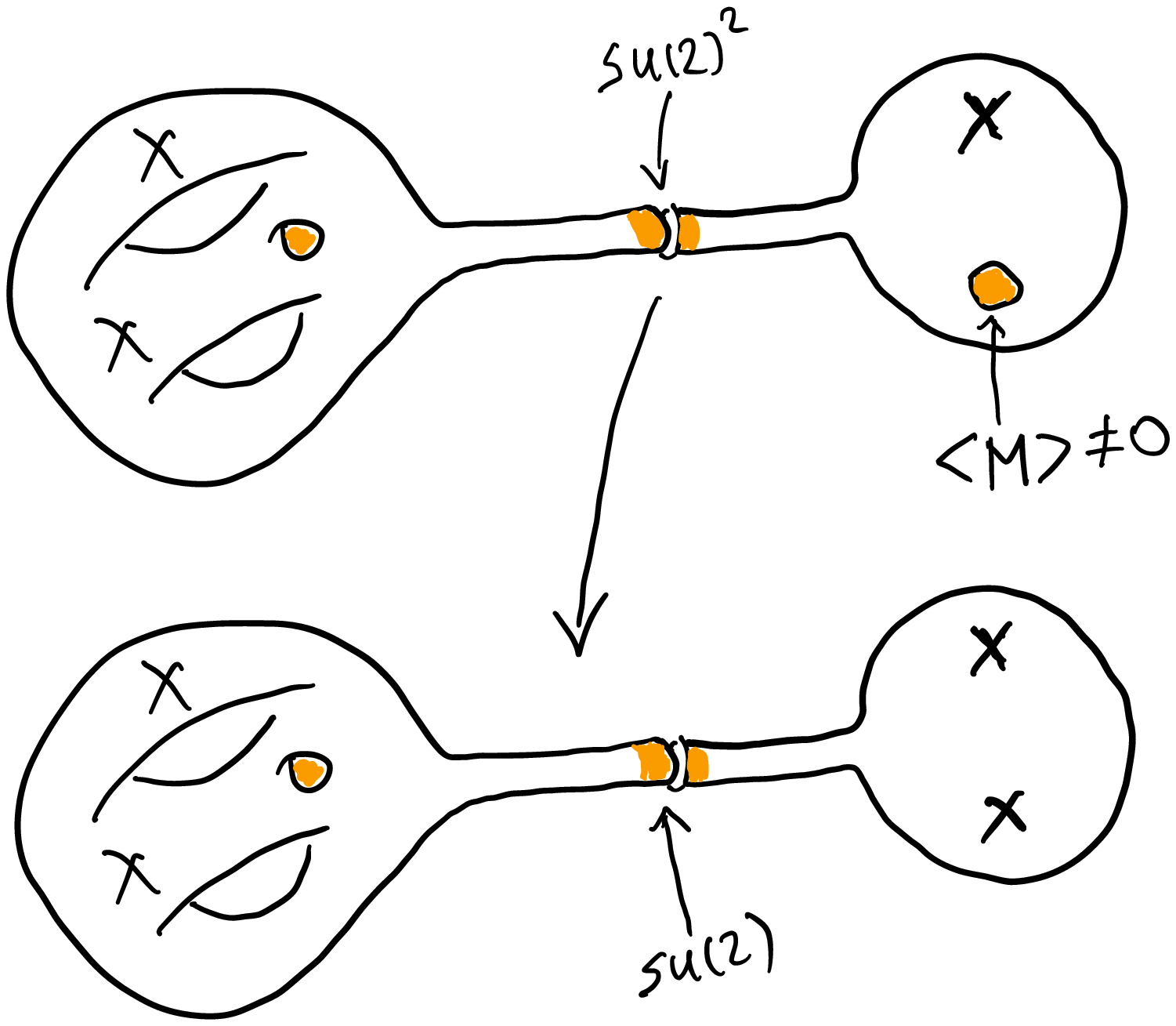}{1.7in}}
\medskip\centerline{\vbox{
\baselineskip12pt\advance\hsize by -1truein
\noindent\footnotefont{\bf Fig.~5:} We can derive a description of a theory in a duality frame where two minimal punctures sit on the same pair of pants. First we start with a theory with one minimal puncture switched for a maximal one, and then give a vacuum expectation value to a mesonic operator corresponding to the maximal puncture. The choice of the pair of pants with two maximal and one minimal puncture gives different dual descriptions of the same model. On the left we take the trinion to be the free one and on the right an interacting trinion of Fig. 6.}} 

\

There are two different colors of maximal punctures for $k$ and $N$ equal to two  (for our choice of $so(5)$ subgroup of $so(7)$). The two trinions 
we will construct have either all three punctures of the same type, $T_A$, or two of the same color and one of a different color, $T_B$. From the two trinions by gauging a subgroup of the flavor symmetry with additional matter one can obtain the  $\Z_2$ orbifold of ${\cal N}=2$ SYM with $su(2)$ gauge group and four flavors.
The trinions belong to theories of type $su(2)_{diag}u(1)^2$ with different choices of fluxes. Trinion $T_A$ has ${\cal F}_A=(\frac14,\frac14,1)$ and trinion $T_B$ has ${\cal F}_B=(-\frac14,\frac14,1)$. 

\

\centerline{\figscale{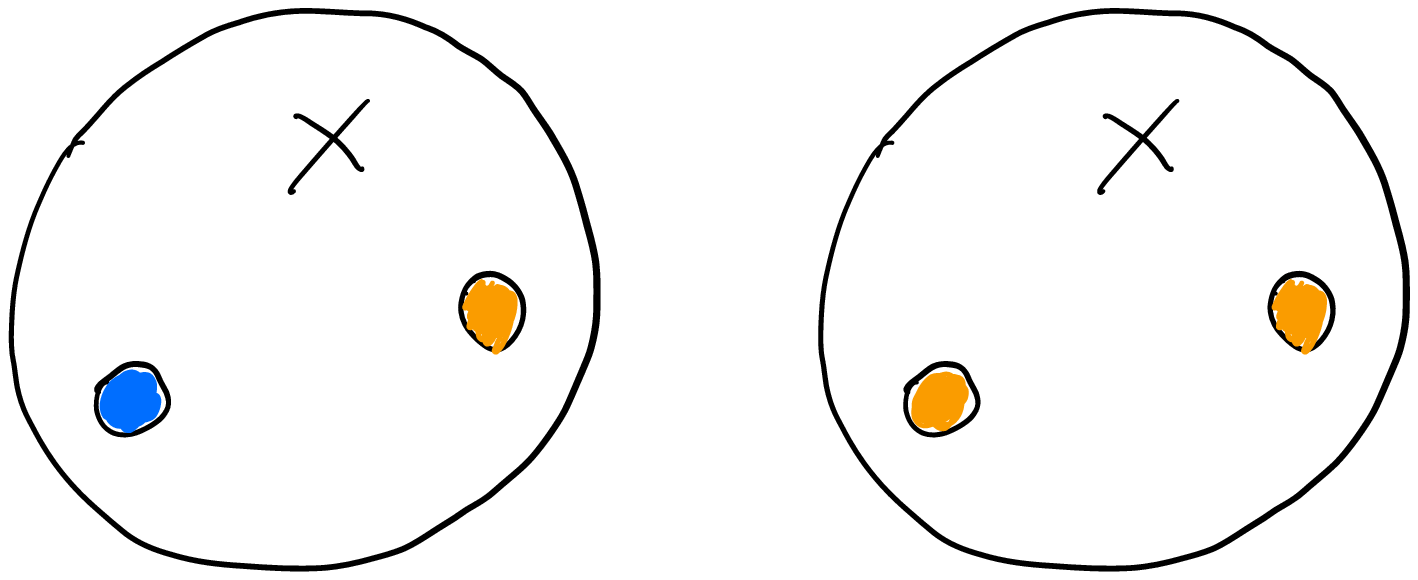}{1.7in}
}
\medskip\centerline{\vbox{
\baselineskip12pt\advance\hsize by -1truein
\noindent\footnotefont{\bf Fig.~6:} Two different trinions with two maximal and one minimal punctures. On the right we have interacting trinion with two maximal punctures of same color. This is $su(2)$ SQCD with four flavors and certain choice of singlets \GaiottoUSA. The trinion on the right is the free one with two maximal punctures of different color.}}

\

We will use these Lagrangians to compute conformal anomalies of theories coming from $k=2$ and two M5 branes, and the supersymmetric index. We will deduce the spectrum of relevant and marginal operators.

\

\centerline{\figscale{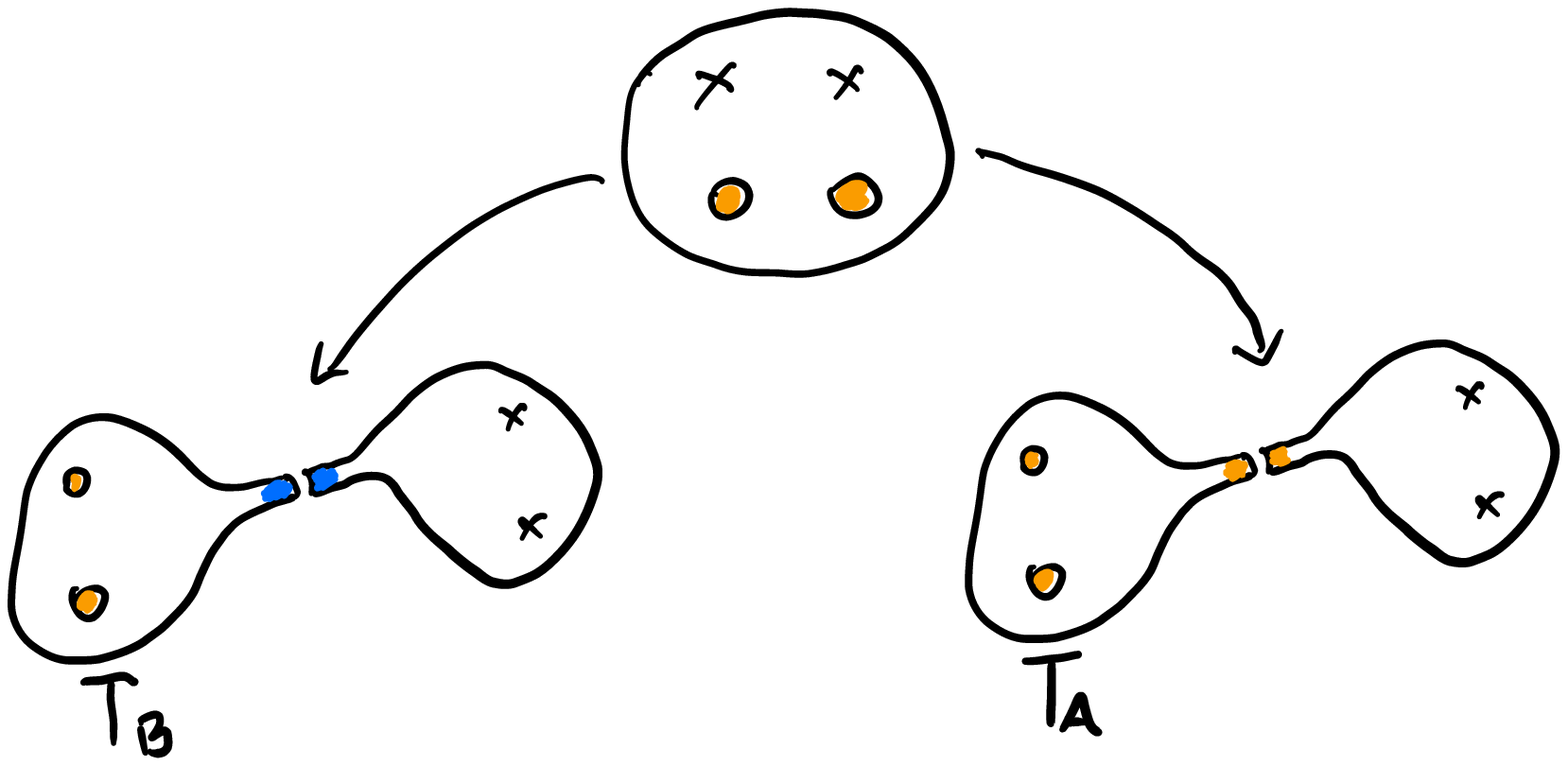}{2.7in}
}
\medskip\centerline{\vbox{
\baselineskip12pt\advance\hsize by -1truein
\noindent\footnotefont{\bf Fig.~7:} Two dualities for the orbifold theory.  The orbifold theory corresponds to four punctured sphere with two maximal puntures of same color and two minimal punctures. Taking the limit of the maximal punctures sitting on same pair of pants  we develop a long tube and decompose the surface into trinions. There are many ways to do so distributing compensating discrete charges between the two trinions. The two trinion theories we discussed are obtained by decomposing the  tube in two different ways. First one corresponding to the procedure detailed in the text using interacting trinion, and second using free trinion.}} 

\

Let  us briefly describe how the dualities used here are argued for in \GaiottoUSA. We can consider a field theory corresponding to compactification on a Riemann surface with $1+S$ maximal and $1+M$ minimal puncture. We can discuss a duality frame of this model where a theory with $S+1$ maximal puncture and $M$ minimal punctures is coupled to free trinion with $\Phi$-gauging, see Fig. 5. We can ask then what happens if we close the maximal puncture  on the free trinion by giving vacuum expectation value to one of the mesonic operators. The resulting theory corresponds to a surface with $S$ maximal and $M+2$ minimal punctures. All the minimal punctures can be exchanged by dualities and are equivalent. The RG flow here triggered by the vacuum expectation value for the meson can be completely traced using standard field theoretic arguments and leads to a description of the model as certain $su(2)$ gauging of a subgroup of the $su(N)^2$ symmetry of the maximal puncture running in the tube with the addition of extra fields. Let us assume now that the theory we started with was a sphere with three maximal and one minimal puncture. The theory we obtain then is a description of the orbifold theory as an $su(2)$ gauging of a subgroup of the symmetry of a three punctured sphere with only maximal punctures. From the six dimensional point of view we decompose the surface into pairs of pants so that the two minimal punctures are on the same pair of pants, see Fig. 7. Instead of free trinion we can use other theories with two maximal and a minimal puncture. It will be useful to use a trinion with two maximal punctures of the same color which is given by $su(2)$ theory with four flavors of \GaiottoUSA. The procedures give us the two different trinions $T_B$ and $T_A$. we could construct more trinions differing by the amount of abelian fluxes, however trinions $T_A$ and $T_B$ would suffice for us to derive a wide variety of theories and match them with six dimensional considerations. The procedure here is an analogue of the appearance of the $E_6$ SCFT in class ${\cal S}$ \ArgyresCN.

\

\subsec{The $\Z_2$ orbifold of the ${\cal N}=2$ SYM}

\

We start our discussion with a $\Z_2$ orbifold of ${\cal N}=2$  $su(2)$ SYM with four flavors. This is an ${\cal N}=1$  superconformal gauge theory with gauge group $su(2)_a\times su(2)_b$ and with the matter content described in the table \theor (see Fig. 4).  
Note that the two gauge groups have six flavors each and thus the one loop gauge beta functions vanish. The theory can be shown to have one exactly marginal 
deformation passing through zero coupling locus, and preserving the full flavor symmetry~\GaiottoUSA. We denote this conformal manifold by ${\cal M}_{conf.}$.
 We also can argue that it is natural to expect that there is an action of a duality group on ${\cal M}_{conf.}$. For example, we can go 
off the conformal manifold and perform a Seiberg duality on one of the gauge nodes. The resulting theory will be an $su(4)$ gauge theory with six flavors coupled to an ${\cal N}=2$ $su(2)$ theory with four flavors and additional gauge singlets. Performing S-duality for the $su(2)$ ${\cal N}=2$ gauge node and then Seiberg duality on the $su(4)$ theory we obtain our original theory above with the $u(1)_\delta$ and $u(1)_\alpha$ symmetries switched~\GaiottoUSA.  

We will also  assume that the conformal manifold has an infinite coupling cusp where the exchange of $u(1)_\alpha$ with $u(1)_\delta$ is a symmetry. We will assume that the symmetry $u(1)_{\delta/\alpha}$ enhances to $su(2)$ at the cusp.

The theory has a number of mesonic chiral operators. Most of them are invariant under exchanging $u(1)_\alpha$ and $u(1)_\delta$ and they map to themselves under the duality.
Some of them are charged  under these symmetries and map to each other, for example $Q^+_i{Q'}^+_i$ and $Q^-_i{Q'}^-_i$. The baryonic operators are charged only under one of the $u(1)_\delta$ or $u(1)_\alpha$ and are exchanged under the duality. We organize the mesons and the baryons in the following table.

\eqn\bames{
\vbox{\offinterlineskip\tabskip=0pt
\halign{\strut\vrule#
&~$#$~\hfil\vrule
&~$#$~\hfil
&~$#$\hfil\vrule
&~$#$\hfil
&~$#$\hfil\vrule
&~$#$\hfil
&~$#$\hfil\vrule
&~$#$\hfil\vrule
&~$#$\hfil\vrule
&\vrule#
\cr
\noalign{\hrule}
&  &  u(1)_{\alpha/\delta}   &u(1)_{\alpha \delta}& su(2)^u_1 & su(2)^u_2 & su(2)^v_1 & su(2)^v_2& u(1)_{\gamma \beta}& u(1)_{\gamma/\beta}& \cr
\noalign{\hrule}
&  M^u_-=Q_1^+Q^-_2         &\quad 0&\quad 0   &\quad {\bf 2} & \quad {\bf 2}  &    \quad  {\bf 1}  & \quad {\bf 1} & \quad 0&\;  -1&\cr
&  M^u_+=Q^-_1 Q^+_2         &\quad 0&\quad 0   &\quad {\bf 2} & \quad {\bf 2}  &    \quad  {\bf 1}  & \quad {\bf 1} & \quad 0&\quad 1&\cr
\noalign{\hrule}
&  M^v_+={Q'}^+_1{Q'}^-_2         &\quad 0&\quad 0   &\quad {\bf 1} & \quad {\bf 1}  &    \quad  {\bf 2}  & \quad {\bf 2} & \quad 0&\quad  1&\cr
&  M^v_-={Q'}^-_1 {Q'}^+_2         &\quad 0&\quad 0   &\quad {\bf 1} & \quad {\bf 1}  &    \quad  {\bf 2}  & \quad {\bf 2} & \quad 0&\; -1&\cr
\noalign{\hrule}
&  B_{\ell;+-}=(Q^{+}_\ell)^2         &\; +1&\; - 1   &\quad {\bf 1} & \quad {\bf 1}  &    \quad  {\bf 1}  & \quad {\bf 1} & \, (-1)^{\ell+1}&\,  (-1)^\ell&\cr
&  B_{\ell;--}=({Q'}^{-}_\ell)^2         &\; - 1&\; - 1   &\quad {\bf 1} & \quad {\bf 1}  &    \quad  {\bf 1}  & \quad {\bf 1} & \, (-1)^{\ell+1}&\,  (-1)^{\ell}&\cr
&  B_{\ell;-+}=(Q^{-}_\ell)^2         &\; - 1&\; + 1   &\quad {\bf 1} & \quad {\bf 1}  &    \quad  {\bf 1}  & \quad {\bf 1} & \, (-1)^{\ell+1}&\,  (-1)^{\ell+1}&\cr
&  B_{\ell;++}=({Q'}^{+}_\ell)^2         &\; + 1&\; + 1   &\quad {\bf 1} & \quad {\bf 1}  &    \quad  {\bf 1}  & \quad {\bf 1} & \, (-1)^{\ell+1}&\,  (-1)^{\ell+1}&\cr
\noalign{\hrule}\noalign{\hrule}
&  T^{111}=Q_1^\pm {Q'}^\pm_1         &\; \pm1&\quad 0   &\quad {\bf 2} & \quad {\bf 1}  &    \quad  {\bf 2}  & \quad {\bf 1} & \quad 1&\quad 0&\cr
&  T^{122}=Q_2^\pm {Q'}^\pm_2         &\; \pm1&\quad 0   &\quad {\bf 1} & \quad {\bf 2}  &    \quad  {\bf 1}  & \quad {\bf 2} & \; -1&\quad 0&\cr
&  T^{212}=Q_1^\pm {Q'}^\mp_2         &\quad 0&\; \mp1   &\quad {\bf 2} & \quad {\bf 1}  &    \quad  {\bf 1}  & \quad {\bf 2} & \quad 0&\quad 0&\cr
&  T^{221}=Q_2^\pm {Q'}^\mp_1         &\quad 0&\; \mp1   &\quad {\bf 1} & \quad {\bf 2}  &    \quad  {\bf 2}  & \quad {\bf 1} &\quad 0&\quad 0&\cr
}\hrule}} The operators in this table have $u(1)_t$ charge one.
We separated the fields in several segments. The first two segments and the last two fields in the last segment are invariant under duality, and these are mesons.
The fields in the third segment are baryons and are exchanged under the duality. The first two fields in the last segment are two pairs of mesons and duality acts within each pair.

\

\subsec{IR dual descriptions A} 

The orbifold theory was conjectured in~\GaiottoUSA\ to have a multitude of IR dual descriptions. A common feature of these descriptions is that they involve a gauging of a strongly-coupled SCFT, a Lagrangian for which is not obviously known. Alternatively, the arguments of ~\GaiottoUSA\  and the resulting dualities can be viewed as definitions of the SCFT's involved in the construction. Let us  consider in detail a couple of such descriptions. 

\

The first description we discuss is as follows. We seek an IR dual of the orbifold theory  which is an $su(2)$ gauging of an SCFT we denote by $T_A$. This theory has, at least an $su(N)^2_{u}\times su(N)^2_{v}\times su(N)^2_{z}\times u(1)_\beta\times u(1)_\gamma\times u(1)_t$ flavor symmetry.  The theory $T_A$ contains operators $M^u_\pm$, $M^v_\pm$, and $M^z_\pm$. These operators are in bi-fundamental representation of the corresponding $su(N)\times su(N)$ symmetry, in fundamental representation of $su(2)_{\gamma/\beta}$ (the charge under the Cartan is the $\pm$ label), have $u(1)_t$ charge $+1$, are not charged under $u(1)_{\gamma\beta}$, and have (the non-conformal) R-charge $\frac43$ in the notations of the previous subsection. The IR dual of the orbifold theory is constructed by gauging $su(2)_{z_1}$ symmetry of $T_A$ while adding the following fields.
\eqn\theori{
\vbox{\offinterlineskip\tabskip=0pt
\halign{\strut\vrule#
&~$#$~\hfil\vrule
&~$#$~\hfil\vrule
&~$#$\hfil\vrule
&~$#$\hfil\vrule
&~$#$\hfil\vrule
&~$#$\hfil\vrule
&~$#$\hfil\vrule
&~$#$\hfil\vrule
&\vrule#
\cr
\noalign{\hrule}
&  &  su(2)_{z}  & u(1)_\delta   &u(1)_\alpha&  u(1)_R & u(1)_\beta & u(1)_\gamma&u(1)_t& \cr
\noalign{\hrule}
&  q^{(\pm)}           & \; \quad {\bf  2}     &\pm1 &\mp1&\quad    0&-1 & -1  &    \quad  0  &\cr
&  {\Phi'}^{(\pm)}    & \; \quad {\bf  2}     &\mp1 &\mp 1&\quad    \frac23&\pm1 & \mp1  &    -1  &\cr
&  B_{1,\pm\pm}   & \; \quad {\bf  1}     &\quad 0 &\pm 2&\quad    \frac43 &1\mp1 & 1\pm1  &    \quad    1&\cr
&  B_{1,\mp\pm}    & \; \quad {\bf  1}     &\pm2  &\quad 0&\quad    \frac43&1\mp1 & 1\pm1  &    \quad  1  &\cr
&  T_0        & \; \quad {\bf  1}     &\quad 0 &\quad 0&\quad    2&\quad 2 & \quad 2  &    \quad  0  & \cr
\noalign{\hrule}
&(M^z_{\mp_2})^{\pm_1}& \; \quad {\bf  2}     &\pm_11 &\pm_11&\quad    \frac43&\pm_21 &\mp_21  &    \quad  1  &\cr
}\hrule}} The extra fields are singlets/have zero charge under other symmetries. The superpotential of the theory contains,

\eqn\supoeigh{
\eqalign{
&W\supset q^{(-)}{\Phi'}^{(+)}B_{1,-+}+q^{(+)}{\Phi'}^{(+)}B_{1,++}+q^{(-)}{\Phi'}^{(-)}B_{1,--}+q^{(+)}{\Phi'}^{(-)}B_{1,+-}+\cr
&\qquad\quad\;\; q^{(+)}q^{(-)} T_0+{\Phi'}^{(+)}(M^z_+)^{+}+{\Phi'}^{(-)}(M^z_-)^{-}\,.
}
} There are additional terms in the superpotential to be discussed shortly once we understand better the protected spectrum of $T_A$.
Under the duality the operator $M^u_\pm$, $M^v_\pm$, and $B_{1,ab}$ map as the names suggest. The baryonic operators $B_{2,ab}$ map as,

\eqn\mapga{
B_{2,+-}\to q^{(-)}(M^z_+)^{-}\,,\quad B_{2,-+}\to q^{(+)} (M^z_-)^{+}\,, \quad  B_{2,++}\to q^{(-)}(M^z_-)^{+}\,,\quad  B_{2,--} \to q^{(+)} (M^z_+)^{-}\, .
} A simple check of the validity of this duality is comparison of  't Hooft anomalies involving $u(1)_{\alpha/\delta}$. The theory $T_A$ does not involve this symmetry with the only charged fields given in the table above. All such anomalies match the orbifold theory. For example, the non-vanishing anomalies are,

\eqn\theorAM{
\vbox{\offinterlineskip\tabskip=0pt
\halign{\strut\vrule#
&~$#$~\hfil\vrule
&~$#$~\hfil\vrule
&~$#$\hfil\vrule
&~$#$\hfil\vrule
&\vrule#
\cr
\noalign{\hrule}
&  &  \;\;\;\;\;\;\;\;\;\;\;Orbifold  &\;\;\;\;\;\;\;\;\;\;\;IR\;\, dual   & A &\cr
\noalign{\hrule}
& u(1)_t\times u(1)_{\alpha/\delta}^2         &\; \;\;\frac12\times4\times(4(\frac12)^2+4(-\frac12)^2)   & \;\;\;1\times 2\times 1^2+1\times2\times (-1)^2&\;\;\;4&\cr
& u(1)_R\times u(1)_{\alpha/\delta}^2         & -\frac13\times4\times(4(\frac12)^2+4(-\frac12)^2)   & -1\times 2\times(1^2+(-1)^2)+\frac13 (2\times 1^2+2\times (-1)^2)&-\frac83&\cr
}\hrule}}  All the rest of the anomalies involving $u(1)_{\alpha/\delta}$ consistently vanish between the two descriptions.

\

\noindent$\bullet${\it The $T_A$ SCFT}

\

Assuming the IR duality is true we can now manipulate consistently both sides to arrive to a description of the theory $T_A$ with no symmetry gauged and no extra matter.
 The procedure is analogous to the one detailed in \RazamatL\ and is as follows.

\noindent $\bullet$ {\it Add matter:} -- First we add the following  fields to both sides of the duality,
\eqn\theord{
\vbox{\offinterlineskip\tabskip=0pt
\halign{\strut\vrule#
&~$#$~\hfil\vrule
&~$#$~\hfil\vrule
&~$#$\hfil\vrule
&~$#$\hfil\vrule
&~$#$\hfil\vrule
&~$#$\hfil\vrule
&~$#$\hfil\vrule
&~$#$\hfil\vrule
&\vrule#
\cr
\noalign{\hrule}
&  &  su(2)_{w}  & u(1)_\delta   &u(1)_\alpha&  u(1)_R & u(1)_\beta & u(1)_\gamma&u(1)_t& \cr
\noalign{\hrule}
&  {\widetilde q}^{(\pm)}           & \; \quad {\bf  2}     &\pm1 &\mp1&\quad    0&1 & 1  &    \quad  0  &\cr
&  b_{1,\pm\pm}   & \; \quad {\bf  1}     &\quad 0 &\mp 2&\quad    \frac23 &-1\pm1 & -1\mp1  &    -    1&\cr
&  b_{1,\mp\pm}    & \; \quad {\bf  1}     &\mp2  &\quad 0&\quad    \frac23&-1\pm1 & -1\mp1  &    -  1  &\cr
&  t_0        & \; \quad {\bf  1}     &\quad 0 &\quad 0&\quad    2&- 2 & - 2  &    \quad  0  & \cr
}\hrule}} We couple the new fields through the superpotential,

\eqn\dsp{
\Delta W= {\widetilde q}^{(-)}{\widetilde q}^{(+)} t_0 + b_{1,\mp\mp}B_{1,\mp\mp}+b_{1,\mp\pm}B_{1,\mp\pm}\,.
} In particular on the IR dual side this superpotential makes the $b_1$ and $B_1$ fields massive, removing them from the theory, while on the orbifold side this is a cubic superpotential term.

\noindent -- {\it Tune to enhance to $su(2)_{\alpha/\delta}$} -- Next, we tune the couplings on the IR dual side such that the $u(1)_{\alpha/\delta}$ symmetry enhances to $su(2)_{\alpha/\delta}$. Note that the $su(2)$ symmetry is only broken through the superpotential terms and thus switching those couplings off will restore the enhanced symmetry. On the orbifold side at no finite coupling such an enhancement is seen in the Lagrangian. The enhancement thus happens at a strong coupling limit of some of the superpotential couplings.

\noindent -- {\it Gauge $su(2)_{\alpha/\delta}$} -- Now, we can gauge the enhanced $su(2)_{\alpha/\delta}$. Note that since we added $\widetilde q$ fields this gauge sector has $N_f=2$ and it is easy to verify that all the symmetries are non anomalous. Gauging this symmetry we can use Seiberg duality to argue that this $su(2)$ sector has a description in terms of gauge invariant mesonic operators parametrizing quantum mechanically deformed moduli space with no point where all the gauge invariant fields have zero vacuum expectation value. Such non-zero vacuum expectation values will Higgs the $su(2)_z$ gauge symmetry and we will be left precisely with theory $T_A$ coupled to the $\Phi'$ fields through  a superpotential. 

\noindent -- {\it Remove extra fields} -- Finally, we remove the fields $\Phi'$ from the IR dual side by adding on both sides of the duality fields $\phi'$ and quadratically coupling the two sets of fields, ${\Phi'}^{(\mp)}{\phi'}^{(\pm)}$. On the orbifold side the fields $\Phi'$  maps to some composite operators.

\

We thus obtain a description of $T_A$ SCFT albeit with a singular superpotential. The flavor symmetry of this theory is at least,

\eqn\flair{
 su(2)_w\times su(2)_{\alpha\delta}\times (su(2))^2_u\times (su(2))^2_v\times u(1)_t\times u(1)_\gamma \times u(1)_\beta\,,
 } and soon we will see that it actually enhances. We can perform several computations using this description. We will denote $w,\alpha\delta\to w_1,w_2$.

\

\subsec{IR dual descriptions B} 

\

Second description we discuss is very similar to the previous one but differs in small but consequential details. We seek an IR dual of the orbifold theory  which is an $su(2)$ gauging of an SCFT we denote by $T_B$. This theory has, at least an $su(N)^2_{u}\times su(N)^2_{v}\times su(N)^2_{z}\times u(1)_\beta\times u(1)_\gamma\times u(1)_t$ flavor symmetry.  The theory $T_B$ contains operators $M^u_\pm$, $M^v_\pm$, and $M^z_\pm$. Operators $M^u$ and $M^v$ are in the bi-fundamental representation of the corresponding $su(2)\times su(2)$ symmetry, in fundamental representation of $su(2)_{\gamma/\beta}$ (the charge under the Cartan is the $\pm$ label), have $u(1)_t$ charge $+1$, are not charged under $u(1)_{\gamma\beta}$, and have (the non-conformal) R-charge $\frac43$ in the notations of the previous subsections. Operator $M^z$ is in bi-fundamental representation of the corresponding $su(2)\times su(2)$ symmetry, in fundamental representation of $su(2)_{\gamma\beta}$ (the charge under the Cartan is the $\pm$ label), has $u(1)_t$ charge $+1$, is not charged under $u(1)_{\gamma/\beta}$, and has (the non-conformal) R-charge $\frac43$.
 The IR dual of the orbifold theory is constructed by gauging the $su(2)_{z_1}$ symmetry of $T_B$ while adding the following fields.
\eqn\theori{
\vbox{\offinterlineskip\tabskip=0pt
\halign{\strut\vrule#
&~$#$~\hfil\vrule
&~$#$~\hfil\vrule
&~$#$\hfil\vrule
&~$#$\hfil\vrule
&~$#$\hfil\vrule
&~$#$\hfil\vrule
&~$#$\hfil\vrule
&~$#$\hfil\vrule
&\vrule#
\cr
\noalign{\hrule}
&  &  su(2)_{z}  & u(1)_\delta   &u(1)_\alpha&  u(1)_R & u(1)_\beta & u(1)_\gamma&u(1)_t& \cr
\noalign{\hrule}
&  q^{(\pm)}           & \; \quad {\bf  2}     &\pm1 &\mp1&\quad    0&\quad1 & - 1  &    \quad  0  &\cr
&  {\Phi'}^{(\pm)}    & \; \quad {\bf  2}     &\mp1 &\mp 1&\quad    \frac23&\mp1 & \mp1  &    -1  &\cr
&  B_{2,\pm-}   & \; \quad {\bf  1}     &-1\mp1 &-1\pm1&\quad    \frac43 &-2 & \quad 0  &    \quad    1&\cr
&  B_{1,\pm+}    & \; \quad {\bf  1}     &1\mp1  &1\pm1&\quad    \frac43&\quad 0 & \quad2  &    \quad  1  &\cr
&  T_0        & \; \quad {\bf  1}     &\quad 0 &\quad 0&\quad    2&-2 & \quad 2  &    \quad  0  & \cr
\noalign{\hrule}
&(M^z_{\mp_2})^{\pm_1}& \; \quad {\bf  2}     &\pm_11 &\pm_11&\quad    \frac43&\mp_21 &\mp_21  &    \quad  1  &\cr
}\hrule}} The extra fields are singlets/have zero charge under other symmetries. The superpotential of the theory contains,

\eqn\supoeigh{
\eqalign{
&W\supset q^{(+)}{\Phi'}^{(+)}B_{1,++}+q^{(-)}{\Phi'}^{(+)}B_{1,-+}+q^{(-)}{\Phi'}^{(-)}B_{2,--}+q^{(+)}{\Phi'}^{(-)}B_{2,+-}+\cr
&\qquad\quad\;\; q^{(+)}q^{(-)} T_0+{\Phi'}^{(+)}(M^z_+)^{+}+{\Phi'}^{(-)}(M^z_-)^{-}\,.
}
} There might be additional terms in the superpotential to be consistent with all the symmetries.
Under the duality the operator $M^u_\pm$, $M^v_\pm$, and $B_{1,a+}$, $B_{2,a-}$ map as the names suggest. The baryonic operators $B_{1,a-}$, $B_{2,a+}$ map as,

\eqn\mapga{
B_{1,+-}\to q^{(-)}(M^z_+)^{-}\,,\quad B_{2,-+}\to q^{(+)} (M^z_-)^{+}\,, \quad  B_{2,++}\to q^{(-)}(M^z_-)^{+}\,,\quad  B_{1,--} \to q^{(+)} (M^z_+)^{+}\, .
}  We can also check the relevant matching of 't Hooft anomalies here.

\

\noindent$\bullet${\it The $T_B$ SCFT}

\

Assuming the IR duality is true we can now manipulate consistently both sides to arrive to a description of the theory $T_B$ with no symmetry gauged and no extra matter.

\noindent $\bullet$ {\it Add matter:} -- First we add the following  fields to both sides of the duality.
\eqn\theord{
\vbox{\offinterlineskip\tabskip=0pt
\halign{\strut\vrule#
&~$#$~\hfil\vrule
&~$#$~\hfil\vrule
&~$#$\hfil\vrule
&~$#$\hfil\vrule
&~$#$\hfil\vrule
&~$#$\hfil\vrule
&~$#$\hfil\vrule
&~$#$\hfil\vrule
&\vrule#
\cr
\noalign{\hrule}
&  &  su(2)_{w}  & u(1)_\delta   &u(1)_\alpha&  u(1)_R & u(1)_\beta & u(1)_\gamma&u(1)_t& \cr
\noalign{\hrule}
&  {\widetilde q}^{(\pm)}           & \; \quad {\bf  2}     &\pm1 &\mp1&\quad    0&-1 & 1  &    \quad  0  &\cr
&  b_{1,\pm-}   & \; \quad {\bf  1}     &1\pm1&1\mp 1&\quad    \frac23 &2 & 0  &    -    1&\cr
&  b_{2,\pm+}    & \; \quad {\bf  1}     &-1\pm1  &-1\mp1&\quad    \frac23&0 & -2  &    -  1  &\cr
&  t_0        & \; \quad {\bf  1}     &\quad 0 &\quad 0&\quad    2&2 & - 2  &    \quad  0  & \cr
}\hrule}} We couple the new fields through the superpotential,

\eqn\dsp{
\Delta W= {\widetilde q}^{(-)}{\widetilde q}^{(+)} t_0 + b_{1,\pm+}B_{1,\pm+}+b_{2,\pm-}B_{2,\pm-}\,.
} In particular on the IR dual side this superpotential makes the $b_1$ and $B_1$ fields massive removing them from the theory, while on the orbifold side this is a cubic superpotential term.

The next stages of the procedure are identical to the previous case. 

\noindent -- {\it Tune to enhance to $su(2)_{\alpha/\delta}$} -- Next, we tune the couplings on the IR dual side such that the $u(1)_{\alpha/\delta}$ symmetry enhances to $su(2)_{\alpha/\delta}$. Note that the $su(2)$ symmetry is only broken through the superpotential terms and thus switching those couplings off will restore the enhanced symmetry. On the orbifold side at no finite coupling such an enhancement is seen in the Lagrangian. The enhancement thus happens at an infinite coupling limit of some of the superpotential couplings.

\noindent -- {\it Gauge $su(2)_{\alpha/\delta}$} -- Now, we can gauge the enhanced $su(2)_{\alpha/\delta}$. Note that since we added $\widetilde q$ fields this gauge sector has $N_f=2$ and it is easy to verify that all the symmetries are non anomalous. Gauging this symmetry we can use Seiberg duality to argue that this $su(2)$ sector has a description in terms of gauge invariant mesonic operators parametrizing quantum mechanically deformed moduli space with no point where all the gauge invariant fields have zero vacuum expectation value. Such non-zero vacuum expectation values will Higgs the $su(2)_z$ gauge symmetry and we will be left precisely with theory $T_B$ coupled to the $\Phi'$ fields through  a superpotential. 

\noindent -- {\it Remove extra fields} -- Finally, we remove the fields $\Phi'$ from the IR dual side by adding on both sides of the duality fields $\phi'$ and quadratically coupling the two sets of fields, ${\Phi'}^{(\mp)}{\phi'}^{(\pm)}$. On the orbifold side the fields $\Phi'$  maps to some composite operators.

\

We thus obtain a description of $T_B$ SCFT albeit with a singular superpotential. The flavor symmetry of this theory is at least,

\eqn\flair{
 su(2)_w\times su(2)_{\alpha\delta}\times (su(2))^2_u\times (su(2))^2_v\times u(1)_t\times u(1)_\gamma \times u(1)_\beta\,,
 } which does not enhance here.  We will denote $w,\alpha\delta\to w_1,w_2$.
 
 \

\subsec{Supersymmetric index and anomalies}

\

The Lagrangians for the trinions can be used to compute different robust quantities. 
First we can compute the conformal anomalies. The above procedure gives a Lagrangian  with infinite couplings, however the anomaly computation is insensitive to these couplings. We can compute the anomalies using a-maximization to obtain that the superconformal R-charge is given by (we refer to appendix D for the derivation),

\eqn\scrc{
\eqalign{
T_A:\qquad R_{c} = R+0.0689 (q_\beta+q_\gamma)-0.044777 q_t\,,\cr
T_B:\qquad R_{c} = R+0.0985 (-q_\beta+q_\gamma)-0.043523 q_t\,,
}} and the anomalies are,

\eqn\anomssi{\eqalign{
T_A:\qquad a = 2.0621\,,\qquad c = 2.56004\,,\cr
T_B:\qquad a = 2.1153\,,\qquad c = 2.61997\,.
}} Note that the mixing of R-charges preserves the $su(2)_{\gamma/\beta}u(1)_{\beta\gamma}u(1)_t$ symmetry in the case $T_A$ and $su(2)_{\beta\gamma}u(1)_{\beta/\gamma}u(1)_t$ in the case of $T_B$. This is consistent with our claim that these theories correspond to compactifications of type $G^{max}=su(2)_{diag}u(1)^2$. 

Next, we can compute the supersymmetric index which turns out to be very informative. The computation is straightforward. The gauging of $su(2)_{\alpha/\delta}$  for the index is equivalent to the inversion formula of Spiridonov-Warnaar \SpirWarnaar\ and the same procedure was applied in different contexts in \refs{\GaddeTE,\RazamatL} .
The result for $T_A$
 is given by,\foot{We remind the reader that the details of the computations of anomalies and indices can be found in appendix D.}

\eqn\indlor{\eqalign{
&{\cal I}_{T_A} = 1+(\frac2{\beta^2\gamma^2t^2}+\frac{\beta^2\gamma^2}{t^2}+t({\bf 2}_{u_1}{\bf 8}_v+{\bf 2}_{v_1}{\bf 8}_s+{\bf 2}_{w_1}{\bf 8}_c) +t\beta^2\gamma^2{\bf 2}_{u_1}{\bf 2}_{v_1}{\bf 2}_{w_1})(p q)^{\frac23}+\cr
&\qquad\qquad (\beta^2\gamma^2({\bf 3}_{u_1}+{\bf 3}_{v_1}+{\bf 3}_{w_1})+\frac1{\beta^2\gamma^2}{\bf 28}
-{\bf 28}-{\bf 3}_{u_1}-{\bf 3}_{v_1}-{\bf 3}_{w_1}-1-1)(p q)+\dots
}}
 Let us explain this expression. First we used the R-symmetry of \theor\ and not the superconformal one to write this expression. Remember that at order $pq$ in the index we have marginal operators minus conserved currents, whence for smaller powers of $pq$ we get relevant operators \BeemYN. Thus, the operators weighed $\beta^2\gamma^2$ at this order above are actually irrelevant and operators weighed by $\beta^{-2}\gamma^{-2}$ are relevant. In particular this SCFT has thus no exactly marginal operators preserving puncture symmetries. 

We also see that the theory has an operator weighed by $\beta^2\gamma^2 p q {\bf 3}_{w_1}$, let us denote it by $\Phi_{w_1}$. This operator can appear in superpotential of the IR dual theory (5.3) as $\Phi_{w_1} q^{(+)} q^{(-)}$ without breaking any symmetries and thus we should add it there. 

Another fact we read off from order $pq$ is that the flavor symmetry is enhanced to \eqn\enso{so(8)\times su(2)_{u_1}\times su(2)_{v_1}\times su(2)_{w_1}\times u(1)_t\times u(1)_{\gamma\beta}.}  The symmetry $u(1)_{\gamma/\beta}$ enhances to $su(2)$ and moreover $su(2)_{\gamma/\beta}\times su(2)_{u_2}\times su(2)_{v_2}\times su(2)_{w_2}$ enhances to $so(8)$. In more detail we have, 

\eqn\eighs{\eqalign{
&{\bf 8}_v = {\bf 2}_{\gamma/\beta} {\bf 2}_{u_2}+{\bf 2}_{v_2}{\bf 2}_{w_2}\,,\qquad 
{\bf 8}_s = {\bf 2}_{\gamma/\beta} {\bf 2}_{v_2}+{\bf 2}_{u_2}{\bf 2}_{w_2}\,,\qquad
{\bf 8}_c = {\bf 2}_{\gamma/\beta} {\bf 2}_{w_2}+{\bf 2}_{v_2}{\bf 2}_{u_2}\,,\cr
&{\bf 28}={\bf 3}_{\gamma/\beta}+{\bf 3}_{w_2}+{\bf 3}_{u_2}+{\bf 3}_{v_2}+{\bf 2}_{w_2}{\bf 2}_{u_2}{\bf 2}_{v_2}{\bf 2}_{\gamma/\beta}\,.
}} Note that the three factors $su(2)_{u_1}\times su(2)_{u_2}$, $su(2)_{w_1}\times su(2)_{w_2}$, and $su(2)_{v_1}\times su(2)_{v_2}$, appear completely symmetrically due to the triality property of $so(8)$.

\

The index for $T_B$ is given by,

\eqn\indlortr{\eqalign{
&{\cal I}_{T_B} = 1+(\frac1{\beta^2\gamma^2t^2}+\frac{\beta^2\gamma^2}{t^2}+\frac{\beta^2}{\gamma^2t^2}+t{\bf 2}_{\beta/\gamma}({\bf 2}_{u_2^1}{\bf 2}_{u_2^2}+{\bf 2}_{u_3^1}{\bf 2}_{u_3^2}) +t{\bf 2}_{\gamma\beta}{\bf 2}_{u_1^1}{\bf 2}_{u_1^2}\cr
&\qquad\quad+t\gamma^2{\bf 2}_{u_1^1}{\bf 2}_{u_3^1}{\bf 2}_{u_2^1}+t\frac1{\beta^2}{\bf 2}_{u_1^1}{\bf 2}_{u_3^2}{\bf 2}_{u_2^2}+t{\bf 2}_{u_1^2}({\bf 2}_{u_3^2}{\bf 2}_{u_2^1}+{\bf 2}_{u_3^1}{\bf 2}_{u_2^2}))(p q)^{\frac23}+\cr
&\qquad\qquad (\beta^2\gamma^2({\bf 3}_{u_2^1}+{\bf 3}_{u_3^1})+\frac1{\beta^2\gamma^2}({\bf 3}_{u_2^2}+{\bf 3}_{u_3^2})
+\frac\gamma\beta{\bf 2}_{u_1^2}(\beta^4 {\bf 2}_{u_3^1}{\bf 2}_{u_2^1}+\frac1{\gamma^4}{\bf 2}_{u_2^2}{\bf 2}_{u_3^2})+\frac{\beta^2}{\gamma^2}{\bf 3}_{u_1^2}+\frac{\gamma^2}{\beta^2}{\bf 3}_{u_1^1}\cr
&\qquad+\beta^4+\frac1{\gamma^4}-\frac{\gamma^2}{\beta^2}-\frac\gamma\beta{\bf 2}_{u_1^2}({\bf 2}_{u_3^1}{\bf 2}_{u_2^1}+{\bf 2}_{u_2^2}{\bf 2}_{u_3^2})-\sum_{i=1}^3\sum_{l=1}^2{\bf 3}_{u_i^l}-1-1-1)(p q)+\dots
}} Here the superconformal R symmetry admixes $u(1)_t$ and $u(1)_{\beta/\gamma}$. Thus we see that the flavor symmetry does not enhance here. We do not have any marginal operators preserving all the symmetries.

\

\subsec{The $G^{max}=su(2)_{diag}u(1)^2$ models}

We can use the trinions obtained here to construct theories corresponding to arbitrary Riemann surfaces. In this section we only glue punctures of the same sign  together. In general such a gauging is not conformal and one needs to perform a-extremization \IntriligatorJJ\ to determine the conformal R charges and anomalies for each theory. We can use the Lagrangians to do this.

\

\noindent  {\it Anomalies} -- Since the gauging is not conformal we have to perform a-maximization for a given theory to determine the superconformal charges and anomalies. For example, the superconfromal R-symmetry and anomalies of a genus $g$ theory with no punctures for theories built from one of the types of trinions (does not matter which one) is,

\eqn\gengnopu{\eqalign{
&R_c= R+0.06591 (q_\beta\pm q_\gamma)-0.02539 q_t\,,\cr & a = 7.99177(g-1)\,\qquad c=8.30369(g-1)\,.}
}  The sign in the mixing of R symmetry depends on whether we use only $T_A$ or only $T_B$ trinions. 
These anomalies precisely match the ones we will obtain by integrating the anomaly polynomial from six dimensions on a Riemann surface.
The fact that we can use either one of the trinions $T_A$ or $T_B$ to obtain the same results, though the theories $T_A$ and $T_B$ are different, is a very important check of our procedure.
Once we have no punctures the theories built from the two types of trinions only differ by a choice of $su(2)_{diag}$ and signs of some fluxes which group theoretically are equivalent, related by a Weyl transformation of the $so(7)$ symmetry. If the claimed models are to come from six dimensions thus it better be the case that the the choice of the trinion $T_A$ or $T_B$ would not matter.

\

Let us discuss the index of a theory corresponding to a general Riemann surface. We can build theories from the two trinions we described by gluing them together in different ways. The most general theory can be obtained by gluing 
together trinions $T_B$ in different ways and then closing punctures by RG flows. We will not proceed in this way here though, but rather focus on special cases exemplifying different types of models one can obtain.
The flows are interesting and important, for example,
trinion $T_A$ can be obtained from a four punctured sphere built from $T_B$ closing one puncture and the free trinion
can be obtained from $T_B$ by closing a maximal puncture to a minimal one. We will discuss these flows in the appendices A and B.

\ 

\noindent$\bullet$ {\it Theories of type $G^{max}=su(2)_{diag}u(1)^2$ constructed from $T_A$}

\

We can compute the index of a general theory (this is for generic genus and number of punctures with expressions for $g=0$ $s=3$ and $g=1$ $s\leq 1$ not fitting the general pattern), 

\eqn\indg{\eqalign{
&{\cal I}_{g,s}=1+ \left(\frac{2g-2+s}{\beta^2\gamma^2}{\bf 3}_{\gamma/\beta}+3g-3+s+ ({\bf 3}_{\gamma/\beta}+1+1) g-{\bf 3}_{\gamma/\beta}-1 -1\right) p q+\cr
&+\left(\sum_{i=1}^s((\gamma^2\beta^2-1){\bf 3}_{u^{(i)}_1}+(\frac1{\gamma^2\beta^2}-1){\bf 3}_{u^{(i)}_2})\right) p q+\cr 
&\left(\frac{2g-2+s}{t^2}\beta^2\gamma^2+\frac{4g-4+2s}{t^2\beta^2\gamma^2}+\frac{3g-3+s}{t^2}{\bf 3}_{\gamma/\beta}+t{\bf 2}_{\gamma/\beta}\sum_{i=1}^s{\bf 2}_{u^{(i)}_1}{\bf 2}_{u^{(i)}_2}\right)(p q)^{\frac23}+\cdots\,
}} First the index is manifestly invariant under the dualities, exchanging the different punctures. We remind again that we use here the R-charge R and not the superconformal R-charge.

 The last line contains relevant operators. The second line contains the currents of symmetries associated to the punctures coming with negative sign. The terms with positive sign are charged under $u(1)_{\gamma\beta}$ and thus become either relevant or irrelevant.
 
Note that in this case all the operators of the form $\Phi_i M_j$ are marginal. In the first line second term contains conserved currents for the internal symmetries $su(2)_{\gamma/\beta}\times u(1)_t\times u(1)_{\gamma\beta}$ coming with negative sign. The term charged under $u(1)_{\beta\gamma}$ is relevant. Finally we have $3g-3+s+ ({\bf 3}_{\gamma/\beta}+1+1) g$ marginal operators. We expect to have $3g-3+s$ exactly marginal couplings corresponding to complex structure moduli of the Riemann surface and we obtain them here. Next we have additional $({\bf 3}_{\gamma/\beta}+1+1)g$ marginal operators, only $3 (g-1)+2g$ of these operators are exactly marginal since turning on an operator in ${\bf 3}_{\gamma/\beta}$ will break the $su(2)_{\gamma/\beta}$ symmetry. Three marginal operators will combine with the conserved currents to become irrelevant.
The dimension of the conformal manifold is then,

\eqn\comani{\eqalign{
&dim {\cal M}^{su(2)_{diag}u(1)^2}_{g,s}=3g-3+s+3(g-1)+2g=\cr
&\;3g-3+s+dim(su(2)_{diag}u(1)^2)\, (g-1+\frac{s}2)-\frac{s}2 dim(su(2)_{diag}u(1)^2) +dim\;u(1)^2\,.}
} Note that  along the complex structure moduli  ($3g-3+s$ moduli) and $2g$ additional directions the symmetry group  is $su(2)\times u(1)^2$.
Along the additional $3(g-1)$ directions the symmetry is $u(1)^2$. This is in agreement with the expectations from previous sections.

\

\noindent$\bullet$ {\it Theories of type $G^{max}=su(2)_{diag}u(1)^2$ constructed with $T_B$}

\

We now consider gluing $T_B$ trinions together. One has here two types of punctures, one type appearing twice in the trinion and another appearing once. We denote the number of punctures appearing twice by $s_2$ and the number of punctures appearing once by $s_1$. The index of a general theory is given by,

\eqn\indgtr{\eqalign{
&{\cal I}_{g,(s_1,s_2)}=1+ \left((2g-2+s_1+s_2)\frac{\beta^2}{\gamma^2}{\bf 3}_{\beta\gamma}+(-\frac{s_2}2){\bf 2}_{\gamma^2/\beta^2}+\right.\cr
&\qquad\quad \left.+\frac{s_2}2{\bf 2}_{(\gamma\beta)^2}+{\bf 3}_{\gamma\beta}(g-1)+2g+(3g+s_2+s_1-3) -1-1\right) p q+\cr
&+\left(\sum_{i=1}^{s_2}((\gamma^2\beta^2-1){\bf 3}_{u^{(i)}_1}+(\frac1{\gamma^2\beta^2}-1){\bf 3}_{u^{(i)}_2})
+\sum_{i=1}^{s_1}((\frac{\gamma^2}{\beta^2}-1){\bf 3}_{\widetilde u^{(i)}_1}+(\frac{\beta^2}{\gamma^2}-1){\bf 3}_{\widetilde u^{(i)}_2})\right) p q+\cr 
&\left(\frac{3g+s_2+s_1-3}{t^2}+\frac{2g+s_1+\frac12s_2-2}{t^2}\frac{\gamma^2}{\beta^2}+(4g+2s_1+\frac32 s_2-4)\frac{\beta^2}{\gamma^2t^2}+\right.\cr &\;\;\;\;\;\left.\frac{3g+s_1+\frac32s_2-3}{t^2}(\gamma^2\beta^2+\frac1{\gamma^2\beta^2})+t{\bf 2}_{\gamma/\beta}\sum_{i=1}^{s_2}{\bf 2}_{u^{(i)}_1}{\bf 2}_{u^{(i)}_2}+t{\bf 2}_{\beta\gamma}\sum_{i=1}^{s_1}{\bf 2}_{\widetilde u^{(i)}_1}{\bf 2}_{\widetilde u^{(i)}_2}\right)(p q)^{\frac23}+\cdots\,
}} 

Several comments are in order. First, If we take $s_2=0$ (this means in particular that genus is higher than zero) the theories so obtained are equivalent to theories obtained from trinion $T_A$ upon taking $\beta\to\frac1\beta$.  Note also that by construction $s_2$ is even. 

We remind the reader that the R symmetry used to write the index is not superconformal. The superconformal R symmetry has admixture of $u(1)_{\gamma/\beta}$ and $u(1)_t$ in it and thus operators appearing at order $p q$ with these charges are actually either relevant or irrelevant.  The number of exactly marginal deformations is then,

\eqn\margefa{\eqalign{
&dim {\cal M}_{g,(s_1,s_2)}^{su(2)_{diag}u(1)^2}=3g-3+s_2+s_1+3(g-1+\frac{s_2}2)-\frac{s_2}2+2g=\cr
&3g-3+s_2+s_1+(g-1+\frac{s_2+s_1}2)dim(su(2)_{\beta\gamma}u(1)_{\beta/\gamma}u(1)_t)\cr
&\qquad\;-\frac{s_2}2dim(su(2)_{\beta\gamma}u(1)_{\beta/\gamma}u(1)_t\cap su(2)_{\beta/\gamma}u(1)_{\beta\gamma}u(1)_t)\cr
&\;\;\;\;\;\;\;-\frac{s_1}2dim(su(2)_{\beta\gamma}u(1)_{\beta/\gamma}u(1)_t\cap su(2)_{\beta\gamma}u(1)_{\gamma/\beta}u(1)_t)+dim\;u(1)^2\,.
}}
When $s_2$ is zero this is equivalent to what we obtained previously.  Note  that along the complex structure moduli and $3g$ additional directions the symmetry group is $u(1)^3$. Along the additional $s_2+2g-3$ the symmetry is broken to $u(1)^2$. Again we find the predicted expression.

Here we have in the notations of section three $A=-C=8$. Note that for the puncture appearing once on the trinion all the operators $M_i\Phi_j$ are marginal and thus the puncture dependence $B$ in \opsf\ is $B=1+\frac{k(k-2)}2=1$. For the puncture appearing twice on the other hand the operators $M_i\Phi_i$ are marginal but $\Phi_jM_{i\neq j}$ are not and thus $B=1+\frac{(k-1)k}2=2$. This is the reason the numbers of the two punctures $s_2$ and $s_1$ appear differently.

\

\subsec{Theories of type $G^{max}=u(1)^3$}

Before giving examples of special types of theories let us comment on the most generic case.
We can construct these by gluing the two trinions $T_A$ and $T_B$ to each other and construct Riemann surfaces from both of them. The resulting theories will have no symmetry relating the $\beta$ and $\gamma$ abelian symmetries as was the case when only one type of trinions was used. In particular the anomalies will have no symmetry exchanging the two factors. This implies that any object charged under these two will in a typical case not have an integer R-charge, and so in particular will not be marginal.  At order $pq$ we will get thus only the currents for puncture symmetries and the three $u(1)$s which are intrinsic to the surface. In addition we will have $3g-3+s_2+s_1+ 3 g$ marginal operators. These operators break no symmetries and will be exactly marginal.  We can construct even more generic theories by introducing minimal punctures through free trinions and subsequently closing minimal punctures and gluing theories with punctures of different signs. Note that  the trinions we constructed have positive punctures, the trinions with negative punctures are obtained by taking $R\to R+2q_t$ and flipping charges under the three abelian symmetries \GaiottoUSA.
This will result in theories coming from compactifications with generic choices of fluxes for the $so(7)$ flavor group. 

\

\newsec{Relations between the models}

In this section we will give examples of theories with enhanced symmetries $G^{max}$. In particular we will derive some more details about theories we have mentioned in section four.

\

\subsec{The $G^{max}=so(5)u(1)$ models from $G^{max}=su(2)_{diag}u(1)^2$ ones}

The general theories we discussed until now correspond to compactifications of M5 branes probing $\Z_{2}$ singularity on a Riemann surface with some general choice of the bundle for the $u(1)^{2}$  Cartan subgroup of $so(5)$ inside $so(7)$. The resulting theories then have as their internal symmetry only the $u(1)^2$ symmetry (times another copy of $u(1)_t$). In special cases this symmetry enhanced as we have observed to $su(2)\times u(1)$. With a trivial choice for the flux for any abelian subgroups of $so(5)$  we expect to have $so(5)$ as a symmetry of the four dimensional theory, at least without punctures and on some loci of the conformal manifold. Let us here derive the theories exhibiting this.

The derivation is guided by several observations and assumptions. First, we assume that the free trinion can be obtained by a compactification on a sphere with two maximal and one minimal punctures with fluxes $(0,0,\frac12)$ as was already stated. This assumption will turn out to give consistent predictions for the map of different four dimensional theories to 6d compactifications. 
Second, it was argued in \GaiottoUSA\ that closing minimal punctures should shift the fluxes of one of the $u(1)$s by half a unit (in certain normalization) while closing a maximal puncture to a minimal one shifts the fluxes  of both $u(1)$s by a quarter. Third, as we show in appendix A, the free trinion is obtained from $T_B$ by closing a maximal puncture to a minimal one. 

From these observations we deduce that if we glue together two appropriate trinions with maximal punctures and two free trinions and subsequently close completely two minimal punctures, we arrive at a theory with the topology of a sphere with four maximal punctures which should correspond to the compactification with trivial fluxes for both symmetries.
Using such a theory we can construct any theory corresponding to genus $g$ surface with even number of maximal punctures. 

Let us detail then a possible construction. We glue together two $T_A$ trinions and two free trinions and close the two minimal punctures. The matter content and the interactions are neatly encoded in the index of the theory. For definitions and details of the computation of supersymmetric indices \refs{\KinneyEJ,\DolanQI} we refer the reader to \RastelliTBZ . We denote the index of the two $T_A$ theories glued together by ${\cal I}_{T}$. The index of the theory with vanishing flux for any $u(1) \subset so(5)$ is,

\eqn\infs{
\eqalign{
&{\cal I}^{so(5)u(1)}_{0,4}({\bf a},{\bf c},{\bf b},{\bf d}) =\cr &\;\;\; (p;p)^2(q;q)^2\Gamma_e(pq\beta^4)\Gamma_e(p q \gamma^4)\Gamma_e(t\frac\beta\gamma a_1^{\pm1}a_2^{\pm1})\oint \frac{dh_1}{4\pi i h_1} 
\frac{\Gamma_e(\frac{p q}t \frac\gamma\beta h_1^{\pm1} a_1^{\pm1})}{\Gamma_e(h_1^{\pm2})}{\cal I}_T({\bf b},{\bf c},{\bf d}, \{h_1,a_1\})\times \cr
&\;\;\;\;\;\oint \frac{dh_2}{4\pi i h_2} \frac{\Gamma_e(\frac{p q}t \gamma\beta h_2^{\pm1} a_1^{\pm1})}{\Gamma_e(h_2^{\pm2})}
\Gamma_e(t \gamma \beta h_2^{\pm1}a_1^{\pm1})\Gamma_e(\beta^{-2}a_2^{\pm1}h_2^{\pm1})\Gamma_e(\gamma^{-2}h_2^{\pm1}h_1^{\pm1})\,.
}} We can compute the anomalies of this theory using a-maximization. We find that the R symmetry \theor\ is actually the superconformal one here. Also we find that gluing two such theories together is conformal. This in particular means 
that the R charges of the $\Phi_i$ fields are $\frac23$. The central charges are given by,

\eqn\anomcv{
a=\frac{125}{24}\,, \qquad\;\;\; c = \frac{71}{12}\,.
} The vector of fluxes for this theory is ${\cal F}=(0,0,2)$. We can glue a theory of genus $g$ and an even number $s$ of punctures from such models.
 The index of these theories with no punctures is,

\eqn\incv{\eqalign{
&{\cal I}^{so(5)u(1)}_{g,0}({\bf u}_i) = 1+(3g-3){\bf 5}_{so(5)}\frac1{t^2}(p q)^{\frac23}+
\left((g-1){\bf 10}_{so(5)}+3g-3+g-1\right)p q+\cdots\,.
}} Here we defined the characters of $so(5)$,

\eqn\soc{\eqalign{
&{\bf 10}_{so(5)}= 1+1+\beta^4+\frac1{\beta^4}+\gamma^4+\frac1{\gamma^4}+(\beta^2+\frac1{\beta^2})(\frac1{\gamma^2}+\gamma^2)\,,\cr
&{\bf 5}_{so(5)}=1+(\beta^2+\frac1{\beta^2})(\frac1{\gamma^2}+\gamma^2)\,.}
} We see that the symmetry that the index exhibits is $so(5)$ for any value of the genus. The number of exactly marginal deformations is given by flat connections for $so(5)$ $(g-1){\bf 10}_{so(5)}$, $3g-3$ complex structure moduli, and $g$ flat connections for $u(1)_t$. The symmetry preserved on a generic point of the conformal manifold is only $u(1)_t$. 

When we add punctures the maximal allowed $so(5)$ symmetry is broken to  its $su(2)_{diag}\times u(1)$ subgroup. The index is given by,

\eqn\incv{\eqalign{
&{\cal I}^{so(5)u(1)}_{g,s}({\bf u}_i) = \cr &\;\;\;1+\left(((3g-3+s){\bf 5}_{so(5)}+\frac{s}2(\beta^2\gamma^2+\frac1{\gamma^2\beta^2}))\frac1{t^2}+t{\bf 2}_{\gamma/\beta}\sum_{j=1}^s (u^{(1)}_j)^{\pm1}(u^{(2)}_j)^{\pm1}\right)
(p q)^{\frac23}+\cr
&\;\;\;\;\;\;\;\;\;\;\;\;\left((g-1+\frac{s}2){\bf 10}_{so(5)}-\frac{s}2({\bf 3}_{\beta/\gamma}+1)+3g-3+s+g-1\right)p q+\cr
&\;\;\;\;\;+(\sum_{i=1}^s((\beta^2\gamma^2-1){\bf 3}_{u^{(1)}_i}+(\frac1{\gamma^2\beta^2}-1){\bf 3}_{u^{(2)}_i})) pq+\cdots\,.
}}

The anomalies of the theories are,

\eqn\ansgt{
a=\frac1{24}((g-1)187+78 s)\,,\qquad\;\;\; c=\frac1{12}((g-1)97+42 s)\,.} This matches the six dimensional computation of the next section.

\

We can consider an alternative construction of the same models starting with $T_B$ trinions and using the relations between different type of trinions. For example, gluing two $T_B$ trinions along the puncture appearing once and calling the index of such a theory ${\cal I}_{T'}$ the index of ${\cal I}^{so(5)u(1)}_{0,4}$ is,

\eqn\infuus{
\eqalign{
&{\cal I}^{so(5)u(1)}_{0,4}({\bf a},{\bf c},{\bf b},{\bf d}) =\cr &\;\;\; (p;p)^2(q;q)^2\Gamma_e(pq\beta^{-4})\Gamma_e(p q \gamma^4)\oint \frac{dh_1}{4\pi i h_1} \oint \frac{dh_2}{4\pi i h_2} 
\frac{\Gamma_e(\frac{p q}t (\frac\gamma\beta) h_1^{\pm1} h_2^{\pm1})}{\Gamma_e(h_1^{\pm2})\Gamma_e(h_2^{\pm2})}
{\cal I}_{T'}({\bf b},{\bf c},{\bf d}, \{h_2,h_1\})\times \cr
&\;\;\;\;\;\;\;\;\;
\Gamma_e(\beta^2  h_1^{\pm1}a_1^{\pm1})\Gamma_e(\gamma^{-2}a_2^{\pm1}h_2^{\pm1})\Gamma_e(t\frac\gamma\beta a_1^{\pm1}a_2^{\pm1})\,.
}} This index matches \infs\ at least in an expansion in fugacities. Such equalities are rather non-trivial mathematical identities and provide checks of the physical arguments, dualities and flows, which we used to deduce them.

We can also consider gluing the $T_B$ trinions along punctures appearing twice. Denoting the index of this theory ${\cal I}_{\hat T}$, we can obtain a theory with two punctures of each type with trivial bundles. The index is,

\eqn\infuusl{
\eqalign{
&{\cal I}^{so(5)u(1)}_{0,(2,2)}({\bf b},{\bf c},{\bf a},{\bf d}) =\cr &\;\;\; (p;p)^2(q;q)^2\Gamma_e(pq\beta^{-4})\Gamma_e(p q \gamma^4)\oint \frac{dh_1}{4\pi i h_1} \oint \frac{dh_2}{4\pi i h_2} 
\frac{\Gamma_e(\frac{p q}t (\frac\gamma\beta)^{\pm1} h_1^{\pm1} h_2^{\pm1})}{\Gamma_e(h_1^{\pm2})\Gamma_e(h_2^{\pm2})}
{\cal I}_{\hat T}({\bf b},{\bf c}; {\bf d}, \{h_2,h_1\})\times \cr
&\;\;\;\;\;\;\;\;\;
\Gamma_e(\beta^2 \beta h_1^{\pm1}a_1^{\pm1})\Gamma_e(\gamma^{-2}a_2^{\pm1}h_2^{\pm1})\Gamma_e(t\frac\gamma\beta h_2^{\pm1}h_1^{\pm1})(\gamma^{-2}a_2^{\pm1}h_2^{\pm1})\Gamma_e(t\frac\gamma\beta a_1^{\pm1}a_2^{\pm1})\,.
}}

\

The anomalies of these theory are exactly the same as the anomalies of the theory with all punctures of the same type. Combining such theories together we get that the index of a general theory with $s_1$ punctures of first type and $s_2$ of second and with genus $g$ is,

\eqn\incvu{\eqalign{
&{\cal I}^{so(5)u(1)}_{g,(s_1,s_2)}({\bf u}_i;{\bf \widetilde u}_i) =\cr&
 \; 1+
\left(((3g-3+s_1+s_2){\bf 5}_{so(5)}+\frac{s_1}2(\beta^2\gamma^2+\frac1{\gamma^2\beta^2})+\frac{s_2}2(\frac{\beta^2}{\gamma^2}+\frac{\gamma^2}{\beta^2}))\frac1{t^2}\right)(p q)^{\frac23}+
\cr &\left(t{\bf 2}_{\gamma/\beta}\sum_{j=1}^{s_1} (u^{(1)}_j)^{\pm1}(u^{(2)}_j)^{\pm1}+t{\bf 2}_{\gamma\beta}\sum_{j=1}^{s_2} (\widetilde u^{(1)}_j)^{\pm1}(\widetilde u^{(2)}_j)^{\pm1}\right)
(p q)^{\frac23}+\cr
&\;\left((g-1+\frac{s_1+s_2}2){\bf 10}_{so(5)}-\frac{s_1}2({\bf 3}_{\beta/\gamma}+1)-\frac{s_2}2({\bf 3}_{\beta\gamma}+1)+3g-3+s_1+s_2+g-1\right)p q+\cr
&\;\;\;\;\;+(\sum_{i=1}^{s_1}((\beta^2\gamma^2-1){\bf 3}_{u^{(1)}_i}+(\frac1{\gamma^2\beta^2}-1)){\bf 3}_{u^{(2)}_i}+\sum_{i=1}^{s_2}((\frac{\gamma^2}{\beta^2}-1){\bf 3}_{\widetilde u^{(1)}_i}+(\frac{\beta^2}{\gamma^2}-1){\bf 3}_{\widetilde u^{(2)}_i})) pq+\cdots\,.
}}  Note that we have obtained here theories only with even number of punctures, and further that the expressions for the index we obtained make sense only for even numbers of punctures.  Here the maximal symmetry is just $u(1)^3$, and on a general point on the conformal manifold it is $u(1)_t$. These models have the same anomalies no matter what type the maximal punctures are. This is suggestive of theories with different types of punctures sitting at different cusps of the conformal manifold, Fig. 8. We can distinguish the types of punctures when the symmetries $u(1)_\gamma$ and $u(1)_\beta$ are not broken, and since those are broken on a general point of the conformal manifold while enhancing at special subloci, it is natural to conjecture that all these theories share the same conformal manifold. In particular one can state that theories with different colors of punctures are dual to each other for $G^{max}=so(5)u(1)$.

\centerline{\figscale{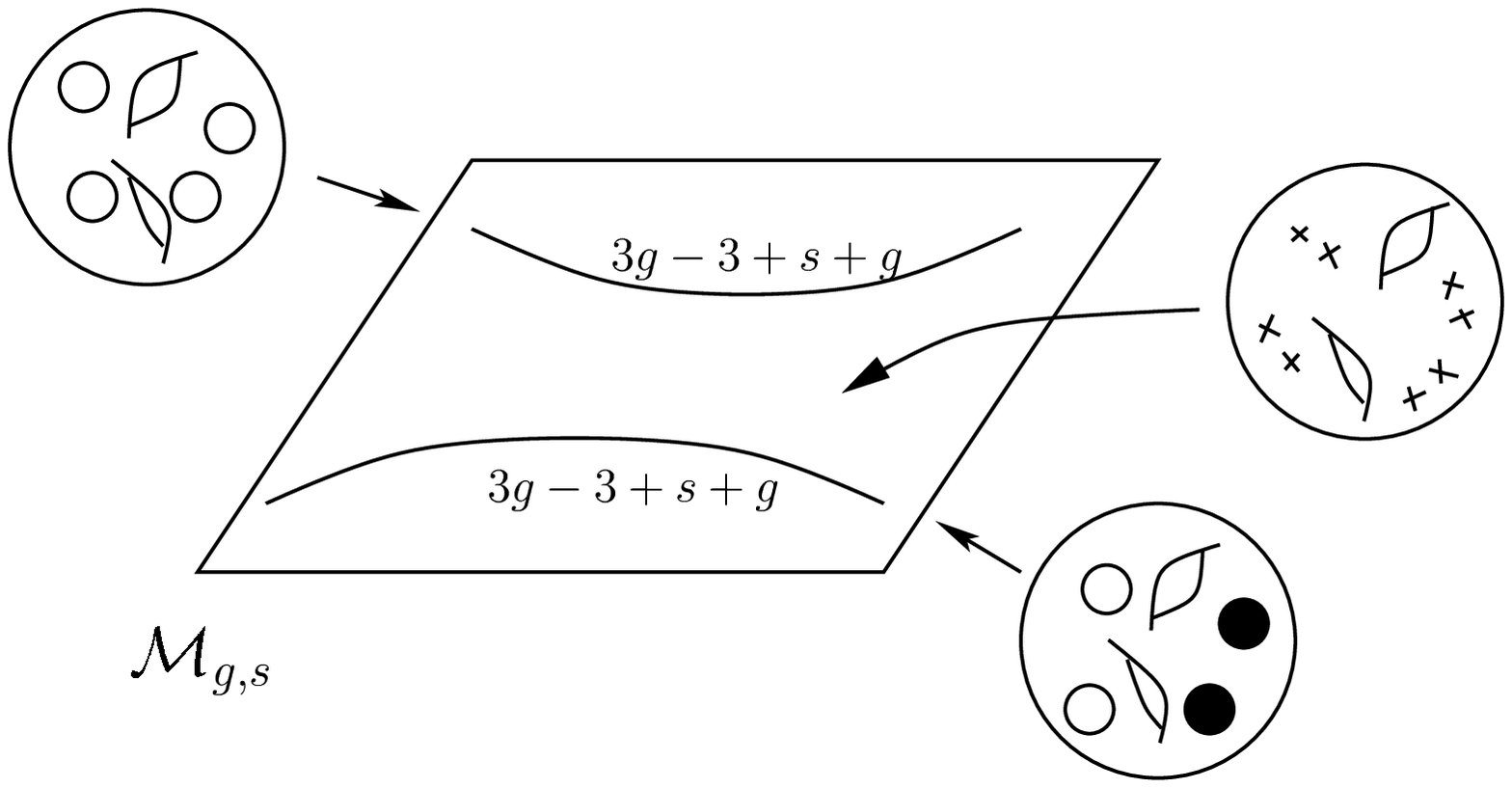}{2.7in}}
\medskip\centerline{\vbox{
\baselineskip12pt\advance\hsize by -1truein
\noindent\footnotefont{\bf Fig.~8:} Conformal manifold of a general theory of $G^{max}=so(5)u(1)$. At general point the notion of type of puncture is non existent since the relevant symmetries are broken. On sub-loci symmetries are enhanced  and the punctures can be distinguished by their color.}} 

\

Although \incvu\ is derived for even number of punctures, unrefining with $u(1)_\beta$ and $u(1)_\gamma$ these make perfect sense also with odd numbers of punctures. In particular, the trinion, as advertised in section four, is obtained by marginal deformations of the orbifold theory breaking a symmetry.  We will momentarily revisit this issue.

\

We can consider adding minimal punctures to the theories. This is done by forming Riemann surfaces also with free trinions. The gaugings here will also be conformal. The index of a theory with $s_1$ punctures of the first type, $s_2$ of the second, $m$ minimal punctures (there are even (or odd) numbers of all types of punctures), and genus $g$ is given by,

\eqn\incvuppe{\eqalign{
&{\cal I}^{so(5)u(1)}_{g,(s_1,s_2,{m})}({\bf u}_i;{\bf \widetilde u}_i,\{\delta_i\}) =
\cr&\;\;\;\; \; 1+
\left(((3g-3+s_1+s_2){\bf 5}_{so(5)}+{m}+\frac{s_1+{m}}2(\beta^2\gamma^2+\frac1{\gamma^2\beta^2})+\frac{s_2+m}2(\frac{\beta^2}{\gamma^2}+\frac{\gamma^2}{\beta^2}))\frac1{t^2}\right)(p q)^{\frac23}+
\cr &\left(t{\bf 2}_{\gamma/\beta}\sum_{j=1}^{s_1} (u^{(1)}_j)^{\pm1}(u^{(2)}_j)^{\pm1}+t{\bf 2}_{\gamma\beta}\sum_{j=1}^{s_2} (\widetilde u^{(1)}_j)^{\pm1}(\widetilde u^{(2)}_j)^{\pm1}+t{\bf 2}_{\beta^2}\sum_{e=1}^{m}\delta^{-2}_i+t{\bf 2}_{\gamma^2}\sum_{e=1}^{m}\delta_i^2\right)
(p q)^{\frac23}+\cr
&\;\;\;\biggl\{(g-1+\frac{s_1}2+\frac{s_2}2){\bf 10}_{so(5)}+\frac{m}2({\bf 5}_{so(5)}-1)-\frac{s_1}2({\bf 3}_{\beta/\gamma}+1)-\frac{s_2}2({\bf 3}_{\beta\gamma}+1)+\cr &\;\;\;\;3g-3+s_1+s_2+g-1\biggr\}p q+\cr
&\;\;\;\;\;+(\sum_{i=1}^{s_1}((\beta^2\gamma^2-1){\bf 3}_{u^{(1)}_i}+(\frac1{\gamma^2\beta^2}-1){\bf 3}_{u^{(2)}_i})+\sum_{i=1}^{s_2}((\frac{\gamma^2}{\beta^2}-1){\bf 3}_{\widetilde u^{(1)}_i}+(\frac{\beta^2}{\gamma^2}-1){\bf 3}_{\widetilde u^{(2)}_i})) pq+\cdots\,.
}} The contribution to the anomaly of the minimal punctures is half the contribution of the maximal punctures. On a general point of the conformal manifold the only internal symmetry remaining is the $u(1)$ and the maximal punctures are broken to have $u(1)\times u(1)$ symmetry. It is informative to consider the index at a generic point turning off fugacities of broken symmetries. We parametrize the $su(2)\times su(2)$ fugacities broken to $u(1)^2$ as $(y_1,y_2)=(\delta_1\delta_2,\delta_1/\delta_2)$, and we treat all the punctures as minimal

\eqn\incvupp{\eqalign{
&{\cal I}^{so(5)u(1)}_{g,{m}}(\{\delta_i\}) = 1+
(3(5g-5+m)\frac1{t^2}+2t\sum_{e=1}^{m}(\delta^{-2}_i+\delta_i^2))
(p q)^{\frac23}+\cr
&\;\;\;\biggl\{(g-1+\frac{m}4)10-\frac{m}4 2+3g-3+m+g-1-m\biggr\}p q+\cdots\,.
}} The minimal punctures behave as half maximal punctures, where a pair of minimal punctures are equivalent here to a maximal one. This can be seen for instance in the conformal manifold, where part of it is that of the complex structure moduli space of a genus $g$ Riemann surface with $m$ minimal punctures, where each maximal puncture contributed as two minimal ones. The other part of the conformal manifold is that of flat connections for the $so(5)$ and $u(1)$ global symmetries where minimal punctures counting as half a maximal one. We have a different way to see that the maximal punctures factorize into pairs of minimal punctures on a general point of the conformal manifold by studying the index as a sum over eigenfunctions of an integrable model \GaiottoUSA. We discuss this in appendix G.

\

Note that although \angst\ was derived  for even number of punctures,  naively considering genus zero with three maximal punctures the anomalies  give us $a=\frac{47}{24}$ and $c=\frac{29}{12}$ which is a very compelling result. These are exactly the anomalies of the orbifold theory, a sphere with two maximal and two minimal punctures.
Now, we understand what is going on. The orbifold theory has a one dimensional manifold of exactly marginal deformations preserving all the symmetries. However, it possessed five more exactly marginal directions along which the symmetries corresponding to maximal punctures are broken to $u(1)^2$ and the symmetry $u(1)_{\beta\gamma}$ is also broken. The maximal punctures give rise to two minimal punctures. At a generic point of the conformal manifold we have a sphere with six minimal punctures. At special sub-loci the $u(1)_{\beta\gamma}$ symmetry is recovered. 
In different other sub-loci different pairs of the $u(1)$ symmetries from minimal punctures combine to maximal punctures.
We can then conjecture that somewhere on this six dimensional conformal manifold all the minimal punctures combine into maximal ones. This is the trinion for the $so(5)$ theories we are considering here. Note that such a theory is non-Lagrangian in a very minimalistic way. We start from the Lagrangian, the orbifold theory, and go to special locus on the parameter space where the symmetry involves three copies of $su(N)^2$. Since we do not have a precise road map to that point this theory for example cannot be put on the lattice, and that is why it is somewhat ``non-Lagrangian''. Note also that some of the symmetries are broken in order to recombine minimal punctures to maximal. We thus observe the fact that theories with odd number of punctures with $G^{max}=so(5)u(1)$ have in general less symmetry than theories with even number of punctures. It will be interesting to understand this better.

\

A different way to obtain theories with $G^{max}=so(5)u(1)$ is to start from  theories with smaller $G^{max}$ and turn on  relevant deformations. For example, theories with $G^{max}=su(2)_{diag}u(1)^2$  built from $T_A$ trinions have, as seen in \indg\ , $2g-2+s$ relevant deformations in ${\bf 3}_{\gamma/\beta}$ and charged $-2$ under $u(1)_{\gamma\beta}$. Turning on these operators breaks these symmetries and makes the operators  marginal. These operators combine with the $(g-1){\bf 3}_{\gamma/\beta}+g-1$ exactly marginal operators of the original theory to give rise to $(g-1+\frac{s}2)10-\frac{s}2 (3+1)$ dimensional component of the conformal manifold of the $G^{max}=so(5)u(1)$ theory associated with flat connections of $so(5)$. 
  Turning on such a deformation washes off the information about the $u(1)_\gamma$ and $u(1)_\beta$ fluxes and leaves us only with the $u(1)_t$ flux. This follows as $u(1)_\gamma$ and $u(1)_\beta$ are broken in the resulting theory. Therefore, assuming there are no accidental symmetries and that we remain within the class ${\cal S}_{k=2}$ theories, the resulting theory must lie on the conformal manifold of a $G^{max}=so(5)u(1)$ theory. For the index switching off fugacities for $u(1)_\beta$ and $u(1)_\gamma$ the indices \indg\ for $G^{max}=su(2)_{diag}u(1)^2$ models and  $so(5)u(1)$ models \incv\ are identical. This is an example of the way \flg\ emerges.

\

\subsec{$so(5)\times u(1)_t$ models with general $u(1)_t$ flux.}

\

The theories above correspond to gluing copies of the same four punctured sphere together. The fluxes of the resulting theory are multiples of the flux of that four punctured sphere. We can consider accessing more general fluxes for $u(1)_t$ by gluing copies of the same four punctured sphere but with the sign inverted. In such a way the flux for $u(1)_t$ will not be proportional to $g-1+\frac{s}2$ any more. This construction is analogous to the class ${\cal S}$ ${\cal N}=1$ theories of \BahDG. For example, to construct a theory of genus $g$ with no punctures and more general values of flux we take a positive number $k$  with $|k|\leq 2g-2$ and  construct $so(5)u(1)$ theory of previous section of genus $k$ and two punctures  and a theory of genus $g-1-k$ with two punctures. Then we flip the sign of the second theory. This is obtained by flipping the signs of the charges under $u(1)_\beta\times u(1)_\gamma\times u(1)_t$ and taking R symmetry to be $R+2q_t$. Then the two theories are glued together along the two maximal punctures with $S$-gluing. Since we glue punctures of different signs we use the vector fields with no bifundamentals and turn on superpotential coupling the mesons of the two models. The flux of the theory obtained in this way is the difference of the fluxes of the two theories, $2k-2+2-2(g-k-1)-2+2=4k-(2g-2)$. See example of Fig. 9.

\centerline{\figscale{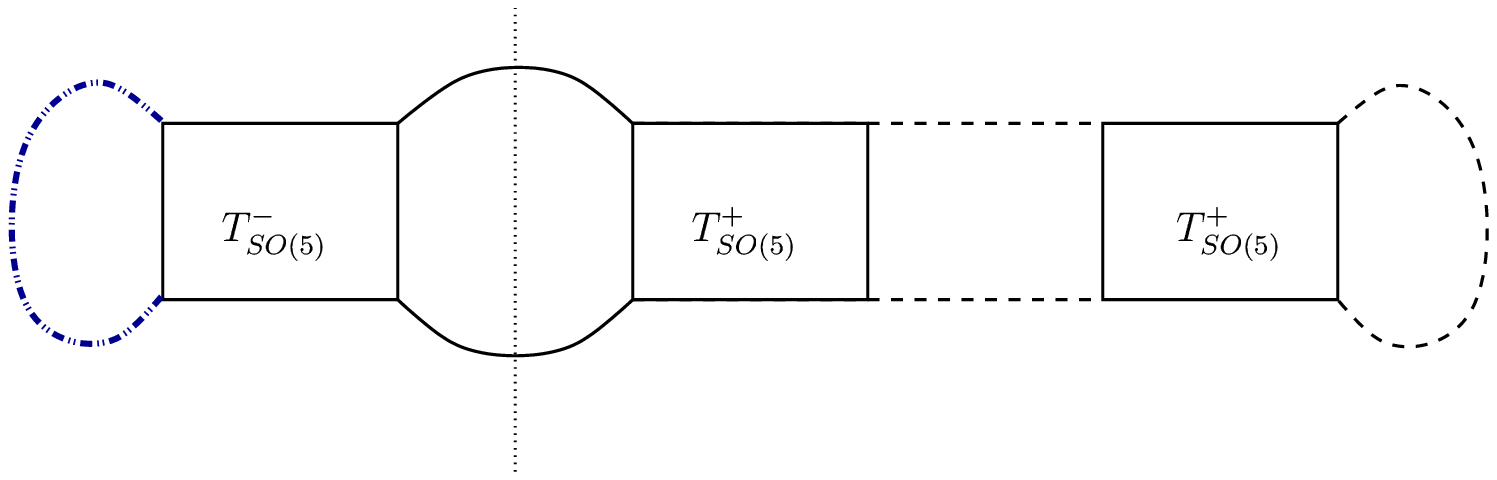}{3.5in}}
\medskip\centerline{\vbox{
\baselineskip12pt\advance\hsize by -1truein
\noindent\footnotefont{\bf Fig.~9:} Construction of a genus four theory with one unit of $u(1)_t$  flux with $G^{max}=so(5)u(1)$. 
The $\pm$ subscript denotes the sign of the theory. The squares are four punctured spheres of the $G^{max}=so(5)u(1)$ theories. Solid lines are connecting punctures of opposite sign and dashed lines punctures of same sign.
Different dashed lines connect pairs of maximal punctures of different colors and/ or signs.}}  

\

We can compute the indices and anomalies of the resulting theories. The indices and anomalies are computed in appendix D, whereas here we only state the results. The theory has a description in terms of a Lagrangian and thus the anomalies are computed using a-extremization. We normalize the flux with the genus to be,

\eqn\nomfl{
z=\frac2{g-1}k-1\,.
} then the result of the a-extremization is that,

\eqn\aextsiof{\eqalign{
&R_c=R-\frac16q_t+\frac{\sqrt{40 z^2 + 9}-3}{12 z}q_t\,,\cr
& a= \frac{8 \left(5 \sqrt{40 z^2+9}+54\right) z^2+9 \left(\sqrt{40 z^2+9}-3\right)}{96 z^2}(g-1)\,,\cr
& c= \frac{\left(44 \sqrt{40 z^2+9}+432\right) z^2+9 \left(\sqrt{40 z^2+9}-3\right)}{96 z^2}(g-1)\,.
}} We will compare this to the computation in six dimension in the next section.
Note that taking the limit $z\to 0$ we expect to obtain the models with $G^{max}$ being $so(7)$. Indeed in this limit $(a,c)=(\frac{51}8(g-1),\, \frac{13}2(g-1))$ which are the anomalies for that case.

\

\subsec{The $G^{max}=so(7)$ models from $G^{max}=so(5)u(1)$ models}

As a special case of the construction with general flux when the number of plus and minus trinions is the same the flux is zero. In such a case we expect to get a $G^{max}=so(7)$ model. Indeed that is the case. For example, in Fig. 10 we have a construction of genus two theory with $so(7)$ symmetry.

\centerline{\figscale{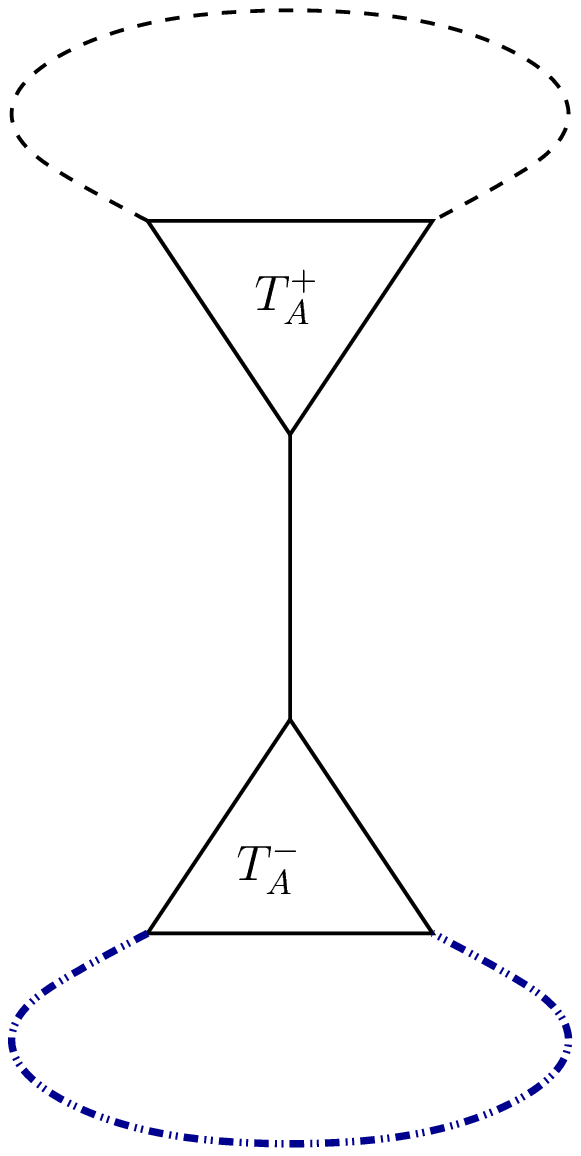}{.9in}}
\medskip\centerline{\vbox{
\baselineskip12pt\advance\hsize by -1truein
\noindent\footnotefont{\bf Fig.~10:} Genus two theory with $G^{max}=so(7)$. Note that we can glue together any plus theory to an identical minus copy and obtain $G^{max}=so(7)$ theory.}}  
We can compute the anomalies here and they match what we have reported in section four. The index here can be explicitly computed and exhibits the full $so(7)$ symmetry when punctures are not present. For example for the theory of Fig. 10.,

\eqn\insosu{
{\cal I}^{so(7)}_{g,0}=1+(\left.(3g-3+(g-1){\bf 21}_{so(7)})\right|_{g=2})p q+\cdots\,.
}  These theories have ${\cal F}=(0,0,0)$. The superconformal R-symmetry is the one used to write \insosu\ . We can construct theories with punctures by gluing equal numbers of copies of plus and minus theories. 

\

We can obtain these theories also by relevant deformations starting from $G^{max}=so(5)u(1)$ models. The theories with $G^{max}=so(5)u(1)$ have relevant deformations, as is seen for example from \incv\ , charged $-2$ under $u(1)_t$. We can turn these operators on and study the flow. The claim is that one ends on the conformal manifold of $G^{max}=so(7)$ models. The relevant operators become marginal at the fixed point. Moreover some of the irrelevant operators of the $G^{max}=so(5)u(1)$ model which have positive charge under $u(1)_t$ also become marginal. For the index setting in \incv\ $t\to (p \, q)^{-\frac16}$, not refining with $u(1)_\beta$ and $u(1)_\gamma$, and keeping track of irrelevant operators (which are missing in \incv\ ) one obtains the index of the $G^{max}=so(7)$ models.

As a special example we can consider the orbifold theory. This theory is a model in class $G^{max}=so(5)u(1)$. Turning on the mentioned relevant operators amounts to adding masses to the $\Phi$ fields. The resulting theory is the quiver of Fig 3.  with general quartic interactions coupling the mesonic operators. This is the theory we mentioned is the starting point, the trinion, for constructing the $G^{max}=so(7)$ models.

\

\subsec{Models with $G^{max}=su(2)\times su(2) \times u(1)$}

\centerline{\figscale{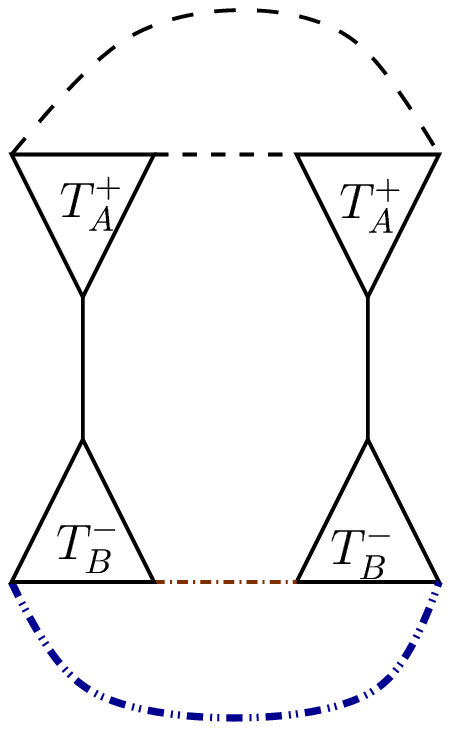}{.9in}}
\medskip\centerline{\vbox{
\baselineskip12pt\advance\hsize by -1truein
\noindent\footnotefont{\bf Fig.~11:} Construction of genus three theory with minimal value of flux with $G^{max}=u(1)su(2)\times su(2)$.}}

We can construct theories from equal number of $T_A$ trinions of one sign and $T_B$ trinions of opposite sign. The resulting theories will have only $u(1)_\gamma$ flux turned on. Thus, they will correspond to compactifications preserving $G^{max}=su(2)\times su(2)\times u(1)$.

 In Fig. 11. we have an example of genus three. The anomalies here are,

\eqn\sofffo{
a=\frac{687}{100}+\frac{61 \sqrt{61}}{75}\,,\;\qquad c=\frac{1}{150} \left(1038+127 \sqrt{61}\right)\,,
} and we will match the above to six dimensional computation.  The vector of fluxes of this theory is twice the difference of ${\cal F}_A$ and ${\cal F}_B$, which is ${\cal F}=(1,0,0)$. 
We can compute the index to give,

\eqn\infdf{\eqalign{&
1+\biggl(\left.(3g-3+g-1+(g-1)({\bf 3}_{\gamma^2}+{\bf 3}_t))\right|_{g=3}  +{\bf 2}_{\gamma^2} {\bf 3}_t(\beta^2+3\frac1{\beta^2})+\frac4{\beta^4}\biggr) pq+\cdots \,.}} States charged under $u(1)_\beta$ are either relevant or irrelevant as the $u(1)_\beta$ is admixed to the superconformal R-symmetry. We shift to the superconformal R-symmetry through 
$t\to (pq)^{\frac{\ell_3}2}t$, $\gamma\to (p\,q)^{\frac{\ell_2}2}\gamma$, and $ \beta\to (p\,q)^{\frac{\ell_1}2}\beta$ with (see appendix D)

\eqn\misuo{
\ell_2=\ell_3=0\,,\qquad\;\ell_1=\frac1{15}(\sqrt{61}-6)\,.
}

\

\subsec{Models with $G^{max}=su(2)u(1)^2$}

We can construct a representative of theories with $G^{max}=su(2)u(1)^2$ by taking two $T_A$ and two $T_B$ trinions and gluing them together to form a genus three surface. Since the two trinions have opposite flux under $u(1)_\beta$ the resulting theory has $u(1)_\beta\to su(2)_\beta$ and corersponds to $G^{max}=su(2)_\beta\times  u(1)_\gamma\times u(1)_t$ models. In general, taking same number of $T_A$ and $T_B$ trinions of same sign will result in such theories. 
We again can compute indices and anomalies. For example for genus three model of Fig. 12. we get,

\eqn\ansosss{
a=15.7822\,,\qquad c=16.3855\,.
} \centerline{\figscale{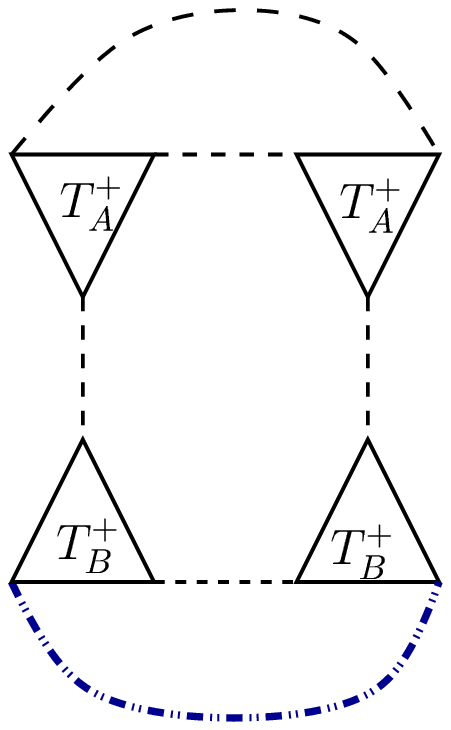}{.9in}}
\medskip\centerline{\vbox{
\baselineskip12pt\advance\hsize by -1truein
\noindent\footnotefont{\bf Fig.~12:} Construction of genus three theory with  $G^{max}=u(1)^2su(2)$.}}   

The vector of fluxes here is the sum of two ${\cal F}_A$ and two ${\cal F}_B$,  ${\cal F}=(0,1,4)$. The index is computed to give,

\eqn\infdfo{\eqalign{&
1+\biggl(\left.(3g-3+2g-2+(g-1){\bf 3}_{su(2)_{\beta^2}})\right|_{g=3}  +\cr&\qquad\;{\bf 2}_{\beta^2} (\gamma^2+\frac3{\gamma^2}-\frac{t^2}{\gamma^2}+\frac{7}{t^2\gamma^2}+\frac{5\gamma^2}{t^2}-3t^2\gamma^2)+\frac6{t^2}+\frac4{\gamma^4}-2t^2\biggr) pq+\cdots \,.}} This gives the expected dimension of the conformal manifold. Note again that operators charged under $u(1)_\gamma\times u(1)_t$ are either relevant or irrelevant. The mixing is determined as before by $s_i$ and here we have,

\eqn\misoi{
\ell_1=0\,,\qquad\, \ell_2=0.0657051\,,\qquad\, \ell_3=0.320467\,.
} The operators charged under $u(1)_t$ with positive charge under $u(1)_t$ in \infdfo\ are irrelevant and with with negative are relevant. In particular the operators giving negative contribution to \infdfo\ are irrelevant as expected from general arguments of \BeemYN.  This is a small but non trivial check of the construction.

\
   
\subsec{Model with $G^{max}=su(3)u(1)$}

We can also construct a model with $G^{max}=su(3)u(1)$. Here we need to correlate the $u(1)_t$ flux with one of the $u(1)_\beta$ or $u(1)_\gamma$ with the flux for the other being zero. Example of such a construction for genus six is depicted in Fig. 13.

\

\centerline{\figscale{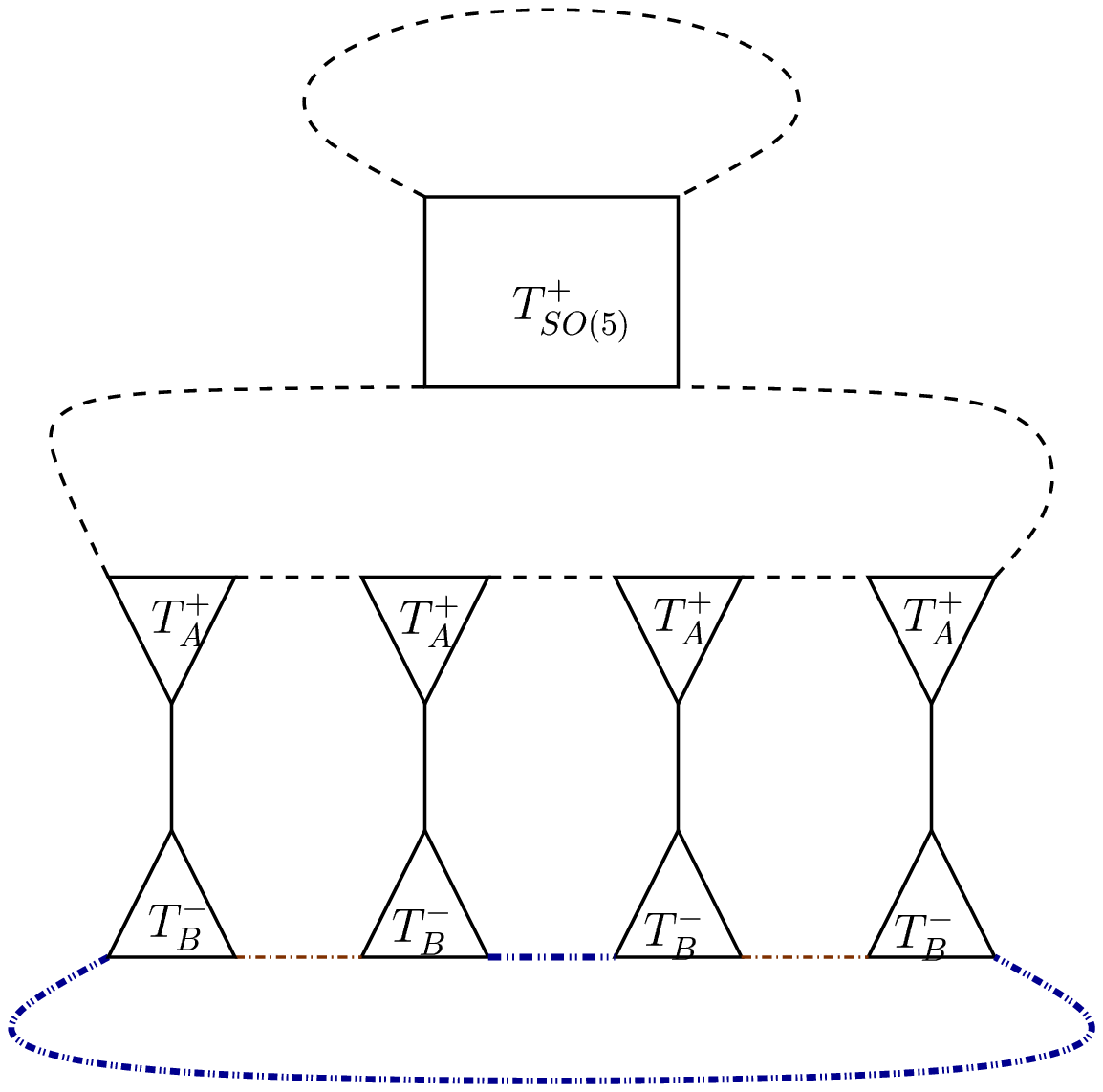}{2.3in}}
\medskip\centerline{\vbox{
\baselineskip12pt\advance\hsize by -1truein
\noindent\footnotefont{\bf Fig.~13:} Construction of genus six theory with minimal value of flux with $G^{max}=u(1)su(3)$.}} We can compute the anomalies from the Lagrangians, 

\eqn\sustt{
a=\frac{35}{512} \left(267+11 \sqrt{385}\right)\,,\;\qquad c= \frac{1}{512} \left(9425+401 \sqrt{385}\right)\,.
} The index of the model in Fig. 13. can be computed to give,

\eqn\insut{\eqalign{&
{\cal I}_{6,0}^{su(3)u(1)}=1+\biggl(\left\{3g-3+s+(g-1+\frac{s}2){\bf 8}_{su(3)}-s+g-1\right\}_{g=6,s=0}+\cr
&\;\;\;3t^{\frac23}\beta^{\frac43}{\bf 3}_{su(3)}+t^{\frac43}\beta^{\frac83}\overline{\bf 3}_{su(3)}+\frac9{t^{\frac43}\beta^{\frac83}}{\bf 3}_{su(3)}+ \frac7{t^{\frac23}\beta^{\frac43}}\overline{\bf 3}_{su(3)}\biggr)p q+\,\cdots\,.}} We have defined,\foot{Note the subtle issue here with identification of fugacities and symmetries. The $so(7)$ decomposes into $su(3)\times u(1)_a$. Although the fugacity for the $u(1)_a$ is $t^{\frac23}\beta^{\frac43}$, as far as the charges go the linear combination of the $u(1)_\beta$ and $u(1)_t$ which gives $u(1)_a$ is $q_a\propto q_\beta+q_t$.}

\eqn\sutfugs{\eqalign{&
{\bf 8}_{su(3)} = 1+1+\gamma^{-4}+\gamma^4+(\frac{\beta^2}{t^2}+\frac{t^2}{\beta^2})(\gamma^{-2}+\gamma^2)\,,\cr& {\bf 3}_{su(3)}= \frac{t^{4/3}}{\beta^{4/3}}+\frac{\beta^{2/3}}{t^{2/3}}(\gamma^{-2}+\gamma^2)\,.}}
The index here is written with non superconformal R- symmetry. To shift to the superconformal one we need to translate $t\to (pq)^{\frac{s_3}2}t$, $\gamma\to (p\,q)^{\frac{s_2}2}\gamma$, and $ \beta\to (p\,q)^{\frac{s_1}2}\beta$ with (see appendix D)

\eqn\misu{
\ell_2=0\,,\qquad\;\ell_1=\ell_3=\frac1{48}(-15+\sqrt{385})\,.
} This renders operators weighed by positive powers of $t\beta^2$ irrelevant and with negative powers relevant.
Vanishing powers give marginal operators minus conserved currents.
The index of this models and similar models in $G^{max}=su(3)u(1)$ determines the conformal manifold to have dimension,

\eqn\dimsy{
dim {\cal M}^{su(3)u(1)}_{g,s}= 3g-3+s+(g-1+\frac{s}2)dimsu(3)-s+g\,.
} This is consistent with our general prediction since $su(3)u(1)\cap su(2)_{diag}u(1)^2=u(1)^3$ and $L=u(1)$ here.
Here the vector of fluxes is constructed from four ${\cal F}_A$ minus four ${\cal F}_B$ and the vector for the $so(5)$ theory of sphere with four maximal punctures (of any color). Adding up all the terms one obtains ${\cal F}=(2,0,2)$.

\

\subsec{Model with $G^{max}=\widetilde{so(5)}u(1)$}

We can construct an example of $G^{max}=\widetilde{so(5)}u(1)$ theory by taking two $T_A^{+}$ trinions and gluing them with the $T^{-}_{SO(5)}$ theory, Fig. 14. The vector of fluxes is ${\cal F}=(\frac12,\frac12,\,0)$. 
The anomalies can be computed to give,

\eqn\ansost{
a=\frac{23 \sqrt{\frac{23}{2}}}{6}\,,\qquad\;\,\,c=\frac{47 \sqrt{\frac{23}{2}}}{12} \,.
} The mixing parameters are,

\eqn\softmo{
\ell_3=0\,,\qquad \ell_2=\ell_1=\frac{\sqrt{46}}{12}-\frac12\,.
} We can compute the index of this model to give,

\eqn\sofhijklvw{\eqalign{
{\cal I}=1+(\widetilde {\bf 5} \beta^2\gamma^2+3\widetilde{\bf 5}\frac1{\gamma^2\beta^2}) p q+(\left.3g-3+g-1+(g-1)\widetilde{\bf 10})\right|_{g=3} p\,q+\dots}
} We have defined the characters of $\widetilde {so(5)  }$,

\eqn\chahg{\eqalign{
&\widetilde {\bf 5}=1+(\frac{\beta t}\gamma+\frac{\gamma}{t \beta})(\frac{\gamma t}\beta+\frac\beta{\gamma t})\,,\cr
&\widetilde {\bf 10}=1+(\frac{\beta t}\gamma)^2+(\frac{\beta t}\gamma)^{-2}+1+(\frac{\gamma t}\beta)^2+(\frac{\gamma t}\beta)^{-2}+(\frac{\beta t}\gamma+\frac{\gamma}{t \beta})(\frac{\gamma t}\beta+\frac\beta{\gamma t})\,.
}}

\centerline{\figscale{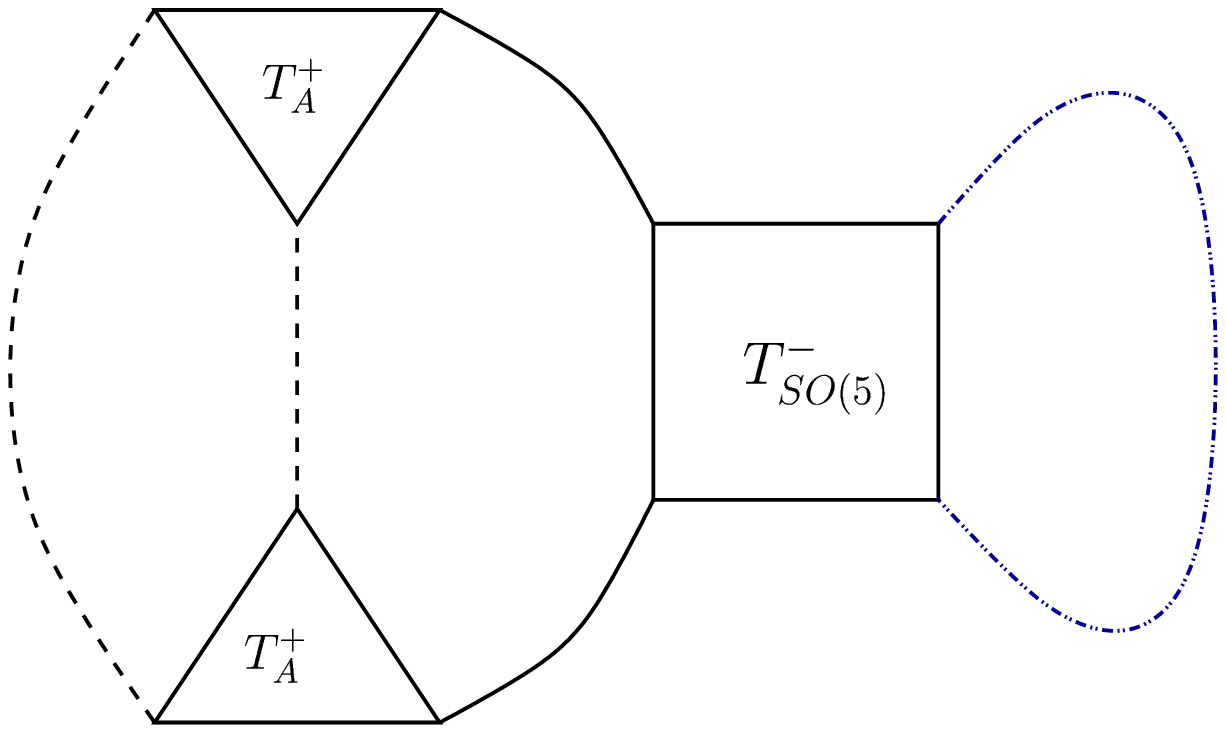}{2.1in}}
\medskip\centerline{\vbox{
\baselineskip12pt\advance\hsize by -1truein
\noindent\footnotefont{\bf Fig.~14:} Construction of genus three theory with $G^{max}=\widetilde{ so(5)}u(1)$. The fluxes are ${\cal F}=(\frac12,\,\frac12,\;0)$.}} 

\

 The contribution $\frac3{\beta^2\gamma^2}\widetilde {\bf 5}p\,q$ in \sofhijklvw\  accounts for relevant operators due to \softmo\ whence  $\widetilde {\bf 5} \gamma^2\beta^2 p q$ is an irrelevant term.
All of this information matches what we have discussed for the $G^{max}=so(5)u(1)$ case upon changing the identification of $so(5)u(1)$ in $so(7)$.  We mentioned in section four that different embeddings of $so(5)$ lead to different maximal punctures, and here we have an example where we can introduce new maximal punctures for which mesons are doublets of $su(2)_{t(\beta/\gamma)^{\mp1}}$ charged under $u(1)_{\beta\gamma}$ and singlets with respect to $su(2)_{t(\gamma/\beta)^{\pm1}}$. We will not discuss this in detail in this work.

\

\

\

In the next section we will study $(1,0)$ theory probing $\Z_k$ singularity and derive the anomalies of theories obtained in four dimensions by integrating anomaly polynomial over a Riemann surface which for simplicity we assume to have no punctures.  We will find that anomalies obtained in six dimensions match the anomalies we derived here using four dimensional field theoretic arguments.

\

\newsec{Anomalies from six dimensions}

We consider the 6d $(1,0)$ SCFT described by a $\Z_2$ orbifold of the $A_1$ type $(2,0)$ theory. This SCFT has a low-energy gauge theory description as $su(2)+4F$ on a generic point on its tensor branch. It is thought that this theory has an $so(7)$ global symmetry where the $8$ half-hypermultiplets transform in the $\bf{8}$ dimensional representation of $so(7)$\refs{\OhmoriPIA,\DelZottoRCA}. The anomaly polynomial for this theory was evaluated in \OhmoriAMP. Alternatively we can evaluate it from the gauge theory description as done in  appendix C. Either way we find\foot{We refer the reader to  appendix C for our conventions regarding characteristic classes.}:

\eqn\AnomPolSOS{\eqalign{
I^{so(7)}_8  = & \frac{11 C^2_2(R)}{12} - \frac{C_2(R) p_1 (T)}{24} + \frac{C_2(so(7))_{\bf{8}} p_1 (T)}{24} - \frac{C_2(R) C_2(so(7))_{\bf{8}}}{2} \cr
+ & \frac{7 C^2_2(so(7))_{\bf{8}}}{48} - \frac{C_4(so(7))_{\bf{8}}}{6} + \frac{29 p^2_1 (T) - 68 p_2 (T)}{2880} \,.
}}
We desire to calculate the central charges of the 4d theory, resulting from the compactification of the above 6d theory on a genus $g>1$ Riemann surface, by integrating the 6d anomaly polynomial. To preserve ${\cal N}=1$ supersymmetry in 4d we must perform a twist. Particularly we decompose the 6d Lorentz group as $so(5,1)\rightarrow so(3,1)\times so(2)_s$ where $so(3,1)$ is the 4d Lorentz group and $so(2)_s$ acts on the Riemann surface. Further we decompose $su(2)_R$ to its Cartan, $u(1)_R$, and twist $so(2)_s \rightarrow so(2)_s - u(1)_R$. Decomposing the supercharges we find:

\eqn\eqq{
{\bf{2}}_R\otimes {\bf{4}}_L\rightarrow (1_{\frac{1}{2}} + 1_{-\frac{1}{2}})\otimes ({\bf{2}}_{\frac{1}{2}} + {\bf{2'}}_{-\frac{1}{2}})\rightarrow{ \bf{2}}_{0} + {\bf{2'}}_{0} + {\bf{2}}_{1} + {\bf{2'}}_{-1} \,,
}
where the subscript states the charges under the twisted $so(2)_s - u(1)_R$ symmetry.

 From the above expression one can see that the twisting $so(2)_s \rightarrow so(2)_s - u(1)_R$ will ensure that half the spinors, those with zero charge in \eqq, are invariant and so will be co-variantly constant. This ensures ${\cal N}=1$ supersymmetry in 4d. 

 Next we proceed with integrating the anomaly polynomial. For this we must first decompose the various characteristic classes into the 4d Pontryagin classes of the tangent bundle, $p_1(T)'$ and $p_2(T)'$, the first Chern class of the Riemann surface, $t$, and the first Chern class of the $u(1)$ R-symmetry bundle, $C_1(F)$. Using the splitting principle we can decompose the Pontryagin classes as\foot{We refer the reader to appendix C for the splitting principle obeyed by the various characteristic classes.}:

\eqn\po{p_1(T) = t^2 + p_1(T)'\,,
}
\eqn\pt{p_2(T) = t^2 p_1(T)' + p_2(T)'
}

For the R-symmetry the splitting principle implies:

\eqn\spp{
1+C_1 (R) x + C_2 (R) x^2 = (1+x n_1)(1+x n_2) \rightarrow C_1 (R)=n_1 + n_2, C_2 (R) = n_1 n_2\,,}
where we use $n_1$ and $n_2$ for the Chern roots. As the R-symmetry is $su(2)$, in the untwisted case the first Chern class must vanish leading to $n_2=-n_1=C_1(F)$ where we have identified the curvature as that for the 4d $u(1)$ R-symmetry. As mentioned to preserve supersymmetry we must perform the twist $so(2)_s \rightarrow so(2)_s - u(1)_R$. This implies we must shift $n_2$ as $n_2\rightarrow n_2 - t$. Thus we set: $n_1 = -C_1 (F), n_2 = C_1 (F) - t$. note that the total first Chern class of the twisted bundle is indeed $-t$ so, for spinors with the appropriate charges, it can cancel the spacetime curvature contribution leading to a co-variantlly constant spinor as desired.   

Thus we get:

\eqn\isoe{
 \int I^{so(7)}_8 = \frac{11 (g-1)}{3} C^3_1 (F) + \frac{g-1}{12} C_1 (F) p_1(T)' + (g-1) C_1 (F) C_2(so(7))_{\bf{8}}\,,}
where we used the Gauss-Bonnet theorem:

\eqn\gaub{
\int t = 2(1-g)\,.}
We need to compare this against the 4d anomaly polynomial. As mentioned in appendix C, a single Weyl fermion contributes to the anomaly polynomial as: $ch(F) \hat{A} (T)$. Expanding the A-roof genus and Chern characters for the $so(7)$ and $u(1)_R$ symmetry bundles we find it contributes:

\eqn\fpl{
I_6 = \frac{Tr(R^3)}{6} C^3_1 (F) - \frac{Tr(R)}{24} C_1 (F) p_1(T)' - Tr(R F^2_{so(7)}) C_1 (F) C_2 (so(7))
\,,}
where we have defined $C_2 (so(7))_{\bf{r}} = T_{\bf{r}} C_2 (so(7))$, for $T_{\bf{r}}$ the second Casimir of the representation $r$, to convert to a Chern class independent of the representation. Comparing this with the 4d anomaly polynomial we get from integrating the 6d one, we find that: $Tr(R^3) = 22(g-1), Tr(R)=-2(g-1)$ and $Tr(R F^2_{so(7)}) = -(g-1)$. For the last result we have used $T_{\bf{8}} = 1$. From here the central charges are,

\eqn\cealk{
a=\frac3{32}(3TrR^3-TrR)=\frac{51}{8}(g-1)\,,\qquad c=\frac1{32}(9TrR^3-5TrR)=\frac{13}2(g-1)\,.}
 These numbers precisely match the QFT computation for theories of type $G^{max}=so(7)$ \anl.

\

\subsec{Models with $G^{max}=so(5)u(1)$}

In this section we consider the compactification with flux for the $so(7)$ global symmetry. Specifically we turn on a non-trivial flux for a $u(1)$ subgroup of $so(7)$ with support on the Riemann surface. This will break $so(7)$ down to the $u(1)$ and its commutant. Here we consider the $u(1)$ whose commutant in $so(7)$ is $so(5)$. Thus we expect the resulting theory to have $u(1)\times so(5)$ global symmetry. We next wish to evaluate the central charges of the resulting 4d theory by integrating the anomaly polynomial of the 6d theory.

First we must determine how the $so(7)$ Chern classes decompose under the new symmetry. Under its $u(1)\times so(5)$ subgroup, the $\bf{8}$ dimensional representation of $so(7)$ decomposes as: ${\bf{8}}\longrightarrow {\bf{4}}^{\frac{1}{2}} +{ \bf{4}}^{-\frac{1}{2}}$. Thus using the splitting principle in appendix C we find:

\eqn\spla{\eqalign{& 8 - C_2(so(7))_{\bf{8}} + \frac{1}{12} \left(C^2_2(so(7))_{\bf{8}} - 2 C_4(so(7))_{\bf{8}} \right) = ch(so(7)_{\bf{8}}) \cr & =  ch\left(u(1)_{\frac{1}{2}}\otimes usp(4)_{\bf{4}} \oplus u(1)_{-\frac{1}{2}}\otimes usp(4)_{\bf{4}}\right) = ch(u(1)_{\frac{1}{2}})ch(usp(4)_{\bf{4}}) \cr & +  ch(u(1)_{-\frac{1}{2}})ch(usp(4)_{\bf{4}}) = \left(1 + \frac{C_1(u(1)_a)}{2} + \frac{C^2_1(u(1)_a)}{8} + \frac{C^3_1(u(1)_a)}{48} + \frac{C^4_1(u(1)_a)}{384} \right)\cr & \left(4 - C_2(usp(4))_{\bf{4}} + \frac{1}{12} \left(C^2_2(usp(4))_{\bf{4}} - 2 C_4(usp(4))_{\bf{4}} \right) \right)\cr + & \left(1 - \frac{C_1(u(1)_a)}{2} + \frac{C^2_1(u(1)_a)}{8} - \frac{C^3_1(u(1)_a)}{48} + \frac{C^4_1(u(1)_a)}{384} \right)\cr & \left(4 - C_2(usp(4))_{\bf{4}} + \frac{1}{12} \left(C^2_2(usp(4))_{\bf{4}} - 2 C_4(usp(4))_{\bf{4}} \right) \right)
\,,}}
where we shall refer to the $u(1)$ as $u(1)_a$ in preparation to the latter sections where we turn on fluxes for several $u(1)$'s.  

Comparing forms of equal dimension we find: $C_2(so(7))_{\bf{8}} = - C^2_1(u(1)_a) + 2 C_2(usp(4))_{\bf{4}}$, $C_4(so(7))_{\bf{8}} = \frac{3 C^4_1(u(1)_a)}{8} - \frac12C^2_1(u(1)_a) {C_2(usp(4))_{\bf 4}} + C^2_2(usp(4))_{\bf{4}} + 2 C_4(usp(4))_{\bf{4}}$.

Inserting these into the anomaly polynomial \AnomPolSOS, we get:

\eqn\AnomPolSOf{\eqalign{
I^{so(5)}_8 & =  \frac{11 C^2_2(R)}{12} - \frac{C_2(R) p_1 (T)}{24} - \frac{C^2_1(u(1)_a) p_1 (T)}{24} + \frac{C_2(R) C^2_1(u(1)_a)}{2} \cr - & C_2(R) C_2(usp(4))_{\bf{4}} + \frac{C_2(usp(4))_{\bf{4}} p_1 (T)}{12} + \frac{C^4_1(u(1)_a)}{12} - \frac{C^2_1(u(1)_a) C_2(usp(4))_{\bf{4}}}{2} \cr + & \frac{5 C^2_2(usp(4))_{\bf{4}}}{12} - \frac{C_4(usp(4))_{\bf{4}}}{3} + \frac{29 p^2_1 (T) - 68 p_2 (T)}{2880} \,.}}

Next we take the flux into account by setting: $C_1(u(1)_a) = -z t + \epsilon_1 C_1(F) + C'_1(u(1)_a)$. The first term is the flux on the Riemann surface whose strength is determined by $z$. The remaining terms are identified with the curvature of the $4d$ $u(1)$ global symmetry. In principle it can now mix with the $6d$ R-symmetry so that the 4d superconformal R-symmetry be: $u(1)_{6d} + \epsilon_1 u(1)_a$, where $\epsilon_1$ is to be evaluated via a-maximization. This is taken into account by the second term in the expansion of $C_1(u(1)_a)$. Finally we denote the curvature of the 4d $u(1)_a$ global symmetry as $C'_1(u(1)_a)$.

The flux $z$ is quantized such that $\int C_1(u(1)_a) = \frac{n}{q}$, where $n$ is an integer and $q$ is the smallest charge in the system. In physical language this just expresses the Dirac quantization condition where we use the expression for the first Chern class in terms of the connection's field strength, $C_1(u(1)_a) = \frac{F_{u(1)_a}}{2 \pi}$. Inserting the expression for $C_1(u(1)_a)$ and integrating over the Riemann surface we find:

\eqn\Quan{
z = \frac{n}{2 q (g-1)}\,.}

This brings us to determining the value of $q$. This depends on whether the global symmetry is $so(7)$ or $Spin(7)$. The minimal $u(1)$ charge for $Spin(7)$ appears in the decomposition of the spinor, and in our conventions is $\frac{1}{2}$. However for $so(7)$, where the spinor is projected out, the minimal charge appears in the decomposition of the vector, and in our conventions is $1$. Thus if the global symmetry is $Spin(7)$, $z$ must be an integer multiplet of $\frac{1}{g-1}$ while for $so(7)$ half-integer multiplets are allowed. Note that while the gauge theory description has a state in the spinor of $so(7)$, and indeed this is why all the Chern classes we use for $so(7)$ are framed in it, this state is gauge variant. All gauge invariant states are made from an even number of this state and so $so(7)$ may still be consistent. 

Returning to integrating the anomaly polynomial, we next insert the decomposition of the Chern classes into \AnomPolSOf\ and perform the integration finding:

\eqn\soff{\eqalign{
&  \int I^{so(5)}_8  = \frac{(2 \epsilon^3_1 z - 6 \epsilon_1 z - 3 \epsilon^2_1 + 11) (g-1)}{3} C^3_1 (F) \cr + & \frac{(g-1)(1-2 z \epsilon_1 )}{12} C_1 (F) p_1(T)' + 2 (g-1) (1- z \epsilon_1 ) C_1 (F) C_2(usp(4))_{\bf{4}} \cr + & 2 (g-1) (\epsilon^2_1 z -z -\epsilon_1)C^2_1 (F) C'_1(u(1)_a) + \frac{2 (g-1) z C'^3_1(u(1)_a)}{3} \cr + & (g-1)(2 z \epsilon_1 - 1) C_1 (F) C'^2_1(u(1)_a) - \frac{(g-1) z C'_1(u(1)_a) p_1(T)'}{6} \cr - & 2 (g-1) z C'_1(u(1)_a) C_2(usp(4))_{\bf{4}}\,.}}
From this we can read of the anomalies. We start with those involving only the R-symmetry, finding:

\eqn\rrrsof{
Tr(R^3) = 2(g-1)(2 \epsilon^3_1 z - 6 \epsilon_1 z - 3 \epsilon^2_1 + 11), Tr(R)=-2(g-1)(1-2 z \epsilon_1 )\,.}
This gives the $a$ central charge:

\eqn\sacsoff{
a= \frac{3(g-1)(6 \epsilon^3_1 z - 20 \epsilon_1 z - 9 \epsilon^2_1 + 34)}{16}\,.}

Performing a-maximization we find that $\epsilon_1=\frac{3-\sqrt{9+40 z^2}}{6 z}$. Inserting this value to a and c we find:

\eqn\amaxx{\eqalign{
 & a = (g-1)\frac{9(-3+\sqrt{9+40 z^2})+8 z^2(54+5\sqrt{9+40 z^2})}{96 z^2}, \cr & c = (g-1)\frac{9(-3+\sqrt{9+40 z^2})+ z^2(432+44\sqrt{9+40 z^2})}{96 z^2} \,.}}
 This matches \aextsiof\ from the four dimensional computation.

For the case of $z=1$, these simplify to:

\eqn\zzo{
a=\frac{187}{24} (g-1),\;\;\;\;\; c=\frac{97}{12} (g-1)\,.}
This matches the computation in four dimensions \angst. 

Another interesting value is $z=\frac{1}{4}$, where the central charges are given by:

\eqn\zzmin{
a=\frac{23}{12} \sqrt{\frac{23}{2}} (g-1),\;\;\;\;\; c=\frac{47}{24} \sqrt{\frac{23}{2}} (g-1)\,.}
For $g=3$, this matches the computation in four dimensions \ansost. Note that for this case $z$ is a half-integer multiple of $\frac{1}{g-1} = \frac{1}{2}$ so this flux choice is consistent only with $so(7)$ and not $Spin(7)$.  This presumably is consistent with the 6d picture where the $8$ half-hypermultiplet doublets of $su(2)$ transform as a spinor of $Spin(7)$ because the only gauge invariant observables we can construct from them will be representations of $so(7)$.

We can also look at other anomalies, particularly those involving the global symmetries. These can all be read from the anomaly polynomial in \soff. We shall not discuss all of them rather mentioning only a selected few. As in the previous case we can consider the contribution of a single Weyl fermion, and expand the Chern characters for the $usp(4)$, $u(1)_a$ and $u(1)_R$ symmetry bundles, where we find:

\eqn\fpls{\eqalign{
& I_6 \supset \frac{Tr(Q^3_a)}{6} C'^3_1(u(1)_a) + \frac{Tr(R Q^2_a)}{2} C_1 (F) C'^2_1(u(1)_a) + \frac{Tr(R^2 Q_a)}{2} C^2_1 (F) C'_1(u(1)_a) \cr & - Tr(R F^2_{usp(4)}) C_1 (F) C_2 (usp(4)) + ...
\,.}}

Comparing this with the the 4d anomaly polynomial we get from integrating the 6d one, we find that:

\eqn\othanom{\eqalign{
 & Tr(Q^3_a) = 4 (g-1) z, \cr & Tr(R Q^2_a) = 2(g-1)(2 z \epsilon_1 - 1), \cr & Tr(R^2 Q_a) = 4 (g-1) (\epsilon^2_1 z -z -\epsilon_1), \cr & Tr(R F^2_{usp(4)})= -(g-1) (1- z \epsilon_1 ) \,,}}
where we have used $T^{usp(4)}_{\bf{4}} = \frac{1}{2}$. 

Using the value for $\epsilon_1$ we got from a-maximization, these become:

\eqn\othanoma{\eqalign{
 & Tr(Q^3_a) = 4 (g-1) z, \cr & Tr(R Q^2_a) = -\frac{2}{3}(g-1)\sqrt{9+40 z^2}, \cr & Tr(R^2 Q_a) = \frac{4 z}{9} (g-1), \cr & Tr(R F^2_{usp(4)})= -\frac{(g-1)}{6} (3 + \sqrt{9+40 z^2} ) \,.}}

For the special case of $z=1$, these further simplify to:

\eqn\othanomb{\eqalign{
 & Tr(Q^3_a) = 4 (g-1), \cr & Tr(R Q^2_a) = -\frac{14}{3}(g-1), \cr & Tr(R^2 Q_a) = \frac{4}{9} (g-1), \cr & Tr(R F^2_{usp(4)})= -\frac{5(g-1)}{3} \,.}}
We will check the anomalies here with the field theory computation in appendix D.

\

\subsec{Models for $G^{max}=su(3)u(1)$}

Besides the $u(1)$ in $so(7)$ whose commutant is $so(5)$, there are other $u(1)$ subgroups of $so(7)$ with smaller commutant groups. Maintaining rank $2$, the other possibilities are $su(3)$ and $su(2)\times su(2)=su(2)^2$. In this section we shall deal with the case of $u(1)\times su(3)$. We shall discuss the second case, $u(1)\times su(2)^2$, later as a limit of a compactification involving two fluxes. 

Under the $u(1)\times su(3)$ subgroup of $so(7)$ the $\bf{8}$ dimensional representation decomposes as: ${\bf{8}}\rightarrow 1^{-3} + {\bf{3}}^1 + 1^{3} + \bar{{\bf{3}}}^{-1}$. Employing the splitting principle, we find:

\eqn\sutt{\eqalign{
& 8 - C_2(so(7))_{\bf{8}} + \frac{1}{12} (C^2_2(so(7))_{\bf{8}} - 2 C_4(so(7))_{\bf{8}}) = ch(so(7)_{\bf{8}}) \cr = & ch\left(u(1)_{-3}\oplus su(3)_{\bf{3}} \otimes u(1)_{1} \oplus u(1)_{3} \oplus su(3)_{\bar{\bf{3}}} \otimes u(1)_{-1} \right) = ch(u(1)_{-3}) \cr + & ch(su(3)_{\bf{3}})ch(u(1)_{1}) + ch(u(1)_{3}) + ch(su(3)_{\bar{\bf{3}}})ch(u(1)_{-1}) \cr = & \left(1 - 3C_1(u(1)_a) + \frac{9 C^2_1(u(1)_a)}{2} - \frac{9 C^3_1(u(1)_a)}{2} + \frac{27 C^4_1(u(1)_a)}{8} \right) \cr + & \left(3 - C_2(su(3))_{\bf{3}} +\frac{C_3(su(3))_{\bf{3}}}{2} + \frac{C^2_2(su(3))_{\bf{3}}}{12} \right) \cr & \left(1 + C_1(u(1)_a) + \frac{ C^2_1(u(1)_a)}{2} + \frac{ C^3_1(u(1)_a)}{6} + \frac{ C^4_1(u(1)_a)}{24}\right) \cr + & \left(1 + 3C_1(u(1)_a) + \frac{9 C^2_1(u(1)_a)}{2} + \frac{9 C^3_1(u(1)_a)}{2} + \frac{27 C^4_1(u(1)_a)}{8}\right) \cr + & \left(3 - C_2(su(3))_{\bf{3}} -\frac{C_3(su(3))_{\bf{3}}}{2} + \frac{C^2_2(su(3))_{\bf{3}}}{12}\right) \cr & \left(1 - C_1(u(1)_a) + \frac{ C^2_1(u(1)_a)}{2} - \frac{ C^3_1(u(1)_a)}{6} + \frac{ C^4_1(u(1)_a)}{24} \right)\,.}}

Matching forms of equal order, we find: $C_2(so(7))_{\bf{8}} = -12 C^2_1(u(1)_a) + 2 C_2(su(3))_{\bf{3}}$, $C_4(so(7))_{\bf{8}} = 30 C^4_1(u(1)_a) - 18 C^2_1(u(1)_a) C_2(su(3))_{\bf{3}} + C^2_2(su(3))_{\bf{3}} - 6 C_1(u(1)_a) C_3(su(3))_{\bf{3}}$. 

Similarly to the previous case we turn on the flux by setting: $C_1(u(1)_a) = -z t + \epsilon_1 C_1(F) + C'_1(u(1)_a)$. Where again $z$ measures the strength of the flux, $\epsilon_1$ takes into account possible mixing with the superconformal R-symmetry and the last term is the actual 4d curvature. Like in the previous cases $z$ is quantized according to equation \Quan. Now the minimal charge of the $u(1)$ for $Spin(7)$ in our convention is $1$, and for $so(7)$ it is $2$. 

Inserting all these into \AnomPolSOS\ and performing the integration we find:

\eqn\sutplk{\eqalign{  & \int I^{su(3)}_8 = \frac{(384 \epsilon^3_1 z - 72 \epsilon_1 z - 36 \epsilon^2_1 + 11) (g-1)}{3} C^3_1 (F) \cr + & \frac{(g-1)(1-24 z \epsilon_1 )}{12} C_1 (F) p_1(T)' + 2 (g-1) (1- 8 z \epsilon_1 ) C_1 (F) C_2(su(3))_{\bf{3}} \cr + & 24 (g-1) (16 \epsilon^2_1 z -z -\epsilon_1)C^2_1 (F) C'_1(u(1)_a) + 128 (g-1) z C'^3_1(u(1)_a) \cr + & 12 (g-1)(32 z \epsilon_1 - 1) C_1 (F) C'^2_1(u(1)_a) - 2(g-1) z C'_1(u(1)_a) p_1(T)' \cr - & 16 (g-1) z C'_1(u(1)_a) C_2(su(3))_{\bf{3}} + 2 z (g-1) C_3(su(3))_{\bf{3}}\,.}}
From this we can read off the anomalies finding:

\eqn\rrrsut{
Tr(R^3) = 2(g-1)(384 \epsilon^3_1 z - 72 \epsilon z - 36 \epsilon^2_1 + 11), Tr(R)=-2(g-1)(1-24 z \epsilon_1 )\,.}
This gives the $a$ central charge:

\eqn\acdutt{
a= \frac{3(g-1)(576 \epsilon^3_1 z - 120 \epsilon_1 z - 54 \epsilon^2_1 + 17)}{8}\,.}
Performing a-maximization we find that $\epsilon_1=\frac{3-\sqrt{9+640 z^2}}{96 z}$. Inserting this value to a and c we find:

\eqn\ansutt{\eqalign{
& a = (g-1)\frac{9(-3+\sqrt{9+640 z^2})+64 z^2(159+10\sqrt{9+640 z^2})}{2048 z^2}, \cr & c = (g-1)\frac{9(-3+\sqrt{9+640 z^2})+ 64 z^2(160+11\sqrt{9+640 z^2})}{2048 z^2} \,.}}

For the case of $z=\frac{1}{10}$, we find:

\eqn\comsutt{
a=\frac{7(267 + 11 \sqrt{385})}{512} (g-1), \;\;\;\;\;\;\;c=\frac{9425 + 401\sqrt{385}}{2560} (g-1)\,.
} This matches the result of \sustt\ obtained in QFT. The total $u(1)$ flux on the surface in this case is $1$. Note that this $u(1)$ differs from the $u(1)$'s used in the preceding and proceeding discussions by a factor of $2$. Thus, even though the flux is $1$, the flux vector is as given in section 6.6. 

We can again consider additional anomalies, specifically those considered in \fpls. From \sutplk we find:

\eqn\othanom{\eqalign{
 & Tr(Q^3_a) = 768 (g-1) z, \cr & Tr(R Q^2_a) = 24(g-1)(32 z \epsilon_1 - 1), \cr & Tr(R^2 Q_a) = 48 (g-1) (16\epsilon^2_1 z -z -\epsilon_1), \cr & Tr(R F^2_{su(3)})= -(g-1) (1- 8 z \epsilon_1 ) \,,}}
where we have used $T^{su(3)}_{\bf{3}} = \frac{1}{2}$.

Using the value for $\epsilon_1$ we got from a-maximization, these become:

\eqn\othanoma{\eqalign{
 & Tr(Q^3_a) = 768 (g-1) z, \cr & Tr(R Q^2_a) = -8(g-1)\sqrt{9+640 z^2}, \cr & Tr(R^2 Q_a) = \frac{16 z}{3} (g-1), \cr & Tr(R F^2_{su(3)})= -\frac{(g-1)}{12} (9 + \sqrt{9+640 z^2} ) \,.}}

For the special case of $z=\frac{1}{10}$, these further simplify to:

\eqn\othanomb{\eqalign{
 & Tr(Q^3_a) = \frac{384}{5} (g-1), \cr & Tr(R Q^2_a) = -\frac{8 \sqrt{385}}{5}(g-1), \cr & Tr(R^2 Q_a) = \frac{8}{15} (g-1), \cr & Tr(R F^2_{su(3)})= -\frac{(g-1)}{60}(45 + \sqrt{385} ) \,.}}
The anomalies here will be compared to four dimensional computation in appendix D.

\

\subsec{Models for $G^{max}=su(2)u(1)^2$}

In this section we consider turning on two fluxes each under a different $u(1)$ inside $so(7)$. We choose the $u(1)$'s so as to preserve the maximal possible symmetry, which is $u(1)^2 \times su(2)$. There are in fact two different ways to do this. In both we start with the previous $u(1)\times so(5)$ case and turn on a flux in a $u(1)$ within $so(5)$. There are two different $u(1)\times su(2)$ subgroups inside $so(5)$. One is the maximal $u(1)\times su(2)$ subgroup, under which the $\bf{4}$ dimensional representation of $so(5)$ decomposes as: $\bf{4}\rightarrow {\bf{2}}^1 + {\bf{2}}^{-1}$. The other is taking the $su(2)\times su(2)$ maximal subgroup and using the Cartan of one of the $su(2)$'s. Under this subgroup the $\bf{4}$ dimensional representation of $so(5)$ decomposes as: ${\bf{4}}\rightarrow \bf{1}^1 + {\bf{1}}^{-1} + {\bf{2}}^{0}$.

We shall first deal with the first case and latter deal with the second. 

\

\noindent $\bullet$ {\it Models with $G^{max}=su(2)_{diag}u(1)^2$}

First we must decompose the $so(5)$ Chern classes to those of the subgroup. For this we again utilize the splitting principle:

\eqn\susus{\eqalign{
& 4 - C_2(usp(4))_{\bf{4}} + \frac{C^2_2(usp(4))_{\bf{4}} - 2 C_4(usp(4))_{\bf{4}}}{12} = ch(usp(4)_{\bf{4}}) \cr = & ch\left(u(1)_{1}\otimes su(2)_{\bf{2}} \oplus u(1)_{-1}\otimes su(2)_{\bf{2}}\right) = ch(u(1)_1)ch(USp(2)_{\bf{2}}) \cr + & ch(u(1)_{-1})ch(USp(2)_{\bf{2}}) = \left(1 + C_1(u(1)_b) + \frac{C^2_1(u(1)_b)}{2} + \frac{C^3_1(u(1)_b)}{6} + \frac{C^4_1(u(1)_b)}{24} \right) \cr  & \left(2 - C_2(su(2))_{\bf{2}} + \frac{C^2_2(su(2))_{\bf{2}}}{12} \right) \cr + & \left(1 - C_1(u(1)_b) + \frac{C^2_1(u(1)_b)}{2} - \frac{C^3_1(u(1)_b)}{6} + \frac{C^4_1(u(1)_b)}{24} \right) \cr & \left(2 - C_2(su(2))_{\bf{2}} + \frac{C^2_2(su(2))_{\bf{2}}}{12} \right)\,,}}
where we have denoted this $u(1)$ as $u(1)_b$.

By matching forms of the same order we find: $C_2(usp(4))_{\bf{4}} = 2C_2(su(2))_{\bf{2}} - 2 C^2_1(u(1)_b), C_4(usp(4))_{\bf{4}} = C^4_1(u(1)_b) + 2 C^2_1(u(1)_b) C_2(su(2))_{\bf{2}} + C^2_2(su(2))_{\bf{2}}$.

The flux is then taken into account by setting: $C_1(u(1)_b) = -x t + \epsilon_2 C_1(F) + C'_1(u(1)_b)$. Like previously, the first term is the flux on the Riemann surface which strength is measured by $x$. The second term takes into account possible mixing with the R-symmetry, while the third is the curvature for the 4d $u(1)_b$ global symmetry. Again $x$ needs to be quantized according to \Quan. Now the minimal charge of the $u(1)$ for $Spin(7)$ in our convention is $1$, and for $so(7)$ it is $2$. 

Next we insert these decompositions in the anomaly polynomial \AnomPolSOf, and integrate over the Riemann surface finding:

\eqn\spoltf{\eqalign{& \int I^{su(2)}_8 = \cr  & \frac{(2 \epsilon^3_1 z + 32 \epsilon^3_2 x - 6 \epsilon_1 z + 12 \epsilon_1 \epsilon^2_2 z + 12 \epsilon_2 \epsilon^2_1 x - 3 \epsilon^2_1 - 12 \epsilon^2_2 - 24 \epsilon_2 x + 11) (g-1)}{3} C^3_1 (F) \cr + & \frac{(g-1)(1-2 z \epsilon_1 -8 x \epsilon_2)}{12} C_1 (F) p_1(T)' + 4 (g-1) (1- z \epsilon_1 - 4 x \epsilon_2) C_1 (F) C_2(su(2))_{\bf{2}} \cr + & 2 (g-1) (\epsilon^2_1 z + 2\epsilon^2_2 z +  4 \epsilon_1 \epsilon_2 x -z -\epsilon_1)C^2_1 (F) C'_1(u(1)_a) + \frac{2 (g-1) z C'^3_1(u(1)_a)}{3} \cr + &
4(g-1)(\epsilon^2_1 x + 8\epsilon^2_2 x - 2 x - 2 \epsilon_2 + 2 z \epsilon_1 \epsilon_2) C^2_1 (F) C'_1(u(1)_b) - \frac{2(g-1) x C'_1(u(1)_b) p_1(T)'}{3} \cr + &
 (g-1)(2 z \epsilon_1 + 4 x \epsilon_2 - 1) C_1 (F) C'^2_1(u(1)_a) - \frac{(g-1) z C'_1(u(1)_a) p_1(T)'}{6} \cr - & 4 (g-1) z C'_1(u(1)_a) C_2(su(2))_{\bf{2}} + 4 x (g-1) C'^2_1(u(1)_a) C'_1(u(1)_b) \cr + & 4 z (g-1) C'^2_1(u(1)_b) C'_1(u(1)_a) + 8 (g-1) (\epsilon_1 x + \epsilon_2 z) C_1 (F) C'_1(u(1)_b) C'_1(u(1)_a) \cr + & \frac{32(g-1) x C'^3_1(u(1)_b)}{3} + 4 (g-1) (8 \epsilon_2 x + \epsilon_1 z - 1) C_1 (F) C'^2_1(u(1)_b) \cr - & 16 x (g-1) C'_1(u(1)_b) C_2(su(2))_{\bf{2}}\,.}}
From these we see that: 

\eqn\rrrthr{\eqalign{& Tr(R^3) = 2(g-1)(2 \epsilon^3_1 z + 32 \epsilon^3_2 x - 6 \epsilon_1 z + 12 \epsilon_1 \epsilon^2_2 z + 12 \epsilon_2 \epsilon^2_1 x - 3 \epsilon^2_1 - 12 \epsilon^2_2 - 24 \epsilon_2 x + 11),\cr & Tr(R)=-2(g-1)(1-2 z \epsilon_1 -8 x \epsilon_2)\,.}}
This gives the $a$ central charge:

\eqn\cctrs{
a= \frac{3(g-1)(34 - 36 \epsilon^2_2 - 80 \epsilon_2 x + 96 \epsilon^3_2 x - 9 \epsilon^2_1 + 36 \epsilon^2_1 \epsilon_2 x + 6 \epsilon^3_1 z - 20 \epsilon_1 z + 36 \epsilon_1 \epsilon^2_2 z)}{16}\,.}

Next we need to perform a-maximization with respect to both $\epsilon_1$ and $\epsilon_2$. In general the solution is quite involved and we won't write it here. However for the specific case of $z=1, x=\frac{1}{4}$, we find: 

\eqn\amstr{
a=7.99177 (g-1),\;\;\;\;\;\;\;\; c=8.30369 (g-1)\,.
} This matches the result in four dimensions \gengnopu. Also note that for $g$ even the flux for $x$ is  consistent only with $so(7)$ and not $Spin(7)$.  

There are a considerable number of other anomalies, which similarly to the previous cases can be calculated from \spoltf, though we shall not consider this here.  

\

\noindent $\bullet$ {\it Models with $G^{max}=su(2)u(1)^2$}

The second subgroup differs in the decomposition of the $\bf{4}$ dimensional representation and thus in the relation between the $usp(4)$ and $u(1)\times su(2)$ Chern classes. Recall that under this subgroup the $\bf{4}$ dimensional representation of $so(5)$ decomposes as: $\bf{4}\rightarrow \bf{1}^1 + \bf{1}^{-1} + \bf{2}^{0}$. Thus using the splitting principle for the Chern classes we find:

\eqn\splsp{
1 + C_2(usp(4))_{\bf{4}} + C_4(usp(4))_{\bf{4}} = (1+C_1(u(1)_b))(1-C_1(u(1)_b))(1 + C_2(su(2))_{\bf{2}})\,.}
Matching forms of the same order we determine that: $C_2(usp(4))_{\bf{4}} = -C^2_1(u(1)_b) + C_2(su(2))_{\bf{2}}, C_4(usp(4))_{\bf{4}} = -C^2_1(u(1)_b) C_2(su(2))_{\bf{2}}$.

Turning on the flux is again taken into account by setting: $C_1(u(1)_b) = -x t + \epsilon_2 C_1(F) + C'_1(u(1)_b)$. Again $x$ is quantized according to \Quan, where now the minimal charge in our convention is $1$ for both cases.

Next we insert all these in the anomaly polynomial \AnomPolSOf\ and integrate over the Riemann surface finding:

\eqn\stttso{\eqalign{& \int I^{su(2)}_8 = \cr  & \frac{(2 \epsilon^3_1 z + 10 \epsilon^3_2 x - 6 \epsilon_1 z + 6 \epsilon_1 \epsilon^2_2 z + 6 \epsilon_2 \epsilon^2_1 x - 3 \epsilon^2_1 - 6 \epsilon^2_2 - 12 \epsilon_2 x + 11) (g-1)}{3} C^3_1 (F) \cr + & \frac{(g-1)(1-2 z \epsilon_1 -4 x \epsilon_2)}{12} C_1 (F) p_1(T)' + 2 (g-1) (1- z \epsilon_1 - x \epsilon_2) C_1 (F) C_2(su(2))_{\bf{2}} \cr + & 2 (g-1) (\epsilon^2_1 z + \epsilon^2_2 z +  2 \epsilon_1 \epsilon_2 x -z -\epsilon_1)C^2_1 (F) C'_1(u(1)_a) + \frac{2 (g-1) z C'^3_1(u(1)_a)}{3} \cr + & 2(g-1)(\epsilon^2_1 x + 5\epsilon^2_2 x - 2 x - 2 \epsilon_2 + 2 z \epsilon_1 \epsilon_2) C^2_1 (F) C'_1(u(1)_b) - \frac{(g-1) x C'_1(u(1)_b) p_1(T)'}{3} \cr + & (g-1)(2 z \epsilon_1 + 2 x \epsilon_2 - 1) C_1 (F) C'^2_1(u(1)_a) - \frac{(g-1) z C'_1(u(1)_a) p_1(T)'}{6} \cr - & 2 (g-1) z C'_1(u(1)_a) C_2(su(2))_{\bf{2}} + 2 x (g-1) C'^2_1(u(1)_a) C'_1(u(1)_b) \cr + & 2 z (g-1) C'^2_1(u(1)_b) C'_1(u(1)_a) + 4 (g-1) (\epsilon_1 x + \epsilon_2 z) C_1 (F) C'_1(u(1)_b) C'_1(u(1)_a) \cr + & \frac{10(g-1) x C'^3_1(u(1)_b)}{3} + 2 (g-1) (5 \epsilon_2 x + \epsilon_1 z - 1) C_1 (F) C'^2_1(u(1)_b) \cr - &  2 x (g-1) C'_1(u(1)_b) C_2(su(2))_{\bf{2}}\,.}}
From these we see that: 

\eqn\rrtrsuttt{\eqalign{& Tr(R^3) = 2(g-1)(2 \epsilon^3_1 z + 10 \epsilon^3_2 x - 6 \epsilon_1 z + 6 \epsilon_1 \epsilon^2_2 z + 6 \epsilon_2 \epsilon^2_1 x - 3 \epsilon^2_1 - 6 \epsilon^2_2 - 12 \epsilon_2 x + 11)\,,\cr & Tr(R)=-2(g-1)(1-2 z \epsilon_1 -4 x \epsilon_2)\,.}}
This gives the $a$ central charge:

\eqn\atrrsutt{
a= \frac{3(g-1)(34 - 18 \epsilon^2_2 - 40 \epsilon_2 x + 30 \epsilon^3_2 x - 9 \epsilon^2_1 + 18 \epsilon^2_1 \epsilon_2 x + 6 \epsilon^3_1 z - 20 \epsilon_1 z + 18 \epsilon_1 \epsilon^2_2 z)}{16}\,.}
Next we need to perform a-maximization with respect to both $\epsilon_1$ and $\epsilon_2$. In general the solution is quite involved and we won't write it here. However we can write the result for specific cases. For instance for $z=1, x=\frac{1}{4}$, we find: 

\eqn\ansutt{
a=7.8911 (g-1),\;\;\;\;\; c=8.19276 (g-1)\,.}
This matches the four dimensional result \ansosss\ for $g=3$ as expected.

Another interesting case is $z=0$. Now we have turned off one of the fluxes and so we expect symmetry enhancement. However the flux forms are different than before leading now to the breaking of $so(7)\rightarrow u(1)\times su(2)^2$ so we expect a 4d theory with $u(1)\times su(2)^2$ global symmetry. In this case we can solve for $\epsilon_1$ and $\epsilon_2$ exactly finding: $\epsilon_1 =0, \epsilon_2 =\frac{3-\sqrt{9+100 x^2}}{15 x}$. 

Inserting these values to a and c we find:

\eqn\ansuttsz{\eqalign{& a = (g-1)\frac{18(-3+\sqrt{9+100 x^2})+25 x^2(117+8\sqrt{9+100 x^2})}{600 x^2}, \cr & c = (g-1)\frac{9(-3+\sqrt{9+100 x^2})+ 10 x^2(147+11\sqrt{9+100 x^2})}{300 x^2} \,.}}
For the case of $x=\frac{1}{4}$, we get:

\eqn\ammsu{
a=\frac{2061 + 244 \sqrt{61}}{600} (g-1)\,, \;\;\;\;\;\; c=\frac{1038 + 127\sqrt{61}}{300} (g-1)\,.}
This matches \sofffo\ taking $g=3$.

\

\newsec{Summary}

In this work we have studied the general structure of ${\cal N}=1$ SCFT's which can be obtained from 6d $(1,0)$ SCFT's.  We have seen that for each such theory, the resulting conformal manifold is enriched, in addition to complex moduli of the Riemann surface, by the structure of the bundle associated with flavor symmetries of the 6d theory.  The diversity of the theories one obtains is further enhanced by the choice of an abelian subgroup of the flavor group where in addition to flat holonomies one can turn on discrete fluxes for them on the Riemann surface.  

We checked the general expectation based on 6d reasoning in detail for compactifications of two M5 branes probing $\Z_2$ singularity on a genus $g>1$ Riemann surface. In particular we have detailed a field theoretic construction of a large set of models from which we read off the anomalies and the conformal manifolds. These objects then were matched to their six dimensional counterparts. This is a special case of a general story but nevertheless it is rather rich. The richness is due to the fact that the compactification outcome is determined, in addition to the number of branes and the type of the singularity, by a choice of flux.  We have obtained field theoretic descriptions of the theories in four dimensions in terms of ``strongly coupled'' Lagrangians. The subtlety here is that in order to build theories corresponding to general Riemann surfaces one has to tune couplings of a theory with standard Lagrangian to a very specific, presumably strongly coupled, point.

There are several directions for further research. First, we have studied the different models by defining building blocks and gluing them together in different ways. A way to generalize the construction is to study more systematically different flows triggered by vacuum expectation values. Such flows geometrically correspond to shifting fluxes and closing punctures. At the very least these should provide non trivial checks of the statements we are making here.

 The six dimensional computations detailed here can be easily generalized to arbitrary $(1,0)$ SCFT's.  In particular we can apply it to M5 branes probing arbitrary singularities, numbers of branes, and choices of flux. For example, the conformal and flavor anomalies for $N$ M5 branes on $\Z_k$ singularity with no fluxes for any abelian subgroup of the $su(k)\times su(k)\times u(1)_t$ symmetry can be computed, and the $a,c$ values are (see appendix F for derivation), 

\eqn\anomsss{\eqalign{&
a=\frac1{32}(N-1)(12+k^2(9N^2+9N-6))(g-1)\,,\,\,\cr &
c=\frac1{32}(N-1)(8+k^2(9N^2+9N-4))(g-1)\,.
}} These anomalies generalize \BeniniMZ\ when $k=1$, as well as the $N=2$ $k=2$ case of this work.
On the field theory side however the construction we used in this work is very fine tuned to the special case of two branes and $\Z_2$ singularity. There are several ways in which one can attempt to address the question from four dimension nevertheless. One is straightforward but technically involved. The indices of the four dimensional theories were shown in \GaiottoUSA\ to be written in terms of eigenfunctions of certain integrable models for any number of branes and $k$ of $\Z_k$ (there should be also a generalization for any ADE type of singularity). See \refs{\MaruyoshiCAF,\ItoFPL} for explicit expressions for the Hamiltonians of these integrable models and \refs{\GaiottoXA,\GaddeIK,\GaddeUV} for the $k=1$ ${\cal N}=2$ case. Moreover knowing the index of a theory one can in principle extract its anomalies \refs{\SpiridonovWW,\ArdehaliBLA,\BobevKZA,\DiPietroBCA}. However, technically finding the eigenfunctions of the integrable models is a difficult task even in the class ${\cal S}$ ($k=1$) case. The different limits in which the problem can be solved more easily \refs{\GaddeUV,\GaiottoUSA} are blind to exactly marginal deformations though capture the relevant deformations.

Another road to generalizations is to understand better the dualities leading to relations between theories with Lagrangians and strongly coupled building blocks. This is the study of the superconformal tails. Such tails were studied for the $\Z_k$ orbifold case in \GaiottoUSA.
 In particular, if relevant dualities can be harnessed to isolate the strongly coupled building blocks one would be able to write ``strongly coupled'' Lagrangians of the type discussed in \RazamatL\ and in this work.  Technically this requires analyzing integral kernels arising from supersymmetric tails and finding procedures to invert them extending the results of \SpirWarnaar. 

The theories with more branes and general singularities will be classified in addition by a choice of an abelian symmetry where we turn on flux  embedded in $G\times G (\times u(1)_t)$ symmetry of the M5 brane setup. In addition we will have a rich choice of punctures, recently addressed 
in \HeckmanXDL, which will further come in different ``colors'' (corresponding to choices of abelian fluxes) classified by the subgroups of $G\times G (\times u(1)_t)$  which they preserve.

Other challenges include making contact with the different limits one can
 consider. Examples are the holographic limits \refs{\GaiottoLCA,\ApruzziZNA}\ and the compactification limits to lower dimensions. One can also consider studying general partition functions. On general grounds we would expect that various indices (lens index~\BeniniNC, $T^2\times \S^2$ index~\refs{\ClossetSXA,\BeniniNOA}) should be derivable in terms of a corresponding TQFT structure (see \refs{\AldayRS,\RazamatJXA}\ for the lens index for $k=1$). The $\S^4$ partition function is expected to pose a more serious challenge. In this work we have concentrated on genus larger than one though studying genus one should be also feasible and it would be interesting to make contact to the results of \refs{\OhmoriPUA,\OhmoriPIA,\DelZottoRCA}.

\

\

\noindent {{\bf Acknowledgmenets}}  We would like to thank Thomas Dumitrescu, Davide Gaiotto, Sergei Gukov, Jonathan Heckman, Patrick Jefferson, Zohar Komargodski, Tom Rudelius, and Amos Yarom for useful discussions. 
The research of CV is supported in part by NSF grant PHY-1067976. GZ is supported in part by the Israel Science Foundation under grant no. 352/13, by the German-Israeli Foundation for Scientific Research and Development under grant no. 1156-124.7/2011, and by World Premier International Research Center Initiative (WPI), MEXT, Japan.  SSR is  a Jacques Lewiner Career Advancement Chair fellow. The research of SSR was also supported by Israel Science Foundation under grant no. 1696/15 and by I-CORE  Program of the Planning and Budgeting Committee.

\

\

\appendix{A}{Free trinion from $T_B$}

Starting from the $T_B$ trinion we can obtain the free trinion by partially closing the maximal puncture, which appears twice, to a minimal one. The puncture is closed by giving a vacuum expectation value to one of the $M$ operators associated to the puncture. Let us choose the component of $M$ with charges $t \frac\gamma\beta z_1^{-1} z_2^{-1}$. This means that a combination of the puncture $su(N)^2$ symmetries and $u(1)_t$ is broken. In terms of fugacities we 
have $t\frac\gamma\beta=z_1z_2$ which can be solved by taking $z_1=t^{\frac12}\gamma \epsilon^{-1}$ and 
$z_2=t^{\frac12}\frac1\beta\epsilon$. We use the Lagrangian we discussed for $T_B$ to understand the IR fixed point of this flow. It is easiest to perform the analysis at the level of the index since it captures all the relevant details of the physics. Here we simply need to figure out what is the residue of the index when $z_1=t^{\frac12}\gamma\epsilon^{-1}$ and $z_2=t^{\frac12}\frac1\beta\epsilon$~\GaiottoXA. The residue gives the index of the theory in the IR. The index is given by,

\eqn\indTbb{\eqalign{
&{{\cal I}^{\bf u}}_{{\bf z}{\bf v}} =\Gamma_e(pq\frac{\beta^2}{\gamma^2})\Gamma_e(t(\beta\gamma u_2)^{\pm1} u_1^{\pm1})(p;p)(q;q)\cr
&\qquad\oint\frac{dz}{4\pi i z}\frac{\Gamma_e(\frac{pq\beta}{t \gamma}(\beta\gamma u_2)^{\pm1} z^{\pm1})}{\Gamma_e(z^{\pm2})}\Gamma_e(\frac\gamma\beta z^{\pm1}u_1^{\pm1}){\cal I}_{{\bf z},{\bf v}, \sqrt{z u_2},\sqrt{u_2/z}}\,.
 }}
Here ${\cal I}_{{\bf z},{\bf v}, \sqrt{z u_2},\sqrt{z/u_2}}$ is the index of the orbifold theory which we will soon write down. We are using here the choice of R symmetry giving the bi-fundamental chirals $\Phi$ charge two in this and next appendix. The function $\Gamma_e(z)$ is the elliptic Gamma function and we refer the reader to \GaiottoUSA\ for all the definitions used here.
 We need to compute the residue in ${\bf z}$ which appears only in the orbifold theory. This index gives us the index of an $su(2)$ SYM with four flavors and a bunch of singlet fields. Let us see how this comes about. First we write the index of this theory in detail,

\eqn\ofrbi{
\eqalign{
&{\cal I}_{{\bf z},{\bf v}, a,b} =(p;p)^2(q;q)^2\oint\frac{dw_1}{4\pi i w_1}\oint\frac{dw_2}{4\pi i w_2}
\frac{\Gamma_e(\frac{pq}t (\beta\gamma)^{\pm1}w_1^{\pm1}w_2^{\pm1})}{\Gamma_e(w_1^{\pm2})\Gamma_e(w_2^{\pm2})}\cr
&\qquad \Gamma_e(t^{\frac12}\beta b^{-1}w_1^{\pm1}z_1^{\pm1})\Gamma_e(t^{\frac12}\gamma^{-1} b w_1^{\pm1}z_2^{\pm1})\Gamma_e(t^{\frac12}\gamma b w_2^{\pm1}z_1^{\pm1})\Gamma_e(t^{\frac12}\beta^{-1} b^{-1}w_2^{\pm1}z_2^{\pm1})\cr
&\qquad \Gamma_e(t^{\frac12}\gamma a w_1^{\pm1}v_1^{\pm1})\Gamma_e(t^{\frac12}\beta^{-1} a^{-1}	 w_1^{\pm1}v_2^{\pm1})\Gamma_e(t^{\frac12}\beta a^{-1}w_2^{\pm1}v_1^{\pm1})\Gamma_e(t^{\frac12}\gamma^{-1} a w_2^{\pm1}v_2^{\pm1})\,.
}
} We see that,

\eqn\cps{
\eqalign{
&\left. \Gamma_e(t^{\frac12}\beta b^{-1}w_1^{\pm1}z_1^{\pm1})\Gamma_e(t^{\frac12}\gamma^{-1} b w_1^{\pm1}z_2^{\pm1})\Gamma_e(t^{\frac12}\gamma b w_2^{\pm1}z_1^{\pm1})\Gamma_e(t^{\frac12}\beta^{-1} b^{-1}w_2^{\pm1}z_2^{\pm1})\right|_{z_1\to t^{\frac12}\gamma \epsilon^{-1},z_2\to t^{\frac12}\frac1\beta\epsilon}\to\cr
&\Gamma_e( b \epsilon w_2^{\pm1})\Gamma_e( t b^{-1}\beta^{-2}\epsilon w_2^{\pm1})\Gamma_e( b^{-1} \epsilon^{-1} w_2^{\pm1})\Gamma_e(t\gamma^2 b \epsilon^{-1} w_2^{\pm1})\cr&\Gamma_e(\frac\beta\gamma b^{-1}\epsilon w_1^{\pm1})\Gamma_e(t\gamma\beta b^{-1}\epsilon^{-1} w_1^{\pm1})\Gamma_e(\frac\beta\gamma b\epsilon^{-1} w_1^{\pm1})
\Gamma_e(t\frac1{\gamma\beta} b \epsilon w_1^{\pm1})\,.
}} By usual pinching of contour integrals logic we have a pole when $w_2=(b \epsilon)^{\pm1}$. The residue is

\eqn\ofrbpi{
\eqalign{
&{\cal I}_{{\bf z},{\bf v}, a,b} \to \Gamma_e(t\beta^{-2}\epsilon^2)\Gamma_e(t\beta^{-2}b^{-2})
\Gamma_e(t\gamma^2b^2)\Gamma_e(t\gamma^2\epsilon^{-2})
 \Gamma_e(t^{\frac12}\gamma^{-1} a (b\epsilon)^{\pm1}v_2^{\pm1})\Gamma_e(t^{\frac12}\beta a^{-1} (b \epsilon)^{\pm1}v_1^{\pm1})\cr&(p;p)(q;q)
\oint\frac{dw_1}{4\pi i w_1}
\frac{\Gamma_e(\frac{pq}t (\beta\gamma b\epsilon)^{\pm1}w_1^{\pm1})}{\Gamma_e(w_1^{\pm2})}\Gamma_e(\frac\beta\gamma (b\epsilon^{-1})^{\pm1} w_1^{\pm1})\Gamma_e(t^{\frac12}\beta^{-1} a^{-1}w_1^{\pm1}v_2^{\pm1})\Gamma_e(t^{\frac12}\gamma a w_1^{\pm1}v_1^{\pm1})\,.
}
} According to the prescription to close maximal puncture one also has to introduce five  singlet fields and couple them through superpotential. In the index this amounts to multiplying the above by  $\Gamma_e(pq\beta^{-2}\gamma^2)\frac{\Gamma_e(t\beta^{-2}\epsilon^{-2})\Gamma_e(t\gamma^2\epsilon^2)}{\Gamma_e(t\beta^{-2}\epsilon^2)\Gamma_e(t\gamma^2\epsilon^{-2})}$.
Although it is not obvious from the expression after this multiplication it is symmetric under the exchange of the $u(1)$ fugacities $a$, $b$, and $\epsilon$. This follows from Seiberg duality. Let us denote the residue above by ${\cal I}_{{\bf z},a,b,\epsilon}$ and use the symmetry to write the residue of \indTbb\ as,

\eqn\indTbbR{\eqalign{
&{{\cal I}^{\bf u}}_{{\bf z}{\bf v}} \to\Gamma_e(pq\frac{\beta^2}{\gamma^2})\Gamma_e(pq\frac{\gamma^2}{\beta^2})\Gamma_e(t(\beta\gamma u_2)^{\pm1} u_1^{\pm1})(p;p)^2(q;q)^2\Gamma_e(t^{\frac12}\gamma^{-1} \epsilon u_2^{\pm1}v_2^{\pm1})\Gamma_e(t^{\frac12}\beta \epsilon^{-1} u_2^{\pm1}v_1^{\pm1})\cr
&\qquad\oint\frac{dz}{4\pi i z}\frac1{\Gamma_e(z^{\pm2})}\Gamma_e(\frac\gamma\beta z^{\pm1}u_1^{\pm1})\cr
 &\oint\frac{dw_1}{4\pi i w_1}
\frac{\Gamma_e(\frac{pq}t (\beta\gamma u_2)^{\pm1}w_1^{\pm1})}{\Gamma_e(w_1^{\pm2})}\Gamma_e(\frac\beta\gamma z^{\pm1} w_1^{\pm1})\Gamma_e(t^{\frac12}\beta^{-1} \epsilon^{-1}w_1^{\pm1}v_2^{\pm1})\Gamma_e(t^{\frac12}\gamma \epsilon w_1^{\pm1}v_1^{\pm1})\,.
 }} We see now that the $z$ $su(2)$ theory has only two flavors and thus is described by the quantum deformed moduli space which in the index implies a delta function identifying $u_1$ with $w_1$. We thus get the index for the fixed point to be,
 
 \eqn\findi{
 \eqalign{
 &{{\cal I}^{\bf u}}_{{\bf z}{\bf v}} \to\Gamma_e(t^{\frac12}\gamma^{-1} \epsilon u_2^{\pm1}v_2^{\pm1})\Gamma_e(t^{\frac12}\beta \epsilon^{-1} u_2^{\pm1}v_1^{\pm1})
\Gamma_e(t^{\frac12}\beta^{-1} \epsilon^{-1}u_1^{\pm1}v_2^{\pm1})\Gamma_e(t^{\frac12}\gamma \epsilon u_1^{\pm1}v_1^{\pm1})\,.
 }} This is the index of a free trinion. We thus deduce that the theory $T_B$ under an RG flow triggered by a vacuum expectation value for one of the mesonic operators associated to the maximal puncture flows to free trinion.

\

\appendix{B}{Trinion $T_A$ from trinion $T_B$}

We can also obtain trinion $T_A$ starting from a four punctured sphere built from  $T_B$ trinions.
As we saw when one considers RG flow triggered by vacuum expectation values for puncture appearing twice on  $T_B$ the fixed point is given by a free trinion. Let us glue together two $T_B$ along the puncture appearing once and then close one of the remaining punctures to minimal one,
 
 \eqn\indTba{\eqalign{
&{{\cal I}_{{\bf w}{\bf c};{\bf u}}}\cdot{{\cal I}^{\bf u}}_{{\bf z}{\bf v}} \to (q;q)^2(p;p)^2\oint \frac{du_1}{4\pi i u_1}\oint \frac{du_2}{4\pi i u_2} \frac{\Gamma_e(\frac{pq}t (\beta\gamma/u_2)^{\pm1}u_1^{\pm1})}{\Gamma_e(u_1^{\pm2})\Gamma_e(u_2^{\pm2})}\cr&
\Gamma_e(t^{\frac12}\gamma^{-1} \epsilon u_1^{\pm1}v_2^{\pm1})\Gamma_e(t^{\frac12}\beta \epsilon^{-1} u_1^{\pm1}v_1^{\pm1})
\Gamma_e(t^{\frac12}\beta^{-1} \epsilon^{-1}u_2^{\pm1}v_2^{\pm1})\Gamma_e(t^{\frac12}\gamma \epsilon u_2^{\pm1}v_1^{\pm1})\cr
&\Gamma_e(pq\frac{\beta^2}{\gamma^2})(p;p)(q;q)\oint\frac{dz}{4\pi i z}\frac{\Gamma_e(\frac{pq\beta}{t \gamma}(\beta\gamma u_2)^{\pm1} z^{\pm1})}{\Gamma_e(z^{\pm2})}\Gamma_e(\frac\gamma\beta z^{\pm1}u_1^{\pm1}){\cal I}_{{\bf c},{\bf w}, \sqrt{z u_2},\sqrt{u_2/z}}\,.
 }} We close the $u(1)$ puncture completely by giving a vacuum expectation value to baryonic operator implying $\epsilon=t^{\frac12}\frac1\beta$. We need to add a single field coupled to the rest through superpotential. The index becomes,
 
  \eqn\indTbaa{\eqalign{
&{{\cal I}_{{\bf w}{\bf c};{\bf u}}}\cdot{{\cal I}^{\bf u}}_{{\bf z}{\bf v}} \to \Gamma_e(t \frac\gamma\beta  v_2^{\pm1}v_1^{\pm1})\Gamma_e(pq\beta^{-4})\Gamma_e(pq\frac{\beta^2}{\gamma^2})\cr&
(q;q)(p;p)\oint \frac{du_1}{4\pi i u_1} \frac{\Gamma_e(\frac{pq}t \frac{v_2}{\beta\gamma}u_1^{\pm1})}{\Gamma_e(u_1^{\pm2})}
\Gamma_e(t\beta^{-1}\gamma^{-1} v_2^{-1} u_1^{\pm1})\Gamma_e(\beta^2  u_1^{\pm1}v_1^{\pm1})
\cr
&(p;p)(q;q)\oint\frac{dz}{4\pi i z}\frac{\Gamma_e(\frac{pq\beta}{t \gamma}(\beta\gamma v_2)^{\pm1} z^{\pm1})}{\Gamma_e(z^{\pm2})}\Gamma_e(\frac\gamma\beta z^{\pm1}u_1^{\pm1}){\cal I}_{{\bf c},{\bf w}, \sqrt{z v_2},\sqrt{v_2/z}}\,.
 }} The $u_1$ $su(2)$ gauge part has three flavors and thus is described in the IR by quadratic gauge invariant composits,
 
   \eqn\indTbaga{\eqalign{
&{{\cal I}_{{\bf w}{\bf c};{\bf u}}}\cdot{{\cal I}^{\bf u}}_{{\bf z}{\bf v}} \to \Gamma_e(t (\frac\gamma\beta  v_2)^{\pm1}v_1^{\pm1})
\Gamma_e(p\,q \frac{1}{\beta^2\gamma^2})
\cr
&(p;p)(q;q)\oint\frac{dz}{4\pi i z}\frac{\Gamma_e(\frac{pq}{t \gamma\beta}(\beta\gamma^{-1} v_2^{-1})^{\pm1} z^{\pm1})}{\Gamma_e(z^{\pm2})}\Gamma_e(\gamma\beta z^{\pm1}v_1^{\pm1}){\cal I}_{{\bf c},{\bf w}, \sqrt{z v_2},\sqrt{v_2/z}}\,.
 }}
This is exactly the index of $T_A$. The trinion $T_A$ can be obtained thus as a fixed point of RG flow  starting from theories built from $T_B$ trinions, and the same of course holds for any theory built from $T_A$ trinions.

The equation \indTbaa\ implies also that $T_A$ trinion is obtainable from $T_B$ trinion by gauging an $su(2)$ subgroup of the latter. There is also an inverse relation giving  $T_B$ from $T_A$.
The operation of gauging an $su(2)$ subgroup is a color-changing operation for a single puncture. Note that if for some reason the $\beta$ $u(1)$ is broken the kernel of \indTbaa\ is trivial and the two theories are the same. 

\

\appendix{C}{Calculating the $6d$ anomaly polynomial}

In this appendix we briefly review and collect the various formulas that we use in this article to calculate and manipulate anomaly polynomials. For a more comprehensive discussion about the calculation of the anomaly polynomial for $6d$ SCFT's see \OhmoriAMP\ and references within. The anomaly polynomial receives contributions from the various fields in the theory. For $6d$ SCFT's the relevant possible fields are the vector, hyper and tensor multiplets. Their contributions can be evaluated using the known contribution of a Weyl fermion \HarveyIT:

\eqn\che{
ch(E) \hat{A} (T)\,,}
where in $6d$ we must project to the $8$ form part.

$\hat{A} (T)$ is the Dirac A-roof genus conveniently given by:

\eqn\attfgd{
\hat{A} (T) = 1 - \frac{p_1 (T)}{24} + \frac{7 p^2_1 (T) - 4 p_2 (T)}{5760} + \dots \,,}
where $p_1 (T), p_2 (T)$ are the first and second Pontryagin classes of the tangent bundle respectively.

$ch(E)$ is the Chern character of the total bundle for any additional local or global symmetries. It is convenient to expand it in terms of the Chern classes of the bundle:

\eqn\cheee{\eqalign{
ch(E) & =  rank(E) + C_1 (E) + \frac{C^2_1 (E) - 2 C_2 (E)}{2} + \frac{C^3_1 (E) - 3 C_1 (E) C_2 (E) + 3 C_3 (E)}{6} \cr & +  \frac{C^4_1 (E) + 4 C_1 (E) C_3 (E) - 4 C^2_1 (E) C_2 (E) + 2 C^2_2 (E) - 4 C_4 (E)}{24} + \dots\,,}}
where $C_i (E)$ stands for the $i$'th Chern class. 

A useful property of the Chern character is its decomposition under the direct sum and product of vector bundles:

\eqn\Spl{
ch(U\oplus V) = ch(U)+ch(v), ch(U\otimes V) = ch(U)ch(v) \,.}
There are similar formulas also for the Chern and Pontryagin classes. These are collectively known as the splitting principle. For the Chern classes, defining the total Chern class as: $C(E) = \sum C_i (E)$, it obeys:

\eqn\cupl{
C(U\oplus V) = C(U)C(v)\,.}
For the Pontryagin classes, given a decomposition of the bundle as a sum of complex line bundles with first Chern classes $e_i$, then:

\eqn\ponm{
p_1 (T) = \sum e^2_i , p_2 (T) = \sum_{i<j} e^2_i e^2_j\,.}
For the theories we consider the additional symmetries we encounter are the $su(2)_R$ R-symmetry, gauge symmetries $G$ and flavor symmetries $F$. Using \Spl\ we see that $ch(E) = ch(R)ch(G)ch(F)$. We next evaluate the contribution to the anomaly polynomial for each multiplet in turn.

\

\noindent $\bullet$ {\it Hyper}

 We consider a single hypermultiplet in a representation $r_G$ of the gauge symmetry and $r_F$ of the flavor symmetry. It contains a single Weyl fermion which is an $su(2)_R$ singlet. Thus its addition to the anomaly polynomial receives contributions from the tangent, gauge and flavor symmetry bundles. Using the formulas presented in this section, it is given by:

\eqn\hyps{\eqalign{ & d_{r_G} d_{r_F} \frac{7 p^2_1 (T) - 4 p_2 (T)}{5760} + d_{r_F} \frac{C^2_2 (G)_{r_G} - 2 C_4 (G)_{r_G}}{12} + d_{r_F}\frac{p_1 (T) C_2 (G)_{r_G}}{24} \cr - & d_{r_G}\frac{p_1 (T)}{48} (C^2_1 (F)_{r_F} - 2 C_2 (F)_{r_F}) - \frac{1}{2} C_2 (G)_{r_G}(C^2_1 (F)_{r_F} - 2 C_2 (F)_{r_F}) + \frac{1}{2} C_1 (F)_{r_F} C_3 (G)_{r_G} \cr + & d_{r_G} \frac{C^4_1 (F)_{r_F} + 4 C_1 (F)_{r_F} C_3 (F)_{r_F} - 4 C^2_1 (F)_{r_F} C_2 (F)_{r_F} + 2 C^2_2 (F)_{r_F} - 4 C_4 (F)_{r_F}}{24}\,,}}
where we use $C_i (G)_{r_G}$ for the $i$'th Chern class of the $G$-bundle with representation $r_G$ whose dimension we denote by $d_{r_G}$. We have also set $C_1 (G)=0$ as all the gauge groups we consider are simple. 

\

\noindent $\bullet$ {\it Vector}

We consider a single vector multiplet. It contains a single Weyl fermion which is an $su(2)_R$ doublet and is in the adjoint representation of the gauge group. Thus its addition to the anomaly polynomials receives contributions from the tangent, gauge and R-symmetry bundles. Note that the chirality of the spinor is opposite to that of the fermion in the hyper and tensor multiplets, and so contribute to the anomaly polynomial with a minus sign. Using the formulas presented in this section, it is given by:

\eqn\vecdf{\eqalign{& -d_{Ad} \frac{7 p^2_1 (T) - 4 p_2 (T)}{5760} - \frac{C^2_2 (G)_{Ad} - 2 C_4 (G)_{Ad}}{12} - d_{Ad} \frac{C^2_2(R)}{24} \cr - & \frac{p_1 (T) C_2 (G)_{Ad}}{24} - d_{Ad}\frac{p_1 (T) C_2(R)}{48} - \frac{C_2(R) C_2 (G)_{Ad}}{2}\,,}}
where we use $C_2(R)$ for the second Chern class of the $su(2)_R$ bundle in the doublet representation.  

\

\noindent $\bullet$ {\it Tensor} 

We consider a single tensor multiplet. It contains a single Weyl fermion which is an $su(2)_R$ doublet. In addition it contains a self dual tensor which is also chiral and thus contribute to the gravitational part of the anomaly, where the exact contribution was evaluated in \AlvarezGaumeIG. Using this and the formulas presented in this section, one finds:

\eqn\poand{
\frac{23 p^2_1 (T) - 116 p_2 (T)}{5760} + \frac{C^2_2(R)}{24} + \frac{p_1 (T) C_2(R)}{48}\,.}
Besides the contributions of the field content one must also add the Green-Schwartz term. This term takes into account the effect of modifying the Bianchi identity for the tensor multiplet. It is a complete square and is chosen so as to make all gauge anomalies vanish.  

\subsec{The anomaly polynomial for the $\Z_2$ orbifold of the $A_1$ $(2,0)$ theory}

Using the above formulas it is now straightforward to calculate the anomaly polynomial of the $\Z_2$ orbifold of the $A_1$ $(2,0)$ theory using its gauge theory description. The matter content includes: a single tensor multiplet, a vector multiplet in the adjoint of $su(2)_G$ and $8$ half-hyper multiplets in the doublet representation of $su(2)_G$. These are rotated with an $so(7)$ global symmetry where they transform as the $\bf{8}$ dimensional spinor representation of $so(7)$.

Using the previous formulas we find after a little algebra:

\eqn\anpolzt{\eqalign{
I^{field}_{8} & =  -\frac{C^2_2(su(2))}{4} - \frac{C_2(R) p_1 (T)}{24} - \frac{C^2_2(R)}{12} - C_2(R) C_2(su(2)) \cr + & \frac{C_2(su(2)) C_2(so(7))_{\bf{8}}}{4} + \frac{C_2(so(7))_{\bf{8}} p_1 (T)}{24} + \frac{C^2_2(so(7))_{\bf{8}} - 2 C_4(so(7))_{\bf{8}}}{12} \cr + & \frac{29 p^2_1 (T) - 68 p_2 (T)}{2880}\,,}}
where we have used $C_2(su(2)_{r}) = T_r C_2(su(2))$, $T_r$ being the second Casimir of the representation $r$, to convert to a second Chern class that is independent of the representation. 

To this we need to add the Green-Schwartz term given by:

\eqn\sgt{
I^{GS}_{8} = (\frac{C_2(su(2))}{2} + C_2(R) - \frac{C_2(so(7))_{\bf{8}}}{4})^2\,.}
Summing both terms we finally get:

\eqn\anpollsmn{\eqalign{
I_8 & =  I^{field}_{8} + I^{GS}_{8} = \frac{11 C^2_2(R)}{12} - \frac{C_2(R) p_1 (T)}{24} + \frac{C_2(so(7))_{\bf{8}} p_1 (T)}{24} - \frac{C_2(R) C_2(so(7))_{\bf{8}}}{2} \cr + & \frac{7 C^2_2(so(7))_{\bf{8}}}{48} - \frac{C_4(so(7))_{\bf{8}}}{6} + \frac{29 p^2_1 (T) - 68 p_2 (T)}{2880}\,.}}

\

\appendix{D}{Computation of anomalies and indices from field theory}

In this appendix we give the details of the computations of the anomalies and indices from the field theory side using the ``strongly coupled'' Lagrangians. Here we use the assignment of R-charges giving R charge $1$ to the $\Phi$ fields. 
The assignments giving other convenient charges, $2$ or $\frac23$, can be obtained from this by admixing a proper multiple of $u(1)_t$ charge.

\

\noindent{\it  $\bullet$ Anomalies}

\

Let us define,

\eqn\defsa{\eqalign{
&a_\chi(R)=\frac3{32}(3(R-1)^3-(R-1))\,,\qquad c_\chi(R)=\frac1{32}(9(R-1)^3-5(R-1))\,,\cr
&a_v(G)=\frac{3}{16}dimG\,,\qquad c_v(G)=\frac1{8}dimG\,,}}
For anomalies of chiral fields of R charge $R$ and vector fields for group $G$.  The superconformal R charge will be denoted,

\eqn\srsc{
R_c(R,q_\beta,q_\gamma,q_t)=R+\ell_1 q_\beta+\ell_2 q_\gamma +\ell_3 q_t\,,
} and is a function of three variables $\ell_i$ to be determined by a maximization for each theory.
We also define,

\eqn\vvbegamma{\eqalign{
&a_v^\pm=a_v(su(2)\times su(2))+4 a_\chi(R_c(1,1,\mp1,-1))+4a_\chi(R_c(1,-1,\pm1,-1))\,,\cr
&c_v^\pm=c_v(su(2)\times su(2))+4 c_\chi(R_c(1,1,\mp1,-1))+4c_\chi(R_c(1,-1,\pm1,-1))\,,}}
as the anomalies for contributions of fields, bifundamental chirals and vectors, introduced when we glue two punctures of the same color. The subscript denotes the color of puncture.

We warm up by computing the anomalies of the orbifold theory. The chiral matter, $Q^\pm_i$ and ${Q'}^\pm_i$, of table \theor\ determine the anomalies to be

\eqn\orbas{\eqalign{&
a_{Orb}=\cr &\;a_v^{-}+8(a_\chi(R_c(\frac12,1,0,\frac12))+a_\chi(R_c(\frac12,-1,0,\frac12))+a_\chi(R_c(\frac12,0,-1,\frac12))+a_\chi(R_c(\frac12,0,1,\frac12)))=\cr &
\;\qquad -\frac38\biggl(3\ell_3^3+9\ell_3^2+36\ell_1\ell_2\ell_3-7\ell_3+18(\ell_1^2+\ell_2^2)-4\biggr)\,.
}} Maximizing $a_{Orb}$ as a function of $\ell_i$ gives us $\ell_i=(0,0,\frac13)$ and  $a_{Orb}=\frac{47}{24}$. For $c$ we define

\eqn\orbassd{\eqalign{&
c_{Orb}=\cr &\;c_v^{-}+8(c_\chi(R_c(\frac12,1,0,\frac12))+c_\chi(R_c(\frac12,-1,0,\frac12))+c_\chi(R_c(\frac12,0,-1,\frac12))+c_\chi(R_c(\frac12,0,1,\frac12)))=\cr &
\;\qquad -\frac18\biggl(9\ell_3^3+27\ell_3^2+108\ell_1\ell_2\ell_3-17\ell_3+54(\ell_1^2+\ell_2^2)-17\biggr)\,.
}} We insert the value of $\ell_i$ maximizing $a$ and obtain $c_{Orb}=\frac{29}{12}$.

From now on we will only consider the $a$ anomaly as $c$ can be obtained in a similar manner. We start from the first non trivial theory, the $T_A$ trinion. It is constructed from the orbifold theory by gauging an $su(2)$ group with chiral fields listed in table (5.6) and the fields $\phi'$ which are flipping the $\Phi'$ fields and thus have opposite charges and their R charges sum up to two. The central charge is then given by,

\eqn\ants{\eqalign{&
a_{T_A}=\cr &\;a_{Orb}+a_v(su(2))+2\biggl(2a_\chi(R_c(0,1,1,0))+a_\chi(R_c(1,1,-1,1))+a_\chi(R_c(1,1,1,1))+\cr&\;\;\;a_\chi(R_c(1,2,0,-1))+a_\chi(R_c(1,0,-2,-1))\biggr)+a_\chi(R_c(2,-2,-2,0))\,.
}} Extremizing this with respect to $\ell_i$ we obtain the result reported in the bulk of the paper \anomssi.
For the $T_B$ trinion we obtain again from the Lagrangian in section five,
\eqn\antf{\eqalign{&
a_{T_B}=\cr &\;a_{Orb}+a_v(su(2))+2\biggl(2a_\chi(R_c(0,-1,1,0))+a_\chi(R_c(1,-1,-1,1))+a_\chi(R_c(1,-1,1,1))+\cr&\;\;\;a_\chi(R_c(1,-2,0,-1))+a_\chi(R_c(1,-2,0,-1))\biggr)+a_\chi(R_c(2,2,-2,0))\,.
}} This upon maximization reproduces \anomssi.

We can compute the anomalies for theories with enhanced symmetry. The four punctured sphere with $G^{max}=so(5)u(1)$  can be constructed as in section six from $T_A$ and free trinions with subsequent closure of minimal punctures. It has the following conformal anomaly,

\eqn\sofad{\eqalign{&
a_{T_{so(5)}}=\cr &\;(2a_{T_A}+a_v^+)+2a_v(su(2))+4\biggl(a_\chi(R_c(1,-1,1,-1))+a_\chi(R_c(1,1,1,-1))+a_\chi(R_c(1,1,1,1))+\cr&\;\;\;+a_\chi(R_c(1,1,-1,1))+a_\chi(R_c(0,-2,0,0))+a_\chi(R_c(0,0,-2,0))\biggr)+\cr&\;\;\;\;\;a_\chi(R_c(2,0,4,0))+a_\chi(R_c(2,4,0,0))=\cr &
\;\;\; -\frac38\biggl(
12\ell_3^3+18\ell_3^2-16\ell_3+36(1+\ell_3)(\ell_1^2+\ell_2^2)+72\ell_3\ell_2\ell_1-11\biggr)\,.
}} Upon maximization it produces the anomalies \anomcv. Moreover the conformal R symmetry here is obtained by $\{\ell_i\}=(0,0,\frac13)$. Gluing such theories together and computing anomalies one discovers that these are consistent with the gauge coupling being exactly marginal. 

For the $so(7)$ theory of Fig. 10 we glue $T_A$ theory with plus sign to $T_A$ theory with minus sign. The contribution to the anomaly of the minus theory is the same as the plus one with $\ell_i$ flipping signs. The anomaly is then

\eqn\sosaf{\eqalign{&
a_{T_{so(7)}}=a_{T_A}(\ell_1,\ell_2,\ell_3)+a_{T_A}(-\ell_1,-\ell_2,-\ell_3)+2a_v(su(2))+a_v^+(\ell_1,\ell_2,\ell_3)+
a^+_v(-\ell_1,-\ell_2,-\ell_3)=\cr&
\;\;\;-\frac{3}{8}  \left(18(2\ell_1^2+ 2\ell_2^2+\ell_3^2)-17\right)\,.
}} Maximizing this expression we get that all $\ell_i=0$ consistently with $so(7)$ symmetry. 

Let us now consider the anomaly for the $su(3)u(1)$ example we discussed in section six. This theory is given by combining together four copies of $T_B^-$, four copies of $T_A^+$, and $T_{so(5)}$. We have different types of punctures glues together and have to be careful about that.  Combining the ingredients we obtain the following anomaly,

\eqn\ssusaa{\eqalign{&
a_{T_{su(3)}}=\cr &\;4(a_{T_A}(\ell_1,\ell_2,\ell_3)+a_{T_B}(-\ell_1,-\ell_2,-\ell_3)+2a_v(su(2)))+a_{so(5)}(\ell_1,\ell_2,\ell_3)+\cr &\;\;\; 6a^+_v(\ell_1,\ell_2,\ell_3)+2a^-_v(-\ell_1,-\ell_2,-\ell_3)+2a^+_v(-\ell_1,-\ell_2,-\ell_3)=\cr &
\;\;\; -\frac38\biggl(120\ell_1^3+24 \ell_3^3+90\ell_3^2-20(\ell_3+2\ell_1)+36(\ell_1^2+\ell_2^2)(5+2\ell_3)+9\ell_1(\ell_3^2+\ell_2^2)-85\biggr)\,.
}}  Maximizing this we obtain the $\ell_i$ given in \misu\ and anomalies of \sustt.

We can compute anomalies for flavor symmetries and mixed R-symmetry flavor symmetry anomalies. Let us give several examples. Since the theories are built iteratively from orbifold theory we start with it. The computation is completely standard and we give the result,

\eqn\oranflavs{\eqalign{&
(TrRu(1)_t^2)_{orb}=-4(1+\ell_3)\,,\qquad\; (TrRu(1)_{t\beta}^2)_{orb}=-4(3+4\ell_2+\ell_3)\,,\cr&
(TrR^2u(1)_t)_{orb}=-4(\ell_3^2+2\ell_3-1+4\ell_1\ell_2)\,,\cr&(TrR^2u(1)_{t\beta})_{orb}=-4(\ell_3^2+2(1+2\ell_2)(\ell_3+2\ell_1)-1)\cr& (Tru(1)_t^3)_{orb}=-4\,,\qquad\; (Tru(1)_{t\beta}^3)_{orb}=-4\,,\cr
&(TrRu(1)_\gamma^2)_{orb}=-8\,.
}} The symmetry $u(1)_{t\beta}$ is the diagonal combination of $u(1)_t$ and $u(1)_\beta$. We have written down the dependence on mixing parameters since we will have to plug in the different values for different theories obtained through the maximization procedure. One should use the expressions we derived for the $a$ anomaly above for various theories and change the functions $a$ for other anomalies for the different ingredients. One point to be cautions about is that when a minus type theory is taken and anomaly of odd number of currents for flavor $u(1)$s is considered then in addition to switching the signature of $\ell_i$ the anomaly has to be taken with a minus sign. Let us quote the results for subset of the different models and we mentioned till now and a sub-set of anomalies.
 For the $T_B$ trinion we obtain,

\eqn\oranflavss{\eqalign{&
(TrRu(1)_t^2)_{T_B}=4(\ell_1-\ell_2-\ell_3-1)\,,\qquad\; (TrRu(1)_{t\beta}^2)_{T_B}=4 (4 (\ell_1-\ell_2)-3)\,,\cr&
(TrR^2u(1)_t)_{T_B}=-4(\ell_1^2+\ell_3^2+\ell_2^2+2\ell_2(\ell_1+\ell_3)-2\ell_3(\ell_1-1)-1)\,,\cr&(TrR^2u(1)_{t\beta})_{T_B}=2 \left(8 \ell_1^2-8 \ell_1 (\ell_2+1)-4 (2 \ell_2+1) \ell_3+1\right)\cr& (Tru(1)_t^3)_{T_B}=-4\,,\qquad\; (Tru(1)_{t\beta}^3)_{T_B}=16\,,\cr
&(TrRu(1)_\gamma^2)_{T_B}=4(\ell_1-\ell_3+5 \ell_2-2)\,.
}} For $T_A$ trinion we have,

\eqn\oranflasvs{\eqalign{&
(TrRu(1)_t^2)_{T_A}=-4(\ell_1+\ell_3+\ell_2+1),,\qquad\; (TrRu(1)_{t\beta}^2)_{T_A}=-4(8 (\ell_1+\ell_2)+4\ell_3+3)\,,\cr&
(TrR^2u(1)_t)_{T_A}=-4(\ell_1^2+\ell_3^2+\ell_2^2+2\ell_3(\ell_2+\ell_1+1)+6\ell_2\ell_1-1)\,,\cr&(TrR^2u(1)_{t\beta})_{T_A}=-2(12\ell_1^2+8\ell_1(1+2\ell_2+\ell_3)+4\ell_3(\ell_3+4\ell_2+1)-3)\,,\cr& (Tru(1)_t^3)_{T_A}=-4\,,\qquad\; (Tru(1)_{t\beta}^3)_{T_A}=-48\,,\cr
&(TrRu(1)_\gamma^2)_{T_A}=-4(\ell_1+\ell_3+5\ell_2+2)\,.
}} The $so(5)$ theory for sphere with four maximal punctures gives,

\eqn\oranflasvss{\eqalign{&
(TrRu(1)_t^2)_{T_{so(5)}}=-8(1+2\ell_3)\,,\qquad\; (TrRu(1)_{t\beta}^2)_{T_{so(5)}}=-8(4 (\ell_1+\ell_2+\ell_3)+3)\,,\cr&
(TrR^2u(1)_t)_{T_{so(5)}}=-8(2((\ell_1+\ell_2)^2+\ell_3^2)+2\ell_3-1)\,,\cr&(TrR^2u(1)_{t\beta})_{T_{so(5)}}=-8(2((\ell_1+\ell_2)^2+\ell_3^2)+2\ell_3(2\ell_2+2\ell_1+1)+4\ell_1-1)\,,\cr& (Tru(1)_t^3)_{so(5)}=-16\,,\qquad\; (Tru(1)_{t\beta}^3)_{T_{so(5)}}=-64\,,\cr&(TrRu(1)_\gamma^2)_{so(5)}=-16(\ell_3+2\ell_2+1)\,.
}} For $G^{max}=so(5)u(1)$ theories the a maximization as we mentioned gives $\ell_i$ to be $(0,0,\frac13)$. We can construct anomalies for genus $g$ surface by gluing $g-1$  four punctured spheres together. Summing the contributions of the spheres and the vectors we derive,

\eqn\ansg{\eqalign{&
(TrRu(1)_t^2)_{T_{so(5),g}}=(g-1)(TrRu(1)_t^2)_{T_{so(5)}}+(2g-2)(-\frac83)=-\frac{56}3(g-1)\,,\cr&(TrR^2u(1)_t)_{T_{so(5),g}}=(g-1)(TrR^2u(1)_t)_{T_{so(5)}}+(2g-2)(-\frac89)=-\frac{8}9(g-1)\,,\cr&(Tru(1)_t^3)_{T_{so(5),g}}=(g-1)(Tru(1)_t^3)_{T_{so(5)}}+(2g-2)(-8)=-32(g-1)\,.}} This agrees with (7.22) up to relative normalization ($-\frac12$) of $u(1)_t$ and $u(1)_a$ charges which is a matter of convention.

Taking all the results and computing the anomalies for the $su(3)$ example of section 6 we obtain,

\eqn\oranflasvses{\eqalign{&
(TrR^2u(1)_{t\beta})_{T_{su(3)}}=-\frac83\,,\qquad\; (TrRu(1)_{t\beta}^2)_{T_{su(3)}}=-8 \sqrt{385}\,,\cr&
 (Tru(1)_{t\beta}^3)_{T_{su(3)}}=-384\,,\qquad\;(TrRu(1)_\gamma^2)_{su(3)}=-\frac43(45+\sqrt{385})\,.
}} This is in perfect agreement with \othanomb\ up to sign for $u(1)_{t\beta}$ which is again a matter of convention.

\

\noindent{\it $\bullet $ Indices}

\

The computation of the indices is structurally identical to the computation of the anomalies. We have already computed the indices of the trinions $T_A$ and $T_B$ in appendices A and B. The indices of the other theories in the paper are obtained by combining these using the usual rules of index computations.  We state here the basic rules of the computations. Gluing two theories, indices of which are given by ${\cal I}_a({\bf u})$ and ${\cal I}_b({\bf u})$, depends on the colors and the signs of the two punctures. If the signs are the same we use the $\Phi$-gluing and the index of the combined  theory is (here we assign R charge one to the $\Phi$ in contrast to two in appendices A and B),

\eqn\coreee{\eqalign{&
{\cal I}_{a\oplus b}=(q;q)^2(p;p)^2\oint\frac{du_1}{4\pi i u_1}\oint\frac{du_2}{4\pi i u_2}
\frac{\Gamma_e((p q)^{\frac12}\frac1{t^{S}} (\beta\gamma^{C})^{\pm1} u_1^{\pm1}u_2^{\pm1})}{
\Gamma_e(u_1^{\pm2})\Gamma_e(u_2^{\pm2})} {\cal I}_a(u_1,u_2){\cal I}_b(u_2,u_1)}} Here the number $C$ takes value in $\pm1$ and denotes the color of the maximal punctures. Number $S$ takes also value in $\pm1$ and denotes the sign of the punctures.
If the signs of the two punctures are opposite then the index of the  combined theory takes the form,

\eqn\og{
{\cal I}_{a\oplus b}=(q;q)^2(p;p)^2\oint\frac{du_1}{4\pi i u_1}\oint\frac{du_2}{4\pi i u_2}
\frac{1}{
\Gamma_e(u_1^{\pm2})\Gamma_e(u_2^{\pm2})} {\cal I}_a(u_1,u_2){\cal I}_b(u_2,u_1)\,.}
If we are given a theory of a particular sign index of which is ${\cal I}({\bf u}_l,\beta,\gamma,t)$ the index of the theory of opposite sign is ${\cal I}({\bf u}^\dagger_l,\frac1\beta,\frac1\gamma,t^{-1})$. Employing these simple rules all the indices reported in the file can be easily derived.

\

\appendix{E}{Conformal manifold of the orbifold theory}

There are several ways to study the conformal manifold of the orbifold theory. We can analyze it for example at the vicinity of free point. At that point the symmetry of the theory is $H=su(8)\times su(8)\times su(2)\times u(1)_t$. The fields ${\cal Q}_1=\{Q^+_1,\,Q_2^-,\,{Q'}^+_1,{Q'}^-_2\}$ and ${\cal Q}_2=\{Q^-_1,\,Q_2^+,\,{Q'}^-_1,{Q'}^+_2\}$ are in the ${\bf 8}$ of one of the two $su(8)$s. The $su(2)$ group rotates the two $\Phi$s. The $u(1)_t$ was defined in section four, and under it the $\Phi$s have charge $-1$ and all the other chirals have charge $\frac12$.  We have two additional abelian symmetries which are anomalous once the gauge interactions are switched on. The marginal operators are, $\lambda \; {\cal Q}_1\cdot \Phi \cdot {\cal Q}_2$. In addition we have two gauge couplings. The coupling $\lambda$ are singlets of $u(1)_t$ and transform as $({\bf 8},{\bf 8},{\bf 2})$ under the non abelian factors. We find the dimension of the conformal manifold by performing the hyper Kahler quotient \GreenDA\ $(\lambda,g_1,g_2)/H$ and computing its dimension. This amounts to counting independent holomorphic invariants built from $\lambda$s and the two gauge $g_1$ and $g_2$. This can be easily done. The invariants under the two $su(8)$s are the ``baryons'' built from the $\lambda$s. These baryons form the ${\bf 9}$ of $su(2)$. There are six independent invariant built from these ``baryons'' and this is the dimension of the conformal manifold.  We have $130$ couplings we start with and the dimension of the group including anomalous symmetries is $132$. We have eight symmetries, all abelian, preserved on a generic point of the manifold. This gives us again conformal manifold of dimension six as the broken symmetries must be consumed by marginal operators.

One of the exactly marginal directions preserves the $su(4)\times su(4)\times u(1)\times u(1)\times u(1)$ symmetry mentioned in the file. At generic point of this manifold we have marginal deformations in $({\bf 15},{\bf 1},0,2)$ and
$({\bf 1},{\bf 15},0,-2)$. This can be deduced for example by writing down the index of the orbifold theory which at order $p q$ gives,

\eqn\orin{
1+(\beta^2\gamma^2-1){\bf 15}_1+(\frac1{\gamma^2\beta^2}-1){\bf 15}_2-1-1-1\,.
} The negative terms are the conserved currents, the first terms is the exactly marginal deformation preserving the symmetries. the rest of the terms are the marginal deformations. We again can perform the quotient looking for holomorphic invariants. We can construct six singlets  of the $su(4)$s charged under the $u(1)$s plus and minus four, six and eight. From the singlets we can build five invariants. With the addition of the deformation we started with this again gives us six dimensional manifold.

\

Let us also count the dimension of the conformal manifold preserving $(su(2)^2)^3$ of the $so(7)$ trinion. The $E_7$ surprise theory has ${\bf 1463}$ marginal deformations forming a single representation of $E_7$. This decomposes into $so(12)\times su(2)_{\gamma/\beta}$,

$${\bf 1463} = ({\bf 66},{\bf 1})+({\bf 77},{\bf 3})+({\bf 352'},{\bf 2})+({\bf 462},{\bf 1})\,.$$
The different representations of $so(12)$ decompose to $su(2)\times su(2)\times so(8)$,

\eqn\decsosuso{\eqalign{&
{\bf 66} = ({\bf 3},{\bf 1},{\bf1})+({\bf 1},{\bf 3},{\bf 1})+({\bf 2},{\bf 2}, {\bf 8}_v)+({\bf 1},{\bf 1},{\bf 28})\,, \cr 
&
{\bf 352'} =
({\bf 2},{\bf 1},{\bf 8}_c)+({\bf 1},{\bf 2},{\bf 8}_s)+({\bf 2},{\bf 3},{\bf 8}_c)+({\bf 3},{\bf 2},{\bf 8}_s)+
({\bf 2},{\bf 1},{\bf 56}_c)+
({\bf 1},{\bf 2},{\bf 56}_s)\cr
& {\bf 462} = ({\bf 1},{\bf 1},  {\bf 28})+({\bf 3},{\bf 1},{\bf 35}_s)+({\bf 1},{\bf 3},{\bf 35}_c)+({\bf 2},{\bf 2},{\bf 56}_v)\cr
&
{\bf 77}=({\bf 1},{\bf 1},{\bf 1})+({\bf 3},{\bf 3},{\bf 1})+({\bf 2},{\bf 2},{\bf 8}_v)+({\bf 1},{\bf 1},{\bf 35}_v)
}} Representation ${\bf 35}_v$ decomposes under $su(2)^4$ as $$({\bf 2},{\bf 2},{\bf 2},{\bf 2})+({\bf 3},{\bf 3},{\bf 1},{\bf 1})+({\bf 1},{\bf 1},{\bf 3},{\bf 3})+({\bf 1},{\bf 1},{\bf 1},{\bf 1})$$
Thus the ${\bf 1463}$ has only two singlets, which we denote as $\lambda$ and $\lambda '$, of $su(2)^6$ decomposition (of $so(12)$) which are both adjoints of $su(2)_{\beta/\gamma}$.  From these singlets we can construct three independent holomorphic invariants of $su(2)^6\times su(2)_{\gamma/\beta}$ ($Tr\lambda^2$, ${Tr\lambda'}^2$, and $Tr\lambda\lambda'$). These give the three deformations preserving the $su(2)^6 $ symmetry. These are the exactly marginal deformations of our interest.

\

\appendix{F}{Anomalies for more general case}

In this appendix we consider the calculation of the $4d$ anomaly polynomial from the $6d$ one in the case of general $N$ and $k$. This differs from the $N=k=2$ calculation performed previously only by the different $(1,0)$ $6d$ SCFT, here being the $\Z_k$ orbifold of the $6d$ $A_{N-1}$ type $(2,0)$ theory. This SCFT has an $N-1$ dimensional Tensor branch along which the theory can be effectively described by an $NF+su(k)^{N-1}+NF$ quiver gauge theory. For general $N$ and $k$ the theory has an $su(k)\times su(k)\times u(1)$ global symmetry. This is enhanced to $su(2k)$ when $N=2$, $su(2)^3$ when $k=2$ and further to $so(7)$ when $N=k=2$.

 We will need the $6d$ anomaly polynomial for this theory to perform the calculation, where for simplicity we shall set the curvature of the flavor symmetries to zero. The anomaly polynomial for this case was derived in \OhmoriAMP\ and reads:

\eqn\anompolgen{\eqalign{&
I_8 = \frac{(N-1) (2 + k^2 (N^2 + N -1)) C^2_2(R)}{24} - \frac{(N-1)(k^2-2) p_1(T) C_2(R)}{48} \cr& + \frac{(7k^2 + 30N - 30)p^2_1(T) - 4 (k^2 + 30 N - 30) p_2 (T)}{5760}\,.
}}

Next we integrate this on the Riemann surface. For this we need to include the twist and decompose $C_2(R)$ and $p_1 (T), p_2 (T)$ exactly as done in section 7. The result is:

\eqn\anompolgen{
I_6 = \frac{(N-1) (2 + k^2 (N^2 + N -1)) (g-1) C^3_1(F)}{6} + \frac{(N-1)(k^2-2) (g-1) p_1(T) C_1(F)}{24}\,.
}

From which we find that $Tr(R^3) = (N-1) (2 + k^2 (N^2 + N -1)) (g-1), Tr(R) = -(N-1)(k^2-2) (g-1)$. Combining these we find:

\eqn\ccgenNk{
a = \frac{1}{32}(N-1)(12 + k^2 (9 N^2 + 9N - 6))(g-1), c = \frac{1}{32}(N-1)(8 + k^2 (9 N^2 + 9N - 4))(g-1) \,.
}

There is one final issue we need to address. For general $N$ and $k$ the global symmetry of the resulting $4d$ theory includes a $u(1)$ coming from the $6d$ flavor one. Thus one might worry it can mix with the R-symmetry and so invalidates \ccgenNk. However, by inspection one can see that the only terms in the $6d$ anomaly polynomial that can contribute to the $4d$ one are proportional to $C_2 (R)$. Furthermore the only term that can appear in the $6d$ anomaly polynomial containing both $C_2 (R)$ and the curvature for the $u(1)$ symmetry, $C_1(u(1)_t)$, is $C_2 (R) C^2_1(u(1)_t)$. Therefore the only non-trivial anomaly for $u(1)_t$ in $4d$ will be $Tr (R u(1)^2_t)$ implying that $a$ will be extremized at zero mixing. 

\

\appendix{G}{Maximal punctures as two minimal punctures and integrable models}

The supersymmetric indices of theories discussed in this paper can be neatly written in terms of eigenfunctions of certain
difference operators~\GaiottoUSA.  This makes the duality properties of the theories manifest through a TQFT structure of the index~\refs{\GaddeKB,\GaddeUV,\GaiottoXA}.
The different properties of the theories are encoded in the index in a duality invariant way and imply certain mathematical identities.
Let us for completeness present this structure here.

We can write the index of the free trinion as~\GaiottoUSA, 

\eqn\feif{
{\cal I}({\bf u},{\bf v},\delta;t,\beta,\gamma)=\sum_\lambda \psi_\lambda({\bf u};t,\beta,\gamma)\psi_\lambda({\bf v};t,\beta^{-1},\gamma) \phi_\lambda(\delta;t,\beta,\gamma)\,.}
Here we use the choice of R charge giving value two for the fields $\Phi$.
Each one of the three functions appearing in this sum is associated to a puncture, with two $\psi_\lambda$s associated to the two different colors of maximal puncture~\GaiottoUSA, and $\phi_\lambda$ to the minimal one.
The functions $\psi_\lambda$ are orthonormal eigenfunctions of a set of difference and integral operators with the index $\lambda$ parametrizing the set.
For example defining,

\eqn\difoppp{
{\cal T}(v_1,v_2;\b,\g,t)=\frac{\theta(\frac{t v_1^{-1} v_2^{-1}}{q}(\frac\g\b)^{\pm1};p)\theta(\frac{t\b v_1}{\g  v_2};p)\theta(\frac{t\b^3\g v_2}{v_1};p)}{\theta(v_1^{2};p)\theta({v_2}^{2};p)}\,,} the functions $\psi_\lambda$ are eigenfunctions of

\eqn\fop{
{\frak S}^{(\b,-)}_{(0,1)}\cdot f(v_1,v_2)=\sum_{a,b=\pm1} {\cal T}(v_1^a,v_2^b;\b,\g,t) f(q^{\frac{a}2} v_1,q^{\frac{b}2} v_2)\,.
} For more details we refer to~\GaiottoUSA. 
The functions $\phi_\lambda$ can be understood as defined by the relation \feif.  
The different functions satisfy some relations. For example, since minimal puncture can be obtained by RG flows from maximal punctures, taking residues of $\psi_\lambda$ one should obtain $\phi_\lambda$ \GaiottoUSA. As additional neat feature let us consider the splitting of maximal punctures into pairs of minimal punctures when $G^{max}$ is $so(5)u(1)$. When we go on the conformal manifold of these models and break the $u(1)_{\beta}\times u(1)_{\gamma}$ symmetry, the minimal punctures, as we have seen, look as half a maximal puncture. Conversely, the deformations breaking $su(2)_{\beta\gamma}$ ($su(2)_{\beta/\gamma}$) split the maximal punctures of first (second) color into two minimal punctures. Turning  deformations splitting both colors breaks $u(1)_\beta$ and $u(1)_\gamma$.  Taking $\beta=\gamma^{-1}$ or $\beta=\gamma$ we expect to obtain relations between functions $\psi$ and $\phi$.  
Here we take take both specializations, $\beta=\gamma=1$, and obtain first,

\eqn\eire{
\phi_\lambda(\alpha;t,1,1)\phi_\lambda(\delta;t,1,1)= C_\lambda(t) \psi_\lambda(\delta\alpha,\delta/\alpha;t,1,1)\,.
} This is the manifestation of two simple punctures combining into a maximal one when $\beta=\gamma=1$.  Explicitly computing the eigenfunction one can verify this property~\GaiottoUSA, at least in some limits of the fugacities ($p$ or $q$ vanish). The proportionality factor $C_\lambda$ can be fixed by studying flows triggered by vacuum expectation values. 
The index of the $G^{max}$ equal $so(5)u(1)$  sphere with three maximal punctures is then,

\eqn\estr{
{\cal I}= \sum_\lambda C_\lambda(t) \psi_\lambda({\bf u};t,1,1)\psi_\lambda({\bf v};t,1,1)\psi_\lambda({\bf w};t,1,1)\,.
} Note that this implies a neat product relation on the eigenfunctions. Defining $\hat \psi_\lambda =C_\lambda \psi_\lambda$ we obtain that,

\eqn\mulrel{\eqalign{
&\hat\psi_\lambda({\bf u};t,1,1)\hat\psi_\lambda({\bf v};t,1,1) =(q;q)^2(p;p)^2\oint\frac{dz_1}{4\pi i z_1}\oint \frac{dz_2}{4\pi i z_2} \cr&\;\;\;\;\;\;\;\;\; \frac{\Gamma_e(\frac{p\,q}t z_1^{\pm1}z_2^{\pm1}\gamma^{\pm2})}{\Gamma_e(z_1^{\pm2})\Gamma_e(z_2^{\pm2})}
{\cal I}_{{\bf u},{\bf v},\sqrt{z_1z_2},\sqrt{z_2/z_1}}(t,1,1)\,
\hat\psi_\lambda({\bf z};t,1,1)\,.}} We can take advantage of this non linear integral equation in principle to solve for the eigenfunctions in similar manner to what was done in \RazamatQFA.
 The index of a general theory with $G^{max}$ being $so(5)u(1)$ is,

\eqn\estri{
{\cal I}_{g,s}= \sum_\lambda (C_\lambda(t,\gamma) )^{2g-2+s}\prod_{j=1}^s\psi_\lambda({\bf u_j};t,1,1)
\,.
}

\

\listrefs
\end